\documentclass[a4paper, amsfonts, amssymb, amsmath, reprint, showkeys, longbibliography, twoside, superscriptaddress, aps, pra]{revtex4-2}
\usepackage{graphicx}
\usepackage[dvipsnames]{xcolor}
\usepackage{txfonts}
\usepackage{comment}
\usepackage{booktabs}
\usepackage{siunitx}
\usepackage{microtype}
\usepackage[export]{adjustbox}

\newcommand{\epolarization}{\lambda}
\newcommand{\polarization}{\gamma}
\usepackage{nicefrac}

\usepackage{hyperref}
\hypersetup{
    colorlinks = true,
    linkcolor  = blue,
    citecolor  = blue,
    urlcolor   = blue,
    pdftitle   = QUOPS
}

\usepackage{mathrsfs}
\usepackage{physics}
\usepackage{quantikz}
\usepackage{tabularray}
\usepackage{booktabs}
\usepackage{multirow}
\usepackage[inline]{enumitem} 
\usepackage[normalem]{ulem}
\usepackage{pifont} 
\newcommand{\cmark}{\ding{51}}%
\newcommand{\xmark}{\ding{55}}%
\usepackage{rotating}

\usepackage{stmaryrd}
\newcommand{\qeccode}[3]{\llbracket {{#1}, {#2}, {#3}} \rrbracket}

\usepackage{xspace}
\usepackage[shortcuts]{extdash} 
\newcommand{\boston}{\texttt{ibm\_boston}\xspace}
\newcommand{\willow}{\texttt{Willow}\xspace}
\newcommand{\helios}{\texttt{Helios\=/1}\xspace}
\newcommand{\htwoone}{\texttt{H2\=/1}\xspace}

\begin{document}
\title{Benchmarking the computational power of quantum computers}
\author{Timothy Proctor}
\thanks{tjproct@sandia.gov}
\affiliation{Quantum Performance Laboratory, Sandia National Laboratories, Livermore, CA 94550, USA}
\author{Oliver Hart}
\affiliation{Quantinuum, Terrington House, 13--15 Hills Road, Cambridge CB2 1NL, UK}
\author{Oliver Widzowski Maupin}
\affiliation{Quantum Performance Laboratory, Sandia National Laboratories, Livermore, CA 94550, USA}
\author{Matthew Girling}
\affiliation{Quantinuum, Terrington House, 13--15 Hills Road, Cambridge CB2 1NL, UK}
\author{Daniel Hothem}
\affiliation{Quantum Performance Laboratory, Sandia National Laboratories, Livermore, CA 94550, USA}
\author{Daniel Mills}
\affiliation{Quantinuum, Terrington House, 13--15 Hills Road, Cambridge CB2 1NL, UK}
\author{Jordan Hines}
\affiliation{Quantum Performance Laboratory, Sandia National Laboratories, Albuquerque, NM 87185, USA}
\author{Karl Mayer}
\affiliation{Quantinuum, 303 S. Technology Ct., Broomfield, Colorado 80021, USA}
\author{Jacob S. Nelson}
\affiliation{Quantum New Mexico Institute (QNM-I), Department of Physics and Astronomy, University of New Mexico, Albuquerque, New Mexico 87131, USA}
\affiliation{Quantum Algorithms and Applications Collaboratory (QuAAC), Sandia National Laboratories, Albuquerque, New Mexico 87185, USA}
\author{Tyler LeBlond}
\affiliation{Quantinuum, 303 S. Technology Ct., Broomfield, Colorado 80021, USA}
\author{Zohim Chandani}
\affiliation{NVIDIA Corporation, 2788 San Tomas Expressway, Santa Clara, 95051, CA, USA}
\author{Diego Forlivesi}
\affiliation{Quantinuum, Terrington House, 13--15 Hills Road, Cambridge CB2 1NL, UK}
\author{Piper C. Wysocki}
\affiliation{Quantum Performance Laboratory, Sandia National Laboratories, Livermore, CA 94550, USA}
\affiliation{Department of Physics and Astronomy, University of New Mexico, Albuquerque, NM 87106, USA}
\author{Boldizsár Poór}
\affiliation{Quantinuum, 17 Beaumont Street, Oxford OX1 2NA, UK}
\author{Joan M. Dreiling}
\affiliation{Quantinuum, 303 S. Technology Ct., Broomfield, Colorado 80021, USA}
\author{Annie Park}
\affiliation{Quantinuum, 303 S. Technology Ct., Broomfield, Colorado 80021, USA}
\author{Adam P. Reed}
\affiliation{Quantinuum, 303 S. Technology Ct., Broomfield, Colorado 80021, USA}
\author{Brian Estey}
\affiliation{Quantinuum, 303 S. Technology Ct., Broomfield, Colorado 80021, USA}
\author{Cameron Foltz}
\affiliation{Quantinuum, 303 S. Technology Ct., Broomfield, Colorado 80021, USA}
\author{Akhil Isanaka}
\affiliation{Quantinuum, 303 S. Technology Ct., Broomfield, Colorado 80021, USA}
\author{M. S. Allman}
\affiliation{Quantinuum, 303 S. Technology Ct., Broomfield, Colorado 80021, USA}
\author{Michael Mills}
\affiliation{Quantinuum, 303 S. Technology Ct., Broomfield, Colorado 80021, USA}
\author{Maxwell D. Urmey}
\affiliation{Quantinuum, 303 S. Technology Ct., Broomfield, Colorado 80021, USA}
\author{Peter E. Siegfried}
\affiliation{Quantinuum, 303 S. Technology Ct., Broomfield, Colorado 80021, USA}
\author{Audrey Faricy}
\affiliation{Quantinuum, 303 S. Technology Ct., Broomfield, Colorado 80021, USA}
\author{Jin-Sung Kim}
\affiliation{NVIDIA Corporation, 2788 San Tomas Expressway, Santa Clara, 95051, CA, USA}
\author{Cristina C\^irstoiu}
\affiliation{Quantinuum, Terrington House, 13--15 Hills Road, Cambridge CB2 1NL, UK}
\author{Andrew D. Baczewski}
\affiliation{Quantum New Mexico Institute (QNM-I), Department of Physics and Astronomy, University of New Mexico, Albuquerque, New Mexico 87131, USA}
\affiliation{Quantum Algorithms and Applications Collaboratory (QuAAC), Sandia National Laboratories, Albuquerque, New Mexico 87185, USA}
\author{Charles H. Baldwin}
\affiliation{Quantinuum, 303 S. Technology Ct., Broomfield, Colorado 80021, USA}
\author{Robin Blume-Kohout}
\affiliation{Quantum Performance Laboratory, Sandia National Laboratories, Albuquerque, NM 87185, USA}
\affiliation{Quantum New Mexico Institute (QNM-I), Department of Physics and Astronomy, University of New Mexico, Albuquerque, New Mexico 87131, USA}

\date{September 2026}
\begin{abstract}
Quantum computing hardware is advancing rapidly toward utility-scale machines that will enable scientific breakthroughs. Many teams are pursuing distinct and difficult-to-compare routes to this goal, using different qubit technologies and logical architectures. Tracking progress toward quantum utility therefore requires rigorous benchmarks that measure computational capability relative to utility-scale challenge problems and enable fair comparison across disparate platforms. Here we demonstrate direct, cross-platform measurement of quantum computational capability using a new benchmark that quantifies the size of the largest computationally relevant quantum circuits that a machine can execute successfully and the speed at which it can execute them. We apply this quantum universal operation performance system (QUOPS) experimentally to leading processors from Quantinuum, Google, and IBM, computing directly on physical qubits.  Translating state-of-the-art resource requirements for recognized challenge problems that represent useful quantum computation into effective QUOPS circuit sizes shows that computational capability must grow by 5 orders of magnitude, motivating fault-tolerant approaches. We use the same benchmark to assess the performance of a simple fault-tolerant logical-qubit processor implemented on up to eight $\llbracket 7,1,3 \rrbracket$-encoded logical qubits using Quantinuum \helios, and project the growth of capability across successive generations of fault-tolerant quantum computers to show how QUOPS can track progress toward quantum scientific utility.
\end{abstract}

\maketitle
\setlength{\parskip}{4pt}
The development of increasingly powerful quantum computers based on several qubit technologies is one of the most active frontiers in science \cite{Arute2019-mk, Google-Quantum-AI-and-Collaborators2025-ad, Daguerre2025-cs, Gupta2024-pr, Bluvstein2023-dp, dasu2026computingencodedlogicalqubits, Tham2026-br, Kim2023-si, Ransford2026-ny}. Existing quantum computers have validated the foundational technologies necessary for further scaling \cite{Arute2019-mk, Google-Quantum-AI-and-Collaborators2025-ad, Daguerre2025-cs, Gupta2024-pr, Bluvstein2023-dp, Tham2026-br,dasu2026computingencodedlogicalqubits} by demonstrating super-classical computation~\cite{Arute2019-mk,DeCrossPRX2025}, beyond-threshold quantum error correction (QEC)~\cite{Google-Quantum-AI-and-Collaborators2025-ad}, and the building blocks of fault-tolerant quantum computation (FTQC)~\cite{Google-Quantum-AI-and-Collaborators2025-ad, Daguerre2025-cs, Gupta2024-pr, Bluvstein2023-dp, Tham2026-br,dasu2026computingencodedlogicalqubits}. However, executing the quantum programs known to deliver transformative applications requires quantum computers whose size and capability exceed those of current systems by orders of magnitude~\cite{Proctor2025-cd, berry2024analyzing, rubin2024quantum, jordan2025optimization, omanakuttan2025threshold, babbush2026grand, Gidney2025-hg, Low2025-nc, CampbellQST22, childs2018toward}. Tracking, guiding, and sustaining progress towards these scientific goals requires the ability to measure the computational power of increasingly sophisticated quantum computers with diverse physical and logical architectures \cite{Proctor2025-cd}.

Most existing methods for assessing quantum computer performance measure physical qubit and logic gate errors rather than computational capability \cite{Proctor2025-cd, Hashim2025-rz}. Randomized benchmarking \cite{Emerson2007-am, Knill2008-jf, Hashim2025-rz} measures mean gate error rates, and gate set tomography \cite{Madzik2022-jh, Blume-Kohout2017-no} characterizes gate errors in detail, but neither addresses the performance of an integrated quantum computer running large circuits. Quantum volume \cite{Cross2019-ku, Hashim2025-rz, BaldwinQuantum2022} was the field’s flagship system-level benchmark, but its classical verification cost scales exponentially with qubit number and its metric is only loosely connected to the requirements of scientific utility. Application-based benchmarks have been proposed \cite{Proctor2025-cd, Lubinski2023-zy}, but they face similar limitations, because the applications executable on contemporary hardware are not representative of utility-scale challenge problems. What is missing is a benchmark of quantum computational capability that is directly comparable to evolving resource estimates for scientific utility, independent of any specific application, and measurable on as-built hardware across diverse current and future architectures.

\begin{figure*}
    \centering
    \includegraphics[width=18cm, valign=c]{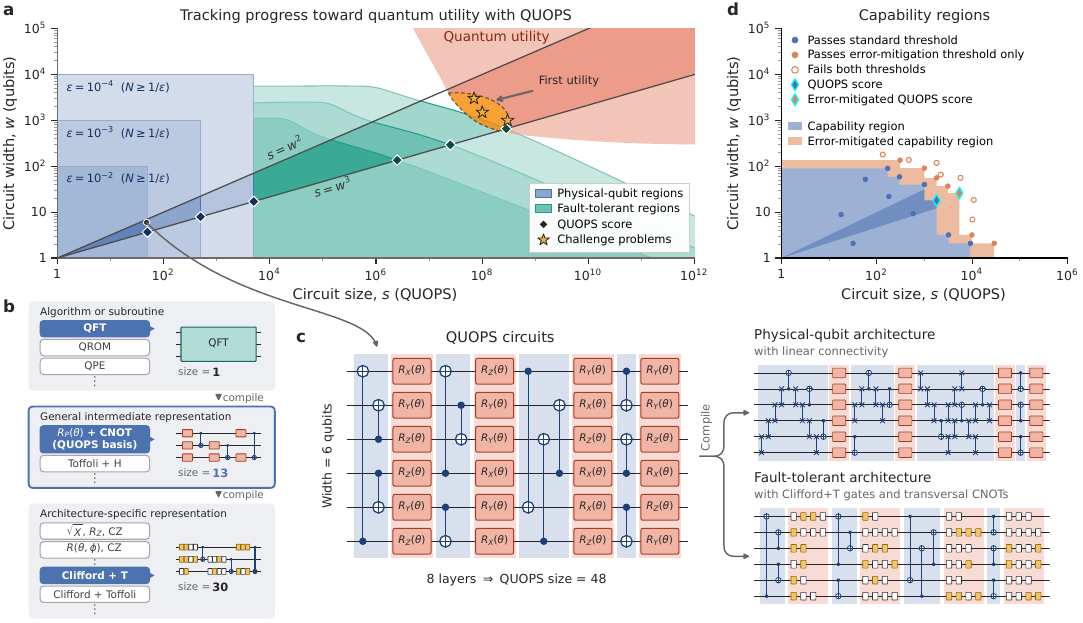}
    \caption{\small{\textbf{QUOPS tracks progress toward quantum utility.} (\textbf{a}) Solving useful and classically intractable problems (red region) requires executing quantum circuits with large width $w$ (number of qubits) and size $s$ (number of gates). Circuits for canonical challenge problems (stars) indicate where first utility is likely to be achieved. A computer’s computational capability is represented by its \textit{capability region} (blue and green regions): the shapes of the circuits it can execute successfully. Its \textit{QUOPS score} is the largest successful circuit size within the cone $w^2 \leq s \leq w^3$. (\textbf{b}) Circuit size depends on the gate basis used. QUOPS uses medium-complexity gates that often appear in algorithms and must be compiled to each architecture. (\textbf{c}) QUOPS circuits consist of alternating layers of random single-qubit $R_P(\theta)$ rotations (orange) and CNOT gates (blue) on $w$ qubits. Executing QUOPS circuits involves architecture-specific compiling and routing. (\textbf{d}) Capability regions are measured by estimating the mean polarization of random QUOPS circuits using mirror circuit fidelity estimation, a technique that enables efficient polarization estimation even for classically hard circuits. Blue circles denote shapes passing a standard threshold, mean polarization $\geq 1/\sqrt{e}$, with 95\% confidence. Hollow circles denote failures.  Error-mitigated capability regions (orange) indicate circuits from which information could be extracted with a predicted sampling overhead of approximately $1/\alpha^2$ (captured by the \textit{QUOPS rate} throughput metric).  They are defined by circuit shapes whose estimated mean polarization exceeds a lower threshold $\alpha <1/\sqrt{e}$ (orange circles).}}
    \label{fig:main:quops}
    \makeatletter      
    \def\@currentlabel{\thefigure{a}}\label{fig:main:quops:schematic}   \def\@currentlabel{\thefigure{b}}\label{fig:main:quops:gatesets}
    \def\@currentlabel{\thefigure{c}}\label{fig:main:quops:circuits}
    \def\@currentlabel{\thefigure{d}}\label{fig:main:quops:capability}
    \makeatother
\end{figure*}

\noindent\textbf{Benchmarking computational capability}---Quantum computation is expected to provide utility for classes of problem that are classically intractable (their solution requires very large classical computations), but quantumly feasible (they can be solved by comparatively small quantum programs). The most famous example is integer factoring~\cite{Shor1994-zh}. The difficulty of solving a particular problem within a class depends strongly on its size: factoring 15 is easy for both classical and quantum computers, whereas factoring RSA-2048 is believed to be infeasible for classical supercomputers and doing it on a quantum computer would achieve quantum utility. It is hard to predict which specific ``utility-scale'' problem will be the first to be solved on a quantum computer, because classical intractability depends on ever-advancing classical algorithms, while quantum feasibility is affected by advances in quantum algorithms and quantum hardware. However, there are compelling and plausible candidates for first quantum utility, corresponding to certain well-studied challenge problems for which detailed logical and physical resource estimates have been developed~\cite{Gidney2025-hg,Low2025-nc,rubin2024quantum}. These include calculating properties of particular lattice models~\cite{CampbellQST22,childs2018toward} and strongly correlated molecules~\cite{Low2025-nc}, and factoring large RSA semiprimes~\cite{Gidney2025-hg}. QUOPS (Fig.~\ref{fig:main:quops}) is designed to benchmark progress toward the computational capabilities required to solve problems of this kind.

The exact \emph{width} (number of qubits) and \emph{size} (number of operations) of the first circuits to achieve utility will depend not just on future innovations in quantum hardware and software, but on how program size is defined.  The ``quop''---a single logical qubit for one logical clock cycle---has been used \cite{GidneyQuantum21,PreskillACM25,Menssen2026} as a \textit{de facto} unit, and our methodology's name is intentional homage.  But the computational utility of a single ``quop'' can be \textit{extremely} architecture-dependent.  Different programs are naturally expressed in different kinds of quantum gates (Fig.~\ref{fig:main:quops:gatesets}), which must be further compiled into each computer's native instructions at widely varying and architecture-dependent cost. Simulations of lattice models based on product formulas, one candidate for early quantum utility \cite{childs2018toward,CampbellQST22}, make heavy use of $R_P(\theta)$ gates, which are gates that apply arbitrary-angle ($\theta$) Pauli ($P$) rotations to single qubits.  In contrast, state-of-the-art approaches to chemical simulation or factoring rely heavily on subroutines that are more naturally expressed in terms of Toffoli gates \cite{Gidney2025-hg, Low2025-nc}. While $R_P(\theta)$ gates are cheap and easy on today’s physical-qubit architectures, it is expected that future FTQCs will have to synthesize each one using dozens of $T$ or Toffoli gates \cite{ross2016optimal,kliuchnikov2023shorter}, each of which will be further compiled into tens or hundreds of native operations.  The overhead of implementing $R_P(\theta)$ gates might be reduced with techniques like synthillation~\cite{campbell2017unified} or Hamming weight phasing~\cite{huggins2025fluidallocationsurfacecode}, but it is generally expected that high-precision arbitrary-angle rotations will be among the most expensive operations in any FTQC architecture.

\begin{figure*}
    \centering
    \includegraphics[width=18cm]{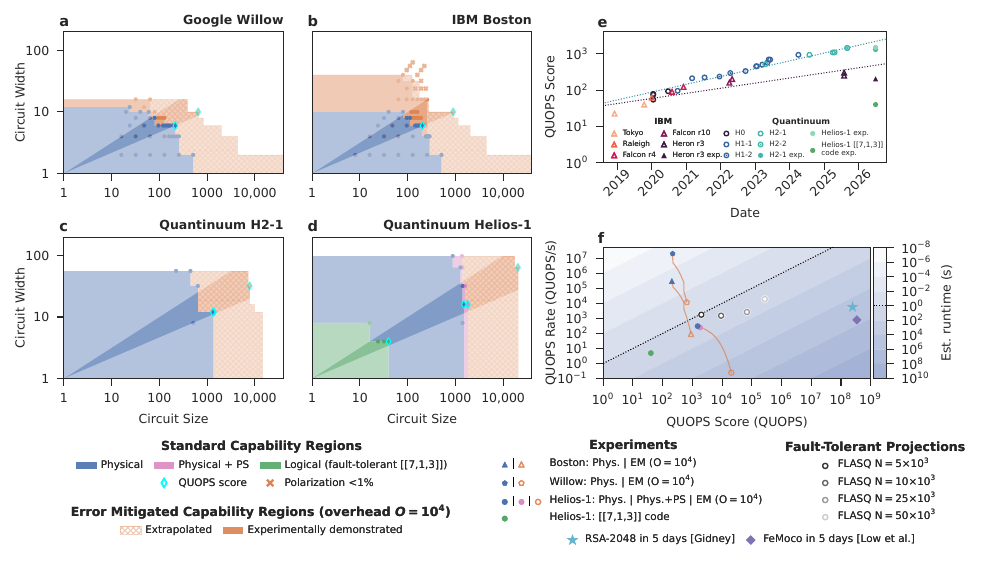}
   \caption{\small{\textbf{Measured capabilities of three state-of-the-art quantum computers.}
(\textbf{a}--\textbf{d}) Experimentally measured QUOPS capability regions for (\textbf{a}) Google's \willow (105 transmon qubits), (\textbf{b}) IBM's \boston (156 transmon qubits), (\textbf{c}) Quantinuum's \htwoone (56 ion qubits), and (\textbf{d}) Quantinuum's \helios (98 ion qubits). Blue regions indicate successful execution of circuits directly on physical qubits, using the standard success threshold of $1/\sqrt{e}$. Pink indicates leakage-postselected performance, and orange indicates error-mitigated capability regions using a $1\%$ success threshold, corresponding to sampling overhead of $10^4$. Pale hatching indicates interpolated or extrapolated capability. The green region in (\textbf{d}) shows the capability of a fault-tolerant architecture using logical qubits encoded in the Steane $\llbracket 7, 1, 3 \rrbracket$ code, chosen for simplicity rather than optimal QUOPS performance. (\textbf{e}) Historical QUOPS scores inferred from quantum volume data for IBM and Quantinuum systems, together with direct QUOPS measurements on \helios, \htwoone, and \boston and the $\llbracket 7, 1, 3 \rrbracket$ fault tolerant processor implemented on \helios. Dashed lines show exponential fits. (\textbf{f}) QUOPS rate versus QUOPS score for the experimentally measured architectures, projected surface-code architectures, and the smallest known requirements for solving RSA-2048 and FeMoco in 5 days. Error mitigation increases QUOPS scores at reduced throughput (orange markers); curves show the continuum of speed--size tradeoffs, achieved by varying the success threshold between $1/\sqrt{e}$ and $1\%$. Dashed lines correspond to a constant number of maximally-sized circuit executions per second.}}
    \label{fig:main:capability}
    \makeatletter
    \def\@currentlabel{\thefigure{a}}\label{fig:main:capability:Willow}
    \def\@currentlabel{\thefigure{b}}\label{fig:main:capability:Boston}
    \def\@currentlabel{\thefigure{c}}\label{fig:main:capability:H2-1}
    \def\@currentlabel{\thefigure{d}}\label{fig:main:capability:Helios}
    \def\@currentlabel{\thefigure{f}}\label{fig:main:capability:quops-vs-rate}
    \def\@currentlabel{\thefigure{e}}\label{fig:main:capability:historical}
    \makeatother
\end{figure*}

\noindent\textbf{The QUOPS benchmark and metrics}---To impose order on this complex landscape, we define a family of randomized benchmark circuits, called QUOPS circuits, built from alternating layers of $R_P(\theta)$ rotations and CNOT gates with independently variable width $w$ and size $s$ (Fig.~\ref{fig:main:quops:circuits}). These circuits are direct proxies for \(R_P(\theta)\)-heavy applications, such as Hamiltonian simulation using product formulas~\cite{childs2018toward}, so successful execution of shape-$(w,s)$ QUOPS circuits provides strong evidence that similarly-shaped Hamiltonian-simulation circuits can be successfully executed. QUOPS circuits can also serve as indirect proxies for applications that primarily use other gates, e.g., Toffoli-heavy circuits---but QUOPS is modular, so other circuit families, such as direct proxies for Toffoli-heavy applications, can also be substituted in.

We capture a computer's computational power by its \emph{QUOPS capability region} (Fig.~\ref{fig:main:quops:capability})~\cite{Proctor2025-cd,Proctor2021-wt}, containing all circuit shapes $(w,s)$ for which random QUOPS circuits can be executed successfully. A shape is deemed successful when the mean process polarization is at least $1/\sqrt{e}\approx 61\%$, as estimated experimentally with 95\% confidence. The random structure and universal gate set of QUOPS circuits make them hard to simulate classically~\cite{Arute2019-mk}, but mirror-circuit techniques~\cite{Proctor2022-zs,Proctor2021-wt} enable efficient and rigorous estimation of their process polarization (see Methods).
 
QUOPS capability regions can track progress toward \emph{any} goal, but we define a single-number \emph{QUOPS score} specifically to track progress toward canonical challenge problems. Analyses \cite{Gidney2025-hg, Low2025-nc} suggest that utility-scale challenge problems require circuits on around $10^3$ qubits with effective QUOPS size between $10^6$ and $10^9$.  We therefore define a machine’s QUOPS score as the QUOPS size $s$ of the \emph{largest} benchmark circuit, within the cone defined by $w^2 \leq s \leq w^3$, that it can execute successfully (Fig.~\ref{fig:main:quops:schematic}). This cone excludes narrow or shallow circuits that admit efficient classical simulation, while capturing the expected shape of (and uncertainty about) first-utility circuits more accurately than quantum volume.  

Circuits with small polarizations can yield computationally useful information when combined with \emph{error mitigation} \cite{Kim2023-si, Cai2023-sj}, but only at the cost of more circuit executions (samples) \cite{Kim2023-si, Cai2023-sj, Takagi2022, Tsubouchi2023-om}. QUOPS captures this capability-vs-throughput trade-off with ``error-mitigated capability'' defined by a lower polarization threshold $\alpha < 1/\sqrt{e}$. The circuit shapes for which the mean process polarization is at least $\alpha$ define an $\alpha$-dependent error-mitigated QUOPS capability region and score (Fig.~\ref{fig:main:quops:capability}). The corresponding sampling overhead, which is approximately $1/\alpha^2$ \cite{Takagi2022, Tsubouchi2023-om}, is reflected in a decreased \emph{QUOPS rate}.

A machine's QUOPS rate measures its throughput, the speed at which it can compute. QUOPS rate is calculated by taking the raw rate at which QUOPS operations are executed and adjusting it for the approximate sampling overhead required to compensate for imperfect circuit polarization and discarded executions when postselection methods, such as error detection, are used. Together, the QUOPS score and rate capture the benefit and the cost of postselection and error mitigation, yielding a more computationally relevant throughput metric than raw rate metrics such as CLOPS~\cite{Wack2021}. The QUOPS rate can be evaluated at any circuit shape $(w,s)$ and it generally varies across the capability region. In our results, we report each QUOPS score together with the QUOPS rate at the circuit shape where that score is achieved. A complete benchmark definition is given in the Methods and Supplementary Information (SI).

\begin{table}[t]
\centering
\begin{ruledtabular}
\begin{tabular}{llrrr}
Platform & Architecture   & QUOPS & QUOPS/s & Width \\
\midrule
\willow
& Physical qubits
& $216$
& $2.0\times 10^{7}$
& $6$
\\
\boston
& Physical qubits
& $204$
& $3.1\times 10^{5}$
& $6$
\\
\htwoone
& Physical qubits
& $1320$
& $353$
& $12$
\\
\htwoone
& Physical qubits + PS
& $1392$
& $231$
& $12$
\\
\helios
& Physical qubits
& $1504$
& $303$
& $16$
\\
\helios
& Physical qubits + PS
& $1824$
& $247$
& $16$
\\
\helios
& $\llbracket 7,1,3\rrbracket$ code
& $40$
& $4.9$
& $4$
\\
\end{tabular}
\end{ruledtabular}
\caption{\textbf{Measured QUOPS scores and rates.}}
\label{tab:measured_extrapolated_quops}
\label{table:results}
\end{table}

\noindent
\textbf{The state of the (physical-qubit) art}---To establish a baseline for the computational capability of current (2026) quantum computers, we applied QUOPS to four of the most advanced quantum processors available---Google's \willow~\cite{Google-Quantum-AI-and-Collaborators2025-ad} (105 transmon qubits, degree-4 connectivity), IBM's \boston (156 transmon qubits, degree-3 connectivity), and Quantinuum's \htwoone and \helios~\cite{Ransford2026-ny} (56 and 98 trapped-ion qubits, respectively, QCCD all-to-all connectivity). QUOPS circuits were executed directly on physical qubits, with transpilation and routing onto subsets of available qubits as necessary to achieve maximum performance. These machines' capabilities are shown by the blue regions of Fig.~\ref{fig:main:capability:Willow}--\ref{fig:main:capability:Helios}. QUOPS scores and rates are given in Table~\ref{table:results}. \willow achieved $216$ QUOPS at $2.0 \times 10^7$ QUOPS/s, \boston achieved $204$ QUOPS at $3.1 \times 10^{5}$ QUOPS/s, \htwoone achieved $1320$ QUOPS at $353$ QUOPS/s, and \helios achieved $1504$ QUOPS at $303$ QUOPS/s.  The capability and/or throughput of \htwoone and \helios can be increased slightly (pink region, Fig.~\ref{fig:main:capability:Helios}) by discarding shots in which leakage errors were detected \cite{Ransford2026-ny}.  On \helios, leakage detection increases throughput at $1504$ QUOPS to $344$ QUOPS/s and enables $1824$ QUOPS at $247$ QUOPS/s. On \htwoone, it increases throughput at $1320$ QUOPS to $360$ QUOPS/s, and enables $1392$ QUOPS at $231$ QUOPS/s.

To explore the value and cost of error mitigation, we used a combination of directly-measured and extrapolated data to identify circuits that are executable with polarization of at least $\alpha = 1\%$ (orange regions in Figs.~\ref{fig:main:capability:Willow}--\ref{fig:main:capability:Helios}). This corresponds to an error mitigation sampling overhead of at most $10^4$. With this significant overhead, \willow achieved $651$ QUOPS at $1.2 \times 10^{4}$ QUOPS/s, while \boston achieved $899$ QUOPS at $91$ QUOPS/s, \htwoone achieved $7796$ QUOPS at $0.09$ QUOPS/s, and \helios reached ${20442}$ QUOPS at $0.29$ QUOPS/s. These 3--11$\times$ increases in QUOPS score come at the cost of 3--4 orders of magnitude reduction in QUOPS rate (Fig.~\ref{fig:main:capability:quops-vs-rate}). However, $\alpha$ can be varied continuously to yield a continuous tradeoff between speed and size (orange lines in Fig.~\ref{fig:main:capability:quops-vs-rate}).

The top-line metrics (Table~\ref{table:results}) and capability regions (Fig.~\ref{fig:main:capability:Willow}-\ref{fig:main:capability:Helios}) reveal the impact of architectural constraints. The qubits in \willow and \boston are arranged on a two-dimensional surface with local connectivity, making CNOT gates between non-adjacent qubits costly and limiting QUOPS capability. Measured capability regions show clearly how maximum circuit size decreases as circuit width is increased. \htwoone and \helios's QCCD trapped-ion architecture has much less of a connectivity penalty, although both of their capability regions exhibit a small dependence of achievable circuit size on width due to ion transport phase errors. These properties are well-known, but understanding their impact on computational performance has required detailed modeling. The QUOPS analysis renders it directly observable. 

Using historical quantum volume data, we estimated the QUOPS scores that would have been achieved by earlier generations of quantum computers. Data from IBM and Quantinuum processors going back to 2018 indicates continual improvement in the QUOPS scores of both trapped-ion and transmon processors (Fig.~\ref{fig:main:capability:historical}). However, the capabilities of physical-qubit architectures are directly limited by the physical-qubit error rate $\epsilon$---achieving approximately $1/(2\epsilon)$ QUOPS, or less in the presence of connectivity constraints---suggesting that this growth will eventually plateau. Furthermore, even continuing the observed trend (doubling QUOPS scores approximately every 1.4 years for Quantinuum and every 2.1 years for IBM) would not reach the capability required for useful challenge problems (see below) until 2050--2070.

\noindent
\textbf{Capability requirements of challenge problems}---To relate QUOPS scores and rates to quantum utility, we converted state-of-the-art resource estimates for two representative challenge problems into approximate QUOPS targets. We analyzed factoring RSA-2048~\cite{Gidney2025-hg,Shor1994-zh} and estimating an energy eigenvalue of the strongly correlated molecule FeMoco~\cite{Low2025-nc}. The most efficient available algorithms for these problems are dominated by Toffoli gates. Current algorithms require width-$1399$ circuits containing $7 \times 10^8$ Toffoli gates for RSA-2048~\cite{Gidney2025-hg}, and width-$1459$ circuits containing $1 \times 10^9$ Toffoli gates for FeMoco~\cite{Low2025-nc}.

To compare these resource estimates to QUOPS scores we used a non-Clifford resource-matching procedure. First, we convert Toffoli gate count to $T$ gate count by assuming that approximately $4\times$ as many $T$ gates as Toffoli gates are needed (exact synthesis of a Toffoli requires 7 $T$ gates, but many of the Toffoli gates in these algorithms are used to compute and uncompute temporary ANDs, which only requires 4 $T$ gates~\cite{gidney2018halving}). Next, we estimate the error rate per logical $T$ gate that would allow the algorithm circuit to be executed with polarization $1/\sqrt{e}$.  Finally, the QUOPS rate attainable by a machine capable of executing $T$ gates with this error rate is estimated by balancing synthesis precision and $T$ gate error in the compilation of $R_P(\theta)$ gates. This analysis is approximate (e.g., it ignores Clifford gate errors in both algorithmic and QUOPS circuits) and partly architecture-dependent (because it assumes all circuits are compiled into $T$ gates), but it enables challenge problem requirements to be compared directly to QUOPS scores.

We find targets of $2.5 \times 10^8$ QUOPS for RSA-2048 and $3.4 \times 10^8$ QUOPS for FeMoco. We also estimate the approximate QUOPS rates required to complete these computations in five days, following Gidney's RSA-2048 runtime~\cite{Gidney2025-hg}. This gives throughputs of $5.7 \times 10^3$ QUOPS/s for RSA-2048, accounting for the multiple circuit executions required for RSA-2048 \cite{Gidney2025-hg}, and $8.0 \times 10^2$ QUOPS/s for FeMoco. Today's physical-qubit processors have (or nearly have) the necessary QUOPS rate. In contrast, current QUOPS scores are insufficient by roughly five orders of magnitude (Fig.~\ref{fig:main:capability:quops-vs-rate}).

\noindent
\textbf{Performance of a fault-tolerant processor}---We benchmark an initial step on the path to utility-scale fault-tolerant computation by applying QUOPS, in experiment, to a fully fault-tolerant architecture implemented on \helios. Up to 8 logical qubits are encoded and protected using a fault-tolerant implementation of the $\llbracket 7, 1, 3 \rrbracket$ Steane code \cite{Steane1996-br}.  QUOPS circuits are compiled into the Clifford+$T$ gate set, with the Clifford gates implemented transversally and the $T$ gates implemented by injecting magic states prepared in a dynamically generated auxiliary Steane code block. The encoded QUOPS circuits are fully fault tolerant at distance three, meaning that no fault at a single circuit location propagates to an error that is uncorrectable by the code.

Logical operations were implemented using Goto’s fault-tolerant logical $|0\rangle$ state preparation circuit~\cite{goto2016minimizing} and an optimized logical $T|+\rangle$ state using flag qubits and repeat until success (RUS). We apply Steane QEC before CNOT gates and measurements where required to ensure fault tolerance, noting that magic state injection circuits couple data and ancillary code blocks with a CNOT gate. This architecture, whose capability is shown by the green region in Fig.~\ref{fig:main:capability:Helios}, achieved $40$ QUOPS at circuit width $4$ with a rate of 4.9 QUOPS/s. The $s=40$ QUOPS circuits were compiled into $20$ or fewer $T$ gates (expected value 10) and subsequently embedded into mirror circuits with double the size.

The Steane-code implementation was developed through three successive experiments. Our initial version achieved $24$ QUOPS. The second used higher fidelity magic states, and per-shot randomization, which almost doubled the QUOPS score to 40. Both of these experiments used adaptive flag QEC after CNOT and $T$ gates, so that any single fault on each gate component remains correctable. For the third implementation, we enforced a stricter notion of composable fault tolerance, enabled by Steane QEC with state preparation circuits that ensure multi-qubit errors propagating through transversal $S$ gates remain correctable when composed with other Clifford gates. In this case we used QEC only where required to ensure fault tolerance of the circuit and improved parallelization. When combined, these improvements increased the QUOPS rate from 1.3 QUOPS/s to 4.9 QUOPS/s. This iterative progress illustrates how the QUOPS benchmark can both capture and guide improvements in software and QEC design that increase the computational power of fixed hardware.

This Steane-code experiment demonstrates QUOPS on a fully fault-tolerant logical architecture but it does not saturate \helios's fault-tolerant capability. A range of techniques have been demonstrated that can increase capability at the cost of complexity, including higher-rate codes~\cite{dasu2026computingencodedlogicalqubits}, higher-fidelity magic states~\cite{dasu2025breakingmagicdemonstrationhighfidelity,Daguerre_2025} that enable higher-precision $R_P(\theta)$ synthesis, and direct fault-tolerant implementation of $\smash{\sqrt{T}}$ gates~\cite{dasu2026flaggingcliffordhierarchyfaulttolerantlogical}.

\begin{figure}
    \centering
    \includegraphics[width=\columnwidth]{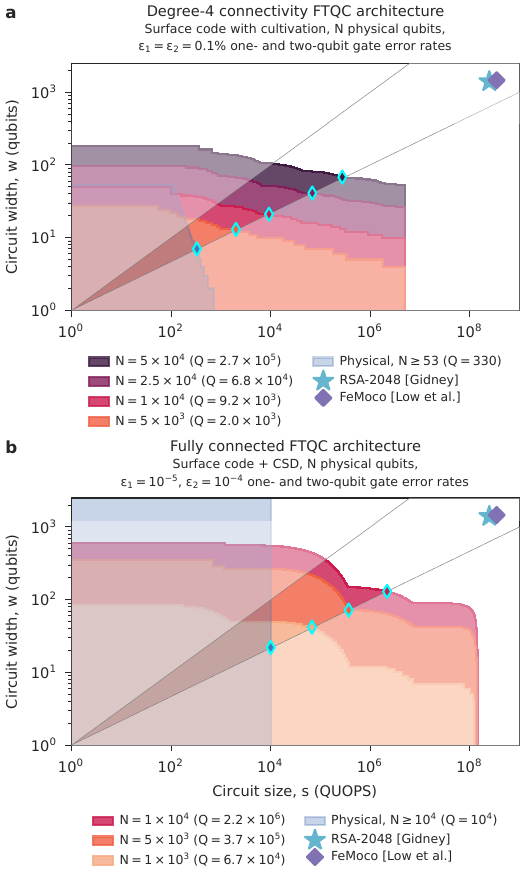}
    \caption{\small{\textbf{Projecting the path to utility.} Projected QUOPS capability regions for two fault-tolerant architecture families: (\textbf{a}) a degree-4 lattice of $N$ qubits arranged into rotated surface code patches, with one- and two-qubit gate error rates $\epsilon_1=\epsilon_2=10^{-3}$, and (\textbf{b}) fully connected physical qubits with $\epsilon_1=10^{-5}$ and $\epsilon_2=10^{-4}$, implementing the optimal choice from surface codes with transversal gates or a CSD code. Capability targets for solving RSA-2048~\cite{Gidney2025-hg} and FeMoco~\cite{Low2025-nc} are shown for comparison. Blue regions show the corresponding capabilities of the same hardware using a physical-qubit architecture, highlighting the much greater capabilities enabled by fault tolerance.}}
    \label{fig:main:projections}
    \makeatletter
    \def\@currentlabel{\thefigure{a}}\label{fig:main:projections:surface_code}
    \def\@currentlabel{\thefigure{b}}\label{fig:main:projections:fully_connected}
    \makeatother
\end{figure}

\noindent
\textbf{Projecting the path to utility}---We projected QUOPS capability regions for two fault-tolerant architecture families and compared them to the QUOPS targets derived from challenge problems (see Fig.~\ref{fig:main:capability:quops-vs-rate}). The first family uses rotated surface code logical qubits on a two-dimensional degree-4 lattice, with lattice surgery CNOT gates and cultivated $T$ states~\cite{Google-Quantum-AI-and-Collaborators2025-ad,Gidney2024-od}. We estimate its capabilities using FLASQ~\cite{huggins2025fluidallocationsurfacecode}. The second family uses fully connected physical qubits and selects the best logical encoding from a small set of candidate codes: surface codes of distance $d=3,5,7,9$ and a \(\qeccode{20}{2}{6}\) concatenated symplectic double code~\cite{berthusen2025simple}.

The FLASQ surface-code architecture with $5\times 10^3$--$5\times 10^4$ physical qubits and a uniform physical gate error rate of $\epsilon=10^{-3}$ achieves QUOPS scores from $2\times 10^3$ to $2.7\times 10^5$ (Fig.~\ref{fig:main:projections:surface_code}). These capability regions cut off around size $10^7$ because the FLASQ architecture relies on $T$-state cultivation. Larger circuit sizes could be reached using distillation, at some cost in width~\cite{Knill2004-me,Litinski2019-nu}. The corresponding QUOPS rates, computed using transmon qubit operation times~\cite{huggins2025fluidallocationsurfacecode}, are shown in Fig.~\ref{fig:main:capability:quops-vs-rate}. The fully connected architecture with $10^3$--$10^4$ physical qubits, single-qubit gate error rates of $\epsilon_1=10^{-5}$, and two-qubit gate error rates of $\epsilon_2=10^{-4}$ achieves QUOPS scores of up to $2.2\times 10^6$ (Fig.~\ref{fig:main:projections:fully_connected}).

These projected capabilities illustrate how effectively fault tolerant architectures can increase computational power. The machines we analyzed here are scaled-up analogues of the extant physical-qubit platforms that we benchmarked experimentally, with similar connectivity and physical error rates but $10\times$--$500\times$ more qubits. Using a physical-qubit architecture would limit the degree-4 connectivity system to approximately $330$ QUOPS and the fully-connected system to about $10^4$ QUOPS, no matter how many physical qubits were available.  Fault tolerant architectures enable projected capabilities orders of magnitude higher, exceeding the capabilities of any plausible physical-qubit architecture. Yet even these projected fault-tolerant capabilities still fall far short of what is needed for RSA-2048 or FeMoco. Closing the remaining gap will require lower error rates, more physical qubits, or more efficient logical architectures than those which we have analyzed.  

\noindent
\textbf{Discussion}---This is, to the best of our knowledge, the first cross-platform benchmarking experiment to directly compare physical- and logical-qubit quantum computations on an identical footing.  The implementation of QUOPS on a fault-tolerant system highlights the overhead cost of fault-tolerant computing---the simple fault-tolerant architecture used here is outperformed by the physical-qubit architecture. But fault-tolerant architectures, unlike physical-qubit implementations, promise rapid growth of capability almost without bound.  Some hardware roadmaps project scientifically transformative capabilities in the early 2030s \cite{USDOE-Office-of-Science-SC-2026-am}, which requires the trend in Fig.~\ref{fig:main:capability:historical} to accelerate by 4$\times$. The path to those capabilities will be costly, creating strong incentives for hype and placing a premium on objective analysis.  Rigorous and well-motivated benchmarking, using QUOPS or similar tools, offers scientists and stakeholders the opportunity to track progress and verify computational capability independently.

The first time a fault-tolerant architecture sets a QUOPS record, beating every physical-qubit architecture, will constitute a major milestone for the field and a clear realization of Gottesman’s criterion for demonstrating fault tolerance~\cite{Gottesman2016-bj}. Achieving it is challenging.  We demonstrated $1500$ QUOPS, and higher scores are attainable, but achieving much beyond $10^4$ QUOPS with physical-qubit operations would be both remarkable and surprising. Our projections suggest that a fault-tolerant architecture encoding $20$--$30$ logical qubits into $1000$--$5000$ physical qubits, using well-understood encodings, could comfortably exceed $10^4$ QUOPS.  Recent advances in QEC and logical architecture \cite{Williamson2026-cb, Yoder2025-ss, Lu2026-un, Yang2026-go, Bhardwaj2026-gr} promise to lower these costs dramatically, and so it is possible this milestone could be rapidly achieved \cite{USDOE-Office-of-Science-SC-2026-am}.  As capabilities grow in the fault-tolerant era, it will become increasingly necessary to understand what computations an as-built (or proposed) quantum computer can perform. QUOPS enables these comparisons, but it need not be the last word. QUOPS is modular and flexible (see SI) to support innovation in tracking and measuring the path to scientifically transformative quantum computation.

\noindent
\textbf{Acknowledgments}---This material is based upon work supported in part by the U.S. Department of Energy, Office of Science, National Quantum Information Science Research Centers, Quantum Systems Accelerator (Award No.~DE-SCL0000121), and by the Laboratory Directed Research and Development program at Sandia National Laboratories. Sandia National Laboratories is a multi-program laboratory managed and operated by National Technology and Engineering Solutions of Sandia, LLC., a wholly owned subsidiary of Honeywell International, Inc., for the U.S.\ Department of Energy's National Nuclear Security Administration under contract DE-NA-0003525. All statements of fact, opinion or conclusions contained herein are those of the authors and should not be construed as representing the official views or policies of the U.S.\ Department of Energy or the U.S.\ Government. This research used IBM Quantum resources of the Air Force  Research Laboratory. OH and MG thank Pablo Andres-Martinez for useful discussions on automated fault-tolerant compilation, as well as Chris N.\ Self for designing and optimizing the magic state preparation circuits used in the final Steane code experiments. DF and OH thank David Amaro for useful discussions on frame tracking and fully fault tolerant encodings. We thank Ryan Byrne and Eliott Rosenberg for running the QUOPS experiments on Google Willow and for providing assistance in the analysis of the Google data. We thank Setso Metodi for programmatic management and helpful comments throughout.

\small
\let\oldaddcontentsline\addcontentsline
\renewcommand{\addcontentsline}[3]{}
\section*{Methods} 
\let\addcontentsline\oldaddcontentsline
\noindent

\noindent
\textbf{QUOPS definition}---A concise summary of the QUOPS methodology used to obtain the main results is given below. Full details can be found in the SI.

\noindent
\emph{Benchmark circuits and shapes}---QUOPS circuits (Fig.~\ref{fig:main:quops:circuits}) are parameterized by their shape $(w,s)$, where $w$ is circuit width and $s$ is the circuit size. A circuit's width is the number of qubits on which it acts. Its size is the number of single-qubit $R_P(\theta)$ gates (rotations about Pauli axes) plus $2\times$ the number of two-qubit CNOT gates that it contains. The circuit shapes tested may be selected freely and typically should be chosen to map out the edge of a system's capability region and find the point that maximizes the QUOPS score.

\noindent
\emph{Circuit sampling and compilation}---For each selected shape, random QUOPS circuits $c$ are sampled and then compiled into architecture-specific circuits $g(c)$ over some unitary gate set, such as Clifford+$T$. Compilation can include routing, gate synthesis and other exact or approximate compilation techniques, serialization of parallel gates, and ancilla use. Compilation may be approximate, so that the unitary superoperator $\mathcal{U}(g(c))$ implemented by $g(c)$ is not exactly equal to $\mathcal{U}(c)$. 

\noindent
\emph{Success criterion}---A quantum computer's success at executing circuit $c$ is measured by the process polarization $\polarization$ \cite{Hashim2025-rz, Proctor2022-zs} between $\mathcal{U}(c)$ and its compiled experimental implementation $\Lambda(g(c))$:
\begin{equation}
    \polarization = \tfrac{1}{4^n -1}\left(4^n \textrm{Tr}\left[\mathcal{U}(c)^{-1}\Lambda\left(g(c)\right)\right] - 1\right).
\end{equation}
Both noise in the implementation of $g(c)$ and imperfect approximate compilation of $c$ into $g(c)$, e.g.,~from synthesizing $R_P(\theta)$ gates using $T$ and Clifford gates, reduce $\polarization$. For each circuit shape $(w,s)$, our success criterion is $\bar{\polarization}_{w,s} \geq \alpha$ where $\bar{\polarization}_{w,s}$ is the mean process polarization of shape-$(w,s)$ circuits and $\alpha$ is a success threshold that is given by $\alpha =  1/\sqrt{e}$ except when quantifying error-mitigated capabilities.

\noindent
\emph{Estimating mean polarization}---The mean polarization $\bar{\polarization}_{w,s}$ is efficiently estimated using mirror circuits~\cite{Proctor2021-wt, Proctor2022-zs}, which enable efficient estimation of $\bar{\polarization}_{w,s}$ even though QUOPS circuits are hard to classically simulate. The mirror circuits are compiled into a system's low-level computational primitives, following strict limitations on permissible compilation strategies imposed by compilation barriers \cite{Proctor2025-cd} throughout the circuits, and then executed. From the mirror circuit data, a point estimate of each $\bar{\gamma}_{w,d}$, denoted $\hat{\bar{\gamma}}_{w,d}$, is produced using the standard mirror circuit data analysis \cite{Proctor2025-cd}. In addition, a one-sided $\beta$ confidence interval that implies $\bar{\polarization}_{w,s} \in [\hat{\polarization}_{\mathrm{lower},\beta,w,s}\,,\,1]$ with $\beta$ confidence is computed. A shape $(w,s)$ is deemed \textit{successful}, also referred to as passing the QUOPS test, with threshold $\alpha$ and confidence $\beta$ if and only if $\hat{\polarization}_{\mathrm{lower}, \beta,w,s}\geq \alpha$, implying $\bar{\polarization}_{w,s} \ge \alpha$ with $\beta$ confidence. One-sided confidence intervals are computed assuming a normal distribution in the polarization estimator, so $\hat{\polarization}_{\mathrm{lower},\beta,w,s} = \hat{\bar{\gamma}}_{w,d} - \Phi^{-1}(\beta)\sigma_{w,s}$ where $\Phi(\beta)$ is the cumulative density function of the standard normal distribution and $\sigma_{w,s}$ is the estimated standard deviation at shape $(w,s)$.

\noindent
\emph{QUOPS capability regions and scores}---A capability region with threshold $\alpha$ is the set of all circuit shapes such that $\bar{\polarization}_{w,s} \geq \alpha$. Setting $\alpha = 1/\sqrt{e}$ yields the standard QUOPS capability region. When $\alpha < 1/\sqrt{e}$ we call the region an \emph{error-mitigated} QUOPS capability region. The QUOPS score $Q$ is the largest successful circuit size $s$ achieved at any shape $(w,s)$ that satisfies $w^2 \le s \le w^3$, called the admissible cone. The QUOPS score must be estimated with at least 95\% confidence unless otherwise stated, meaning that a system's QUOPS score is overestimated with a probability of at most 5\%. Any statistical procedure that satisfies this criteria can be used. In our analysis we used gated hypothesis tests \cite{Mascha2012-vg}: a sequence of circuit shapes $\mathcal{S} = (w_1,s_1), (w_2,s_2),\dots $, within the admissible cone and with increasing size (i.e., $s_{i+1} > s_{i}$), is chosen to test, in order, at 95\% confidence. The procedure terminates when a circuit shape is reached such that $\hat{\polarization}_{\mathrm{lower},0.95,w,s} < \alpha$, and the estimated QUOPS score is the size of the preceding circuit shape. 

The QUOPS capability region must be estimated with at least 90\% confidence unless otherwise stated, meaning that the estimation procedure produces capability regions that contain any point that is outside the system's true capability region with probability at most 10\%. The QUOPS capability region can be estimated using any procedure that satisfies this statistical criteria. Our procedure includes all circuit shapes from $\mathcal{S}$ that passed the QUOPS test, when estimating the QUOPS score, in the capability region. It also tests all circuit shapes that were not in $\mathcal{S}$, using a procedure with 5\% family-wise error rate (FWER). The simplest such procedure tests all remaining $\zeta$ circuits with confidence $\beta = 1 - 0.05/\zeta$ (a Bonferroni correction). Our analysis uses a slightly more powerful test: the Hochberg procedure \cite{Hochberg1988}. The estimated capability region then contains all circuit shapes that are smaller than or equal to any circuit shape for which the hypothesis tests conclude that $\bar{\gamma}_{w,s} \geq \alpha$. This is a 90\% confidence estimate of the QUOPS capability region assuming that $\bar{\gamma}_{w,s}$ is non-increasing with increasing circuit width or size. 

Error-mitigated QUOPS capability regions and scores ($\alpha < 1/\sqrt{e}$) can be estimated or extrapolated, with extrapolation necessary if circuits with low polarizations were not run. The extrapolation procedure used is based on maximum likelihood fitting of the mean polarization data to $\bar{\gamma}_{w,s} = \exp( -r_{w} s)$ for a $w$-dependent rate $r_{w}$. Details are given in the SI. Note that extrapolated capability regions and scores might not be achievable in practice, for reasons including that they can contain circuit sizes that are larger than any executed circuit. Estimated error-mitigated QUOPS scores are denoted $Q^*_{\alpha}$ where $\alpha$ is the corresponding threshold. When error-mitigated QUOPS scores are reported they should be stated alongside the $\alpha$ used, the corresponding sample overhead $\mathcal{O} = 1/\alpha^2$, and the QUOPS rate at the circuit shape at which $Q^*_{\alpha}$ is obtained.

\noindent
\emph{QUOPS rates}---The theoretical QUOPS rate at circuit shape $(w,s)$ is defined by $\Omega(w,s)=s\bar{\polarization}_{w,s}^2/\tau$, where $\tau$ is the execution time of a QUOPS circuit of shape $(w,s)$. $\Omega$ attenuates the raw QUOPS rate $s/\tau$ by $\bar{\polarization}_{w,s}^2$, which is the estimated sampling overhead required to mitigate unheralded errors using standard methods \cite{Takagi2022, Tsubouchi2023-om}. In practical use, we must adjust this definition because the QUOPS benchmark executes mirror circuits created from QUOPS circuits, which are approximately twice the size of the QUOPS circuits, and because the definition of $\tau$ is complicated by various overheads. The operational definition for the QUOPS rate is 
\begin{equation}
\Omega(w,s) = \frac{2 s \, \hat{\bar{\polarization}}_{w, s}^2 }{\tau_{\text{wall}}} \sqrt{\kappa_\text{total} \kappa_\text{kept}}
\end{equation}
where $\tau_{\text{wall}}$ is the time taken running all mirror circuits for QUOPS shape $(w,s)$, $\kappa_\text{total}$ is the total number of circuit executions, and $\kappa_\text{kept}$ is the number of samples kept. This formula accounts for the larger size of the mirror circuits and includes other time overheads, such as the time taken to switch between circuits. The QUOPS rate should be reported alongside the QUOPS score achieved at the same circuit shape. However, note that the QUOPS rate might be maximized at a different circuit shape to the QUOPS score, and QUOPS score and rate pairs can be reported from any point within the intersection of the capability region and the admissible cone.

\noindent
\textbf{Data and code availability}---QUOPS has been implemented in CUDA-Q \cite{cudaq}, Guppy~\cite{Koch2024guppy}, pytket \cite{Sivarajah_2020}, and pyGSTi \cite{pygsti,Nielsen2020-rd}. All four implementations are open-source. All experimental and simulated data presented in this paper and associated code will be released publicly at \cite{PRAQTICE} prior to publication.

\let\oldaddcontentsline\addcontentsline
\renewcommand{\addcontentsline}[3]{}
\bibliographystyle{apsrev4-2-trunc}
\bibliography{bib}
\let\addcontentsline\oldaddcontentsline

\newpage
\setlength{\parskip}{2pt}
\let\oldaddcontentsline\addcontentsline
\renewcommand{\addcontentsline}[3]{}
\section*{SUPPLEMENTARY INFORMATION}
\let\addcontentsline\oldaddcontentsline
\tableofcontents
\setlength{\parskip}{4pt}

\section{Definitions}\label{sec:definitions}
This section provides formal definitions of terms, notation, and mathematical objects that are used throughout the Supplementary Information (SI), the main text, and the Methods. 

\subsection{Quantum circuits}\label{ssec:circuits_definitions}
In this paper, a quantum circuit $c$ is a sequence of layers $l_1$, \dots, $l_d$ that each consist of quantum gates from some gate set $G$. When a circuit is written as a list of layers, the right-most layer is the first to be performed, in order to reflect the rules of matrix multiplication. So the sequence of layers $l_1$ to $l_d$, with $l_1$ performed first, is denoted by $c = l_d\cdots l_2l_1$. Each pair of gates in a layer act on disjoint sets of qubits. Some qubits may not be acted on by any gate. If the ideal action of every gate in an $n$-qubit circuit $c$ is unitary, then the ideal action of $c$ is a unitary that we denote by $U(c) \in \text{SU}(2^n)$ where 
\begin{equation}
 U(c)=U(l_d)\cdots U(l_1).   
\end{equation}
Note that a circuit $c$ is not the same as the unitary $U(c)$ that it implements, but is instead defined by the sequence of layers it contains. The superoperator representation of $U(c)$ is denoted by $\mathcal{U}(c)$, i.e., $\mathcal{U}(c)[\rho] = U \rho U^{\dagger}$.

We refer often to a circuit's \textit{depth}, \textit{width}, and \textit{size}. In this work, a circuit's width is the number of distinct qubits appearing in it, and its \emph{depth} is the number of layers in the circuit. Note that other definitions of ``depth'' are used elsewhere, and our definition does not correspond to the minimum number of layers needed to implement the circuit's unitary $U(c)$. A circuit's \emph{size} is determined by the number of gates appearing in it, but can be usefully defined in different ways by attributing different sizes to different gates. We formalize this idea with a function $\kappa(g)$ that defines the size of each gate $g$ in a gate set $G$. A circuit's $\kappa$-\emph{size} is then
\begin{equation}
    s_{\kappa}(c) = \sum_{g \in G} \kappa(g) \mathcal{N}_g(c)
    \label{eq:circuit-kappa-size}
\end{equation}
where $\mathcal{N}_g(c)$ is the number of times that $g$ appears in $c$. In this work, the primary definition for circuit size is given by $\kappa(g) =k$ if $g$ is a $k$-qubit gate. Empty circuit locations (i.e., an idling qubit) are not considered to be a gate. Circuit size without further qualification refers to this particular definition of size. It is also sometimes useful to count the instances of one gate $g$ in a circuit (e.g., the number of $T$ gates in a circuit), using $\kappa(g') = \delta_{g,g'}$. We call this the circuit's $g$-size or $g$-count.

\subsection{Process fidelity and polarization}\label{ssec:fidelity_definitions}
QUOPS uses \emph{process polarization}, a rescaling of process fidelity, to quantify how well a circuit is executed. The process fidelity (also called the entanglement fidelity) between an $n$-qubit unitary superoperator $\mathcal{U}$ and an $n$-qubit CPTP superoperator $\Lambda$ is defined by \cite{Hashim2025-rz, Nielsen2002-iu, Horodecki1999-rk}
\begin{equation}
    F(\mathcal{U}, \Lambda) =  \langle \Psi_e| (\Lambda \mathcal{U}^{\dagger} \otimes I)  [ | \Psi_e \rangle \langle \Psi_e|]  | \Psi_e \rangle,
\end{equation}
where $I$ is the $n$-qubit identity superoperator (i.e., $I(\rho)=\rho$) and $| \Psi_e \rangle$ is any state in $\mathcal{H}(2^n) \otimes \mathcal{H}(2^n)$ that is maximally entangled between these two systems, where $\mathcal{H}(2^n)$ is the $n$-qubit Hilbert space. Here and throughout, composition of superoperators is denoted by multiplication, so $\Lambda$ composed with $\mathcal{U}^{\dagger}$ is given by $\Lambda \mathcal{U}^{\dagger}$. Process fidelity is closely related to average gate fidelity \cite{Nielsen2002-iu, Horodecki1999-rk}, another widely used quality metric \cite{Hashim2025-rz}, and the difference between them for $n$ qubits is $O(1/4^n)$.

Process polarization is a rescaling of process fidelity that is defined by \cite{Hashim2025-rz}
\begin{equation}
    \polarization(\mathcal{U}, \Lambda) = \frac{4^n F(\mathcal{U}, \Lambda) - 1 }{ 4^n -1}.
\end{equation}
As long as $\Lambda$ is CPTP, this can also be written as
\begin{equation}
    \polarization(\mathcal{U}, \Lambda) = \frac{4^n \textrm{Tr}(\mathcal{U}^{-1}\Lambda) - 1 }{ 4^n -1},
\end{equation}
which is the expression given in the Methods. Process polarization is more useful than process fidelity in our context because, as circuit depth grows, $\polarization \to 0$ (whereas $F\to 1/4^n$). In particular, if $\Lambda=\mathcal{D}_{p}\mathcal{U}$, where $\mathcal{D}_{p}[\rho] = p \rho + (1-p) \mathbb{I}/2^n$ is an $n$-qubit depolarizing channel, then $\polarization(\mathcal{U},\Lambda) = p$. Note, however, that the difference between $\polarization$ and $F$ is negligible except for small $n$.

\subsection{Random circuits}\label{ssec:random_circuits_definitions}
The circuits used in QUOPS are \emph{random} and so we use notation for random circuits and expectation values over ensembles of random circuits. In particular, it is often useful to use a \emph{circuit-valued random variable} to denote a random circuit. We use upper-case letters to denote random variables. If $C$ is a circuit-valued random variable it is defined by a distribution $P$ over some set of circuits $\mathbb{C}$. Functions of these random variables are themselves random variables, e.g., $\mathcal{U}(C)$ is a superoperator-valued random variable. We use $\mathbb{E}\{\cdot\}$ to denote the expectation value of random variables, e.g., $\mathbb{E}\{F(\mathcal{U}(C),\Lambda(C))\}$ is the mean process fidelity between the noisy and ideal implementation of the random circuit $C$. When an expression contains multiple random variables and an expectation value is taken only over some of those random variables this is denoted by a subscript on $\mathbb{E}$, e.g., $\mathbb{E}_A\{A +B\}$ is an expectation value only over $A$.

\section{QUOPS circuits}\label{sec:quops_circuits}
This section contains definitions of the QUOPS gate set (Section~\ref{ssec:gate_set}), QUOPS circuits (Section~\ref{ssec:quops_circuits}), and the kinds of compilation of these circuits that are permitted in the benchmark (Section~\ref{ssec:compilation}). The QUOPS benchmark is intended to be executable on many different architectures, to summarize the computational capabilities of a specific system, and to enable direct comparisons between different architectures. These principles underpin the choices made for the gate set, circuit structure, and compilation rules.

\begin{figure}
    \centering
    \includegraphics[width=\linewidth]{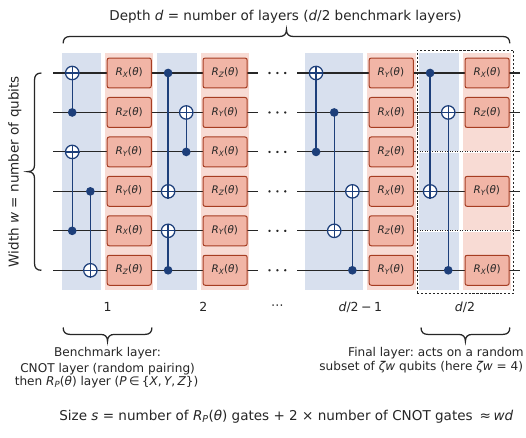}
    \caption{\small{\textbf{QUOPS circuits.} A schematic of QUOPS circuits, which are the random circuits used in the QUOPS benchmark. A complete definition of the QUOPS circuits is provided in Section~\ref{ssec:quops_circuits}.}}
    \label{fig:quops_circuits}
\end{figure}

\subsection{QUOPS gate set}\label{ssec:gate_set}
The QUOPS benchmark is designed to be executable on many different quantum computing architectures and produce scores that can be directly compared across architectures. We therefore define QUOPS circuits using an architecture-independent gate set. A QUOPS circuit (see Fig.~\ref{fig:quops_circuits}) contains gates from the gate set $G_{\textrm{QUOPS}}$ consisting of CNOT gates between all pairs of qubits (``all-to-all CNOTs'') and $R_P(\theta)$ single-qubit gates on each qubit, where $R_P(\theta)$ is a rotation by angle $\theta$ around the Pauli axis $P \in \{X, Y, Z\}$, corresponding to the unitary
\begin{equation}
    U\left(R_P(\theta)\right) = \exp(-i\theta P/2), 
\end{equation}
where $P$ denotes the corresponding Pauli matrix (e.g., $Z = \ket{0}\bra{0} -\ket{1}\bra{1}$).

The QUOPS gate set was chosen to reflect the operations used in algorithms rather than the constraints of any particular hardware. Translating circuits from one gate set to another generally incurs overheads, and for many architectures compiling circuits over $G_{\textrm{QUOPS}}$ into native operations incurs significant ones. On most physical-qubit architectures the single-qubit gates in $G_{\textrm{QUOPS}}$ are cheap, because, e.g., $\pi/2$ rotations about $X$ and arbitrary rotations about $Z$ are typically available, but restricted two-qubit connectivity (e.g., a square lattice) means that CNOT gates between distant qubits require SWAP chains (or CNOT ladders). These overheads are an intentional feature of QUOPS, because they approximately reflect the overheads of compiling realistic algorithms, although all-to-all CNOT circuits are likely more expensive to compile than a typical algorithmic circuit expressed in $G_{\textrm{QUOPS}}$. We therefore encourage also reporting properties (e.g., the size) of compiled QUOPS circuits, which could be compared with algorithm resource estimates expressed in the same architecture-specific gate set where such estimates exist.

\subsection{QUOPS circuits}\label{ssec:quops_circuits}
A QUOPS circuit (see Fig.~\ref{fig:quops_circuits}) is a random circuit over $G_{\textrm{QUOPS}}$ that can be described using three variables: \emph{width} ($w$, an integer), depth ($d$, an even integer), and \emph{final layer filling fraction} ($\zeta \in \frac{1}{w}, \frac{2}{w},\dots,1$). A QUOPS circuit acts on a set of $w$ \emph{computational qubits}, and it consists of $d/2$ pairs of layers of gates. We call each pair of layers a \emph{benchmark layer}. Each of the first $d/2-1$ benchmark layers of gates consist of:
\begin{enumerate}
\item A layer applying CNOT gates between an independent and uniformly sampled pairing of the qubits (leaving one uniformly random qubit out if $w$ is odd). \item A layer consisting of an independent and uniformly sampled $R_P(\theta)$ gate on each qubit, where each $P$ is an i.i.d.~sample from $\{X, Y, Z\}$ and each $\theta$ is an i.i.d.~sample from the uniform distribution on $[0,2\pi)$.   
\end{enumerate}
The final layer of a QUOPS circuit consists of uniformly sampling $\zeta w$ of the qubits to include in the final layer, and then applying the above benchmarking layer sampling procedure to only those qubits, which results in a partially filled final layer if $\zeta < 1$.

The size of a QUOPS circuit ($s$) is the sum of the number of single-qubit gates it contains and twice the number of CNOT gates it contains (corresponding to the definition of circuit size given in Section~\ref{ssec:circuits_definitions}). Allowing a partially filled final circuit layer ($\zeta < 1$) enables for more fine-grained variation of the size of QUOPS circuits. The size of a QUOPS circuit is $s = s_{\textrm{bulk}} + s_{\textrm{final}}$ where $s_{\textrm{bulk}}$ is the size of circuit excluding the final layer and $s_{\textrm{final}}$ is the size of the final layer, given by
\begin{align}
    s_\text{bulk} &= \left( 2\left\lfloor \frac{w}{2} \right\rfloor + w \right)(d/2-1) \\
    s_\text{final} &= 2\left\lfloor \frac{\zeta w}{2} \right\rfloor + \zeta w
\end{align}

For a fixed $w$, there are discrete possible values for $s$, each of which corresponds to one value for $d$ and $\zeta$. Therefore, the QUOPS circuit distribution can be defined by the two variables $w$ and $s$ (where the allowed values of $s$ are implicitly defined above). This is how the QUOPS distributions are parameterized within the QUOPS benchmark. We use $P_{w,s}$ to denote the distribution from which a width-$w$ and size-$s$ QUOPS circuit is drawn, and $C_{w,s}$ to be a (circuit-valued) random variable that is $P_{w,s}$-distributed. 

The QUOPS circuits are dense (i.e., filled with gates), random, and disordered. The dense and disordered structure of QUOPS circuits---i.e., i.i.d.~layers of gates---is not representative of quantum algorithms. There is, however, no universal structure that appears in the circuits of all quantum algorithms. Circuit structure impacts the difficulty of circuit compilation (including sub-tasks within compilation like qubit routing and magic state scheduling) and it is likely that the dense, disordered, randomized QUOPS circuit are at least as hard to compile as most quantum algorithm circuits. This choice therefore likely prevents over-estimating computational capabilities. Disordered circuits are known to prevent the systematic addition of coherent errors---whereas highly structured circuits can have long-range coherent addition or cancellation of coherent errors \cite{Proctor2021-wt}---so a typical disordered circuit will have a higher fidelity than the \emph{worst-case} structured circuit \cite{Proctor2021-wt} in the presence of coherent errors. However, this effect is unlikely to have an important impact on computational capabilities as (i) coherent error on computational qubits can, if necessary, be converted into stochastic errors using randomized compilation \cite{Knill2005-xm, Wallman2016-rd, Hashim2023-qk}, (ii) coherent errors are expected to be strongly suppressed when implementing computations on logical qubits protected by fault-tolerant QEC \cite{Beale_2018, huang2019coherent, Iverson_2020}.

\subsection{QUOPS circuit compilation}\label{ssec:compilation}
Executing QUOPS circuits on a specific system requires compiling them into that system's primitive operations. The compilation rules are a key part of what a benchmark measures \cite{Proctor2025-cd}. In full, compilation converts a circuit into a schedule of the system's lowest-level operations (gate pulses, qubit measurements, ion shuttling, and so on). It is convenient here, however, to describe compilation in two stages. The rules of QUOPS are stated in terms of these stages.

In the first stage of compilation, a QUOPS circuit $c$ is transformed into a circuit $c' = g(c)$ over an architecture-specific gate set $G_{\textrm{arch}}$ containing only unitary gates. Here, $g(\cdot)$ is a function used to encompass this compilation step, and note that $g(\cdot)$ could create a randomized circuit (e.g., containing Pauli frame randomization) but this possibility will not be explicitly denoted. For example, for a physical qubit architecture $G_{\textrm{arch}}$ might consist of general single-qubit unitaries and CNOT gates between connected qubits, whereas for a surface code architecture $G_{\textrm{arch}}$ might consist of $T$ gates, CNOT gates, and single-qubit Clifford gates. This architecture-specific gate set can be freely chosen, but this is the only stage of the compilation that is permitted to be approximate---so it must be possible to (in the absence of errors) implement these unitaries to exactly (in the absence of errors) using the primitive operations of that architecture.

The rules for the compiler $g(\cdot)$ are as follows. The compiler $g(\cdot)$ must transform a QUOPS circuit $c$, expressed in the QUOPS gate set $G_{\textrm{QUOPS}}$, into a circuit $c' = g(c)$ that (i) acts on $w' \geq w(c)$ qubits, (ii) contains only logical operations from $G_{\textrm{arch}}$, and (iii) implements a unitary that is \emph{approximately} equal to the unitary implemented by $c$, i.e., 
\begin{equation}
    U(c') \approx U(c) \otimes V_{w' -w}, \label{eq:approximation_compilation}
\end{equation}
where $V_{w'-w}$ is some unitary operator on any ancillary qubits used in $c$. In condition (iii) there is no specified minimum level of approximation accuracy so it does not put any formal constraint on $g(\cdot)$. However, it is included because the QUOPS benchmark assesses how close the (imperfect) implementation of $c'$ is to implementing $U(c)$---so any approximation error in Eq.~\eqref{eq:approximation_compilation} will negatively impact the QUOPS score. Allowing approximate compilation is important because if $G_{\textrm{arch}}$ is a discrete gate set then  implementing $U(c)$ exactly is impossible for a finite-depth circuit $c'$ for almost every QUOPS circuit $c$ (because $c$ contains continuous single-qubit rotations which can only be synthesized to finite precision).

The compiler $g(\cdot)$ is permitted to compile across the layers in $c$ and to implement a unitary that differs from the target by a compiler-specified permutation of the qubits. This compilation freedom is allowed in some other benchmarks, including the quantum volume benchmark \cite{Cross2019-ku}, but not all benchmarks, such as randomized benchmarking. The possibility of compilation across layers means that it is possible that $g(c)$ is shallower and/or smaller than $c$, even if $G_{\textrm{arch}}$ is a less expressive gate set than $G_{\textrm{QUOPS}}$. An illustrative example is as follows: if $G_{\textrm{arch}}$ contains all-to-all CNOT gates and arbitrary single-qubit unitaries, then $g(\cdot)$ can compile any width-$1$ QUOPS circuit (i.e., a circuit $c_{1,s}$ sampled from $P_{w,s}$ for any $s$) into a depth-1 circuit over $G_{\textrm{arch}}$. Similarly, any width-$2$ QUOPS circuit (i.e., a circuit $c_{2,s}$ sampled from $C_{2,s}$ for any $s$) can be compiled into a circuit over this $G_{\textrm{arch}}$ of at most size 13 (3 CNOTS and 7 single-qubit gates) and depth 7 \cite{Shende2004-ng} (so, if CNOT gates can be implemented with a fidelity of 95\%, and single-qubit gates with a fidelity of 99\%, QUOPS circuit of width-2 and \emph{any} size will pass the process polarization threshold of $1/\sqrt{e}$, resulting in a QUOPS score of at least 8, corresponding to the boundary of $s=w^3$ with $w=2$).

The compiled QUOPS circuit $g(c)$ is typically not directly executable on a system: the fundamental operations on a system are lower-level operations like gate pulses, ion shuttling, qubit measurements, etc., and in an FTQC architecture there is an intermediate layer of FTQC subroutines with which logical gates are implemented (e.g., syndrome extraction, magic state cultivation or distillation, etc.). However, we do not immediately further compile a compiled QUOPS circuit $g(c)$ into lower-level operations. Instead, a QUOPS circuit is first embedded within fidelity estimation ``proxy'' circuits which are also described using the $G_\textrm{{arch}}$ gate set (which are mirror circuits in our implementation of QUOPS). We detail this in Section~\ref{sec:quops_operational}. The compilation of those circuits (i) must be exact---i.e., in the absence of noise the unitary for each gate in those mirror circuits is implemented exactly, and (ii) must not compile across circuit layers (often called \emph{compilation barriers}) except when such compilation is necessary due to the operations available in the architecture constraints (e.g., $R_Z(\theta)$ rotations in physical qubit architectures are typically implemented ``in software'', meaning updating the rotation axes of gates in the following layers). In such cases where some compilation across barriers is required it is necessary to show that the ``proxy'' circuit fidelity estimation method works correctly (e.g., the existing theory for that method might still be applicable).

In summary, QUOPS permits approximate compilation of its circuits, including compilation across the layers of a circuit and the use of ancilla qubits. Approximate compilation is typically necessary for a benchmark if it is to be executable on architectures with different discrete gate sets, because two such gate sets typically cannot implement the same unitaries exactly with finite-depth circuits. Compilation across the whole circuit, as in the quantum volume benchmark, and the use of ancillae reflect the intent of QUOPS as a measure of \emph{computational} capability. Because QUOPS quantifies performance by the fidelity between the noisy evolution of the qubits and the ideal unitary, it cannot be spoofed, without explicitly cheating, by classically sampling from the error-free output distribution.

\section{A foundational definition of QUOPS}\label{sec:quops_foundational}
In this section we provide a foundational definition of QUOPS and its constituent elements, building on the definition of the QUOPS circuits in Section~\ref{sec:quops_circuits}.  These definitions are \emph{foundational} in the sense that they use quantities that can only be estimated in experiment (e.g., probabilities) and/or that are defined using idealized models of reality (e.g., superoperator representations of noisy circuit executions).  In Section~\ref{sec:quops_operational} we provide protocols (i.e., experimental procedures) for estimating QUOPS capability regions, scores, rates, and all their associated quantities. Those methods are reliable estimators of the quantities defined in this section under certain assumptions. They also constitute a complementary \emph{operational} definition for QUOPS that is largely assumption-free.

This section provides foundational definitions of the success criteria for an individual QUOPS circuit (Section~\ref{ssec:success_metric}), the success criteria for an ensemble of equal-shape QUOPS circuits (Section~\ref{ssec:ensemble_success_metric}), the QUOPS capability regions (Section~\ref{ssec:capability_regions}), the QUOPS score $Q$ (Section~\ref{ssec:quops_score}),  the QUOPS rate (Section~\ref{ssec:foundational_quops_per_s}), and
error-mitigated QUOPS scores $Q^*_{\alpha}$ (Section~\ref{ssec:conditional_quops}).

\subsection{QUOPS circuit success}\label{ssec:success_metric}
QUOPS finds the largest QUOPS circuits that can be run with low error, which requires a metric for quantifying the error in circuit executions. Here, this metric is defined and the noise model on which it is predicated is stated. Before a QUOPS circuit $c_{w,s}$ is executed on a system, it is compiled into a circuit $c'_{w,s} = g(c_{w,s})$ over the architecture-specific gate set $G_{\textrm{arch}}$ (see Section~\ref{ssec:compilation}), before being further compiled in a low-level circuit described in the system's lowest-level operations. QUOPS uses process polarization (see Section~\ref{ssec:fidelity_definitions}) to quantify how close the noisy implementation of $c'_{w,s}$ is to performing the ideal unitary implemented by $c_{w,s}$.

Making this definition precise requires a model for the noisy implementation of $c'_{w,s}$. The circuit $c'_{w,s}$ acts on $w' = w + a$ qubits, where $a \geq 0$ is the number of ancilla qubits used, and
\begin{equation}
\mathcal{U}(c'_{w+a,s}) \approx \mathcal{U}(c_{w,s}) \otimes \mathcal{V}_{a},
\end{equation}
where $\mathcal{V}_{a}$ is some unitary superoperator on the ancillae, which is Eq.~\eqref{eq:approximation_compilation} expressed as superoperators. The foundational definition for QUOPS assumes that the noisy implementation of $c'_{w,s}$ is a $w'$-qubit superoperator of the form
\begin{equation}
\Lambda_{w+a}(c'_{w,s})  = \Lambda(c'_{w,s}) \otimes \Lambda_a(c'_{w,s}) \label{eq:markovian_assumption}
\end{equation}
where $\Lambda_a(c'_{w,s})$ is an $a$-qubit superoperator acting on the ancillas and $\Lambda(c'_{w,s})$ is a $w$-qubit superoperator acting on the $w$ computational qubits. 
The quality of the execution of the QUOPS circuit $c_{w,s}$ is quantified by the process polarization between $\Lambda(c'_{w,s})$ and $\mathcal{U}(c_{w,s})$, i.e.,
\begin{equation}
\polarization(c_{w,s}) \equiv \polarization(\mathcal{U}(c_{w,s}), \Lambda(g(c_{w,s})).
 \label{eq:circuit_process_polarization}
\end{equation}
The process polarization quantifies both (i) noise in the execution of $g(c_{w,s})$ and (ii) approximation error in $g(c_{w,s})$ (e.g., due to synthesizing $R_P(\theta)$ gates using $T$ and Clifford gates). The process polarization is not impacted by errors in the initial preparation of the computational qubits or final readout of the computational qubit (but it \emph{is} impacted by any measurements used within the computation, e.g., measurements for syndrome extraction). This choice is made because the errors in gates will dominate utility-scale quantum circuits. However, note that the QUOPS procedure (see Section~\ref{sec:quops_operational}) enables also quantifying initial state preparation and final readout errors.

Equation~\eqref{eq:markovian_assumption} is an idealization that is similar to the ``Markovianity'' assumptions often made in benchmarking and characterization theory \cite{Hashim2025-rz, Proctor2025-cd}. There are common noise processes in physical quantum computing systems that cannot be modeled by Eq.~\eqref{eq:markovian_assumption}. This includes leakage and many forms of non-Markovian noise \cite{Hashim2025-rz}. Furthermore, if $c'_{w,s}$ is implemented by encoding each computational qubit into a logical qubit protected by QEC then even if the noise on physical qubits can be represented by superoperators (i.e., is Markovian) this does not imply that the noise on the logical qubits can be represented by a superoperator \cite{ziyad2026emergent, kwiatkowski2025approximate}. The assumption of Eq.~\eqref{eq:markovian_assumption} is, however, not required for an \emph{operational} definition of QUOPS (see Section~\ref{sec:quops_operational}). The operational definition for QUOPS is such that the procedure accurately estimates the foundational definition of QUOPS when Eq.~\eqref{eq:markovian_assumption} is approximately valid, and is still an operationally well-motivated benchmark when it is not.

\subsection{QUOPS circuit ensemble success}\label{ssec:ensemble_success_metric}
QUOPS circuits are random, i.e., each specific QUOPS circuit $c_{w,s}$ is a sample from the distribution $P_{w,s}$. Therefore, the process polarization of QUOPS circuits is a random variable $\polarization(C_{w,s})$ with a distribution $P_{\polarization,w,s}$ over $[\frac{-1}{4^w -1}, 1]$ (the range of the process polarization). Two different QUOPS circuits of shape $(w,s)$ sampled from $P_{w,s}$ will typically have different process polarization, when compiled and executed on the same system, i.e., $P_{\polarization,w,s}$ is not a zero-variance distribution. Differences can arise because two compiled shape-$(w,s)$ QUOPS circuits can have different sizes once compiled for the target architecture (e.g., because of CNOT routing) and because they can interact differently with the noise in the system (e.g., because of coherent addition or cancellation of errors \cite{Proctor2021-wt}).

The $(w,s)$-parameterized distribution $P_{\polarization,w,s}$ contains interesting information about system performance, but this parameterized distribution is unwieldy to represent, and it is expensive to learn its details (e.g., estimating the first $k$ moments of $P_{\polarization,w,s}$ with low uncertainty). We therefore summarize the performance of shape-$(w,s)$ QUOPS circuits by the mean of $P_{\polarization,w,s}$, i.e., by
\begin{equation}
 \bar{\polarization}_{w,s}  \equiv  \mathbb{E} \left\{  \polarization (\mathcal{U}(C_{w,s}), \Lambda(g(C_{w,s})) \right\}.
 \label{eq:mean_process_polarization}
\end{equation}
The median of $P_{\polarization,w,s}$ is arguably better-motivated (it has the simple interpretation that a randomly sampled QUOPS circuit will have a polarization at or above the median with probability 50\%). However, the mean is simple to estimate and is unlikely to be very different from the median for realistic $P_{\polarization,w,s}$.

\subsection{QUOPS capability regions}\label{ssec:capability_regions}
A key output of QUOPS is a QUOPS \emph{capability region} \cite{Proctor2021-wt, Proctor2025-cd}. This capability region separates circuit shapes into ``success'' and ``fail'' shapes, where the system is deemed to have succeeded or failed on the QUOPS circuits. A circuit shape $(w,s)$ is designated ``success'' if
\begin{equation}
    \bar{\polarization}_{w,s} \geq \alpha,
\end{equation}
where $\alpha$ is a threshold value for the mean process polarization. The \emph{standard success threshold} in QUOPS is
\begin{equation}
\alpha = \frac{1}{\sqrt{e}}.\label{eq:pass_threshold}
\end{equation}

The capability region uses a threshold to turn a continuous value into a binary ``success'' or ``fail'' score. An $\mathcal{O}(1)$ threshold is chosen because this corresponds to an $\mathcal{O}(1)$ error rate for the circuits, and if an algorithmic circuit has $\mathcal{O}(1)$ probability of error the correct result can typically be extracted with only $\mathcal{O}(1)$ overhead (see Section~\ref{ssec:conditional_quops}). The particular value of $1/\sqrt{e}$ is essentially arbitrary, but it does have a convenient interpretation. Under a simplistic error model in which each circuit location has an error rate $\epsilon$, this threshold is achieved by a circuit of size $s \approx 1/(2\epsilon)$. So we can define the effective error rate per QUOPS gate at the threshold as $\epsilon_{\textrm{eff}} \approx 1/(2s)$. If a threshold of $1/e$ was instead used, this would remove the factor of two in the above interpretations. However, $1/\sqrt{e}$ enables estimating success probabilities near the success threshold with fewer samples. The mirror circuits used to estimate $\bar{\polarization}_{w,s}$ typically have success probabilities of approximately $\bar{\polarization}_{w,s}^2$, and so near the threshold these success probabilities are $\sim 1/e \approx 37\%$. Success probabilities near this boundary can be estimated to fixed precision with less data than required for a success probability threshold of $1/e^2 \sim 14\%$ \cite{Proctor2022-zs}.

The QUOPS capability region with threshold $\alpha$ is the set of all circuit shapes at which the mean polarization of QUOPS circuits of that shape is above the threshold $\alpha$ of Eq.~\eqref{eq:pass_threshold}, i.e.,
\begin{equation}
    Q_{\textrm{region}, \alpha} = \left\{ (w,s) \mid \bar{\polarization}_{w,s} > \alpha \right\}.
\end{equation}
The standard QUOPS capability region, $ Q_{\textrm{region}}$ corresponds to $\alpha = 1/\sqrt{e}$.  When $\alpha <  1/\sqrt{e}$, $ Q_{\textrm{region}, \alpha}$ is called an error-mitigated capability region, with threshold $\alpha$. These regions are discussed further in Sec.~\ref{ssec:conditional_quops}.

An illustration of a QUOPS capability region is shown in Fig.~\ref{fig:main:quops:capability}. The QUOPS capability region is a coarse-graining of the circuit-shape parameterized probability distribution $P_{\polarization,w,s}$, turning a probability distribution at each shape $(w,s)$ into a binary ``success'' or ``fail''. There are two challenges to measuring capability regions: (1) it is infeasible to run circuits with every possible circuit shape $(w,s)$, and (2) it is only possible to estimate $\bar{\polarization}_{w,s}$ (or estimate if $\bar{\polarization}_{w,s} \geq \alpha$) at each shape rather than measure it exactly. Methods for estimating QUOPS capability regions are presented in Section~\ref{sec:quops_operational}.

\subsection{QUOPS score}\label{ssec:quops_score}
QUOPS supplements its capability region with a scalar performance metric---the \emph{QUOPS score}---which is defined from the capability region. This performance metric is intended to be reported together with the QUOPS rate (Section~\ref{ssec:foundational_quops_per_s}).
The QUOPS score $Q$ is the maximum size of the QUOPS circuits that were run successfully \emph{and} that fall within a \emph{cone} of circuit-shape space defined by
\begin{equation}
Q_{\textrm{admissible}} = \{(w,s) \mid w^2 \leq s \leq w^3\}, \label{eq:cone}
\end{equation}
which we refer to as the admissible cone. That is, $Q$ is defined by
\begin{equation}
    Q = \textrm{max}(\{ s \mid (w,s) \in Q_{\textrm{region}} \cap Q_{\textrm{admissible}}\}).
\end{equation}
The circuit width at which $Q$ is achieved is denoted $w_Q$, that is, $Q$ is achieved at circuit shape $(w_Q,Q)$.
The cone, a QUOPS capability region, and the $Q$ extracted from the capability region are shown in Fig.~\ref{fig:main:quops} in the main text.

The admissible cone is intended to include the shapes of the circuits that are currently expected to be the first to provide quantum utility (see Fig.~\ref{fig:main:quops:schematic} and discussion in the main text). In contrast, quantum volume constrains circuit shape by considering only square circuits ($s=w^2$ in the quantum volume gate set). A machine might therefore achieve a high quantum volume while still falling far short of the circuit size required for useful quantum computations. QUOPS strongly encourages the reporting of the full capability region while defining its scalar score $Q$ as the largest successful circuit size within the admissible cone. This cone is intended to  preserve continuity with square-circuit benchmarks, via the $s = w^2$ boundary. The admissible cone excludes deep but narrow (i.e., few-qubit) circuits and wide (i.e., many-qubit) but shallow circuits from the evaluation of the scalar QUOPS score. For example, with $w = 2$ the admissible cone's lower and upper bounds on $s$ are $s = 4$ and $s = 8$, so the maximum QUOPS obtainable with two qubits is $Q = 8$.

The admissible cone can also be approximately stated in term of depth $d$. For QUOPS shapes at which the QUOPS circuits are dense (even $w$ and $s$ an even multiple of $w$), $s = wd$ and so
\begin{equation}
w \leq d \leq w^2. \label{eq:cone-depth}
\end{equation}
The lower bound on $d$ (i.e., $d=w$) corresponds to square circuits, i.e., circuits with equal depth and width, as used in the quantum volume benchmark. For all other circuit shapes, the inequalities of Eq.~\eqref{eq:cone-depth} are approximate. This means that the lower-bound on circuit size in the admissible cone enables comparisons between QUOPS scores (and capability regions) and quantum volume measurements.

\subsection{QUOPS rate}\label{ssec:foundational_quops_per_s}
The QUOPS score $Q$ is intended to be reported alongside a \emph{QUOPS rate} $\Omega$ that quantifies the effective rate at which universal quantum operations (elements of $G_{\textrm{QUOPS}}$) can be applied. Whereas $Q$ measures the size of the largest circuit that can be run with low probability of error, $\Omega$ measures a processor's rate of \emph{useful} operations, accounting for the additional circuit executions that circuit errors make necessary. It is defined so that $K \kappa Q / \Omega$ is a reliable proxy for the time required to obtain data equivalent to $K$ error-free samples from each of $\kappa$ circuits of shape $(w_Q, Q)$.

There are many factors that complicate an operational definition of $\Omega$. Here, we provide a definition based on a simplified model for execution times in a quantum computer. In our simplified model, one complete execution of a quantum circuit $c$ takes time $\tau_{\textrm{accept}}(c)$, and no time is required to change between circuits or implement other tasks such as recalibration. In QUOPS error detection is permitted, which consists of using ancillary information during or at the end of the circuit to discard the outcome of a circuit execution. $\Omega$ takes this ``wasted'' time into account. Let $p_{\textrm{accept}}(c)$ and $p_{\textrm{reject}}(c)= 1 -p_{\textrm{accept}}(c) $ be the probability that an execution of circuit $c$ is accepted or rejected, respectively, by the error detection method and $\tau_{\textrm{reject}}(c)$ be the expected time spent in execution if a circuit execution is rejected. Then, in our model, generating $K$ samples (i.e., accepted circuit executions) from the quantum circuit $c$ requires, in expectation, time 
\begin{equation}
    K\tau_{\textrm{circ}}(c) = K\left(\tau_{\textrm{accept}}(c) +  \frac{p_{\textrm{reject}}(c)}{p_{\textrm{accept}}(c)}\tau_{\textrm{reject}}(c)\right),
\end{equation}
and generating $K$ samples from the circuits $c_1,\dots,c_{\kappa}$ takes time 
\begin{equation}
\tau = K\sum_{i=1}^{\kappa}\tau_{\textrm{circ}}(c_i).
\end{equation}

Given the random variable $\tau_{\textrm{circ}}(g(C_{w,s}))$,
we define a ``raw'' QUOPS rate $\Omega_{\textrm{raw}}(w,s)$ for circuits of shape $(w,s)$:
\begin{align}
    \tau(w,s) &= \textrm{Median}\{\tau_{\textrm{circ}}(g(C_{w,s}))\} \\
        \Omega_{\textrm{raw}}(w,s)& = \frac{s}{ \tau(w,s)}.
\end{align}
The quantity $\Omega_{\textrm{raw}}(w,s)$ enables computation of the typical time to generate $K$ samples from $\kappa$ randomly sampled QUOPS circuits of shape $(w,s)$. This time is $K \kappa s /\Omega_{\textrm{raw}}(w,s)$. However, $\Omega$ is intended to quantify the time to generate the amount of data needed to extrapolate error-free outcomes (if error mitigation succeeds). The multiplicative sample overhead of error mitigation is approximately $O(w,s) = 1/\bar{\polarization}_{w,s}^2$ (see Section~\ref{ssec:conditional_quops}), and so we define
\begin{align}
        \Omega(w,s)& = \frac{\Omega_{\textrm{raw}}(w,s)}{O(w,s)} = \frac{s \bar{\polarization}_{w,s}^2}{ \tau(w,s)}. 
\end{align}
The QUOPS rate $\Omega$ to be reported alongside a QUOPS score of $Q$ is then simply 
\begin{align}
\Omega &= \Omega(w_Q, Q) \label{eq:omega_foundational}.
\end{align}

The definition of $\Omega$ given here is an idealization that does not account for important practical factors, resulting in a definition that is ambiguous for real systems. There are often significant additional time costs when running quantum circuits---examples include the time required to load a different quantum circuit into control hardware, time required to for periodic recalibration experiments, and time performing operations for which it is ambiguous whether they are part of the time required to execute a circuit (e.g., loading ions in an ion trap)---and the definition of $\Omega$ in Eq.~\eqref{eq:omega_foundational} does not specify whether these times should be included or not. Furthermore, the circuits executed in the QUOPS benchmark are proxy circuits---which estimate the process polarization of QUOPS circuits---rather than simply the compiled QUOPS circuits. The time to execute those circuits might differ from the time to execute the compiled QUOPS circuits. These issues are addressed in Section~\ref{ssec:operational_quops_per_second}, where an operational definition for $\Omega$ is provided that is specific to the mirror circuit approach to estimating QUOPS.

\subsection{Error-mitigated QUOPS regions and scores}\label{ssec:conditional_quops}
Error mitigation \cite{Kim2023-si} allows a quantum algorithm to succeed even when an error has probably corrupted the output of a circuit, at the cost of additional samples or time. Two kinds of technique must be distinguished. Techniques that use \emph{heralded} errors, such as error detection and postselection, are permitted in QUOPS, and the executions they discard are accounted for in the QUOPS rate. Techniques that address \emph{unheralded} errors, such as probabilistic error cancellation, which we refer to as error mitigation, are not applied to QUOPS data. Indeed they cannot be, because error mitigation acts on the classical outcomes of circuits whereas the QUOPS success metric is a property of the quantum evolution itself. Instead, the potential benefit of error mitigation is captured by the error-mitigated capability regions and scores defined below.

The error-mitigated QUOPS capability region $Q_{\textrm{region},\alpha}$ and score $Q^*_{\alpha}$ are parameterized by a threshold $\alpha < 1/\sqrt{e}$ and are defined by
\begin{align}
    Q_{\textrm{region},\alpha} &= \{ (w,s) \mid \bar{\polarization}_{w,s} \geq \alpha\} \\
    Q^*_{\alpha} &= \textrm{max}(\{ s \mid (w,s) \in Q_{\textrm{region},\alpha} \cap Q_{\textrm{admissible}}\}),
\end{align}
and the width at which $Q^*_{\alpha}$ is achieved is denoted $w_{Q^*_{\alpha}}$. The sample overhead $O = 1/\alpha^2$ should be reported alongside $Q^*_{\alpha}$. This is also incorporated into the QUOPS rate at $(w_{Q_{\alpha}^*},Q_{\alpha}^*)$, which should also be reported alongside the error-mitigated QUOPS score:
\begin{align}
\Omega^{*}_{\alpha} &= \Omega(w_{Q_{\alpha}^*},Q_{\alpha}^*) \label{eq:omega_foundational-2}.
\end{align}

The sample overhead is an approximate estimate of the multiplicative increase in the number of circuit executions required to extract a noise-free estimate of an observable for the QUOPS circuits of this size. This is based on the following. If each layer in a depth $l$ quantum circuit experiences local depolarizing noise with strength $p$, then the sample overhead for any error mitigation technique scales as:
\begin{equation}
    O \geq \frac{1}{(1-p)^{2l}} = \frac{1}{\polarization_p^2},
\end{equation}
where $\polarization_p = (1-p)^l$ is the circuit's process polarization \cite{Takagi2022}. This bound is an approximation of the sampling cost of error mitigation for a noisy implementation of a circuit that has process polarization of $\polarization_p$. It is an approximation because error mitigation might fail---the error mitigation approach might not correctly infer the noise-free value as the amount of data increases, i.e., the error mitigated estimator is biased---and it does not take into account the cost of any noise model learning necessary to implement the error mitigated data analysis.

\begin{figure*}
    \centering
    \includegraphics[width=\linewidth]{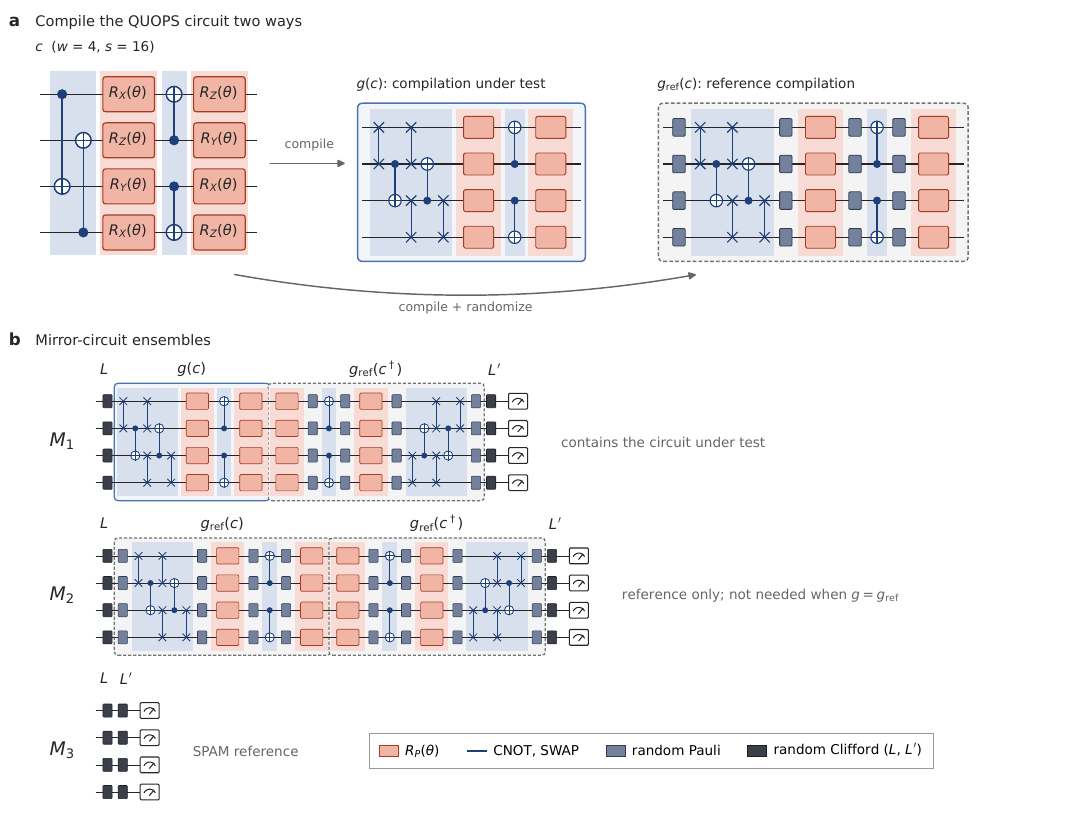}
    \caption{\textbf{Mirror circuits for estimating the process polarization of QUOPS circuits.}
    (\textbf{a})~A QUOPS circuit $c = c_{w,s}$ (here $w=4$, $s=16$) is compiled in two ways. A compilation $g$ chosen for optimal performance (maximizing circuit polarization) produces the compilation $g(c)$. In this example, relevant to physical-qubit architectures with linear connectivity, $g$ routes each CNOT gate onto a line of qubits using SWAP gates. The reference compiler $g_{\rm ref}$ also compiles $c$ but must do so in a way that produces, in expectation, stochastic errors. This is typically a randomized compilation. The one shown here implements standard Pauli frame randomization (Section~\ref{sec:reference_compiler}): a uniformly random Pauli layer (grey) is inserted before every two-qubit-gate layer and a compensating Pauli layer after it, which is then compiled into the single-qubit gates. The compiler $g$ may itself serve as the reference compiler ($g = g_{\rm ref}$) if it satisfies the requirements on $g_{\textrm{ref}}$. (\textbf{b})~The three mirror-circuit ensembles of Section~\ref{ssec:mcfe} [see Eq.~\eqref{eqn:2xreference-ensemble-definition} and the surrounding text]. $M_1 = L'\,g_{\rm ref}(c^\dagger)\,g(c)\,L$ contains the compiled circuit whose polarization is being estimated. $M_2 = L'\,g_{\rm ref}(c^\dagger)\,g_{\rm ref}(c)\,L$ contains only reference compilations. $M_3 = L'L$ is a state preparation and measurement (SPAM) reference. Here $c^\dagger$ is the inverse of $c$, $L$ is a layer of independent, uniformly random single-qubit Clifford gates, and $L'$ implements $L^{-1}$ followed by an independent, uniformly random Pauli on each qubit. Every mirror circuit is a definite-outcome circuit: in the absence of errors it returns a known target bit string $x_c$ when all qubits are measured in the computational basis. Executing sampled circuits from each ensemble and recording the Hamming distances of their outputs from $x_c$ yields their effective polarizations (Section~\ref{ssec:mcfe}), from which $\polarization(c)$ [Eq.~\eqref{eq:circuit_process_polarization}] is estimated via Eqs.~\eqref{eq:mcfe_equation} and \eqref{eq:mcfe_estimator}: the data from $M_3$ circuits remove the contribution of SPAM error, $L$ and $L'$ errors, and data from $M_2$ circuits remove that of the reference half $g_{\rm ref}(c^\dagger)$, isolating the error of $g(c)$. When $g = g_{\rm ref}$, $M_1$ and $M_2$ are identically distributed and only $M_1$ and $M_3$ circuits need to be run [Eq.~\eqref{eq:mcfe_equation-simplified}].}
    \label{fig:mcfe}
\end{figure*}

\section{An operational definition of QUOPS}\label{sec:quops_operational}
This section explains how QUOPS capability regions, scores, and rates are estimated in experiment. It can be read as both a procedure for estimating the quantities defined in Section~\ref{sec:quops_foundational} and as an operational definition of these quantities. The key to implementing QUOPS is a method for efficiently estimating the polarization of a QUOPS circuit, as this is central to estimating all of the QUOPS metrics. This is the topic of Section~\ref{ssec:mcfe}--\ref{ssec:ensemble_mcfe}. Those sections cover mirror circuit fidelity estimation (MCFE) \cite{Proctor2022-zs} and its application and adaption to QUOPS circuits, including circuits executed on logical qubits. Section~\ref{ssec:mcfe} reviews how MCFE can be used to estimates the process polarization of a single QUOPS circuit. Section~\ref{sec:reference_compiler} specifies how, in the case of QUOPS circuits, we can create the randomized \emph{reference compilations} that MCFE requires. Section~\ref{ssec:ensemble_mcfe} gives the estimator for $\bar{\polarization}_{w,s}$ that combines the mirror circuit data from many sampled QUOPS circuits of one shape, together with its uncertainty. 

The remainder of this section is independent of whether MCFE is used to estimate QUOPS circuit polarizations. Section~\ref{ssec:estimating_regions_scores} specifies how mean polarization estimates at each circuit shape are turned into an estimated QUOPS capability region and QUOPS score with stated statistical confidence. Section~\ref{ssec:operational_quops_per_second} gives the operational definition of the QUOPS rate.

\subsection{Estimating QUOPS circuit success: MCFE}\label{ssec:mcfe}
In this section we provide a method for efficiently estimating the success metric for QUOPS circuits. The QUOPS benchmark's circuit quality metric is the mean process polarization of
randomly sampled QUOPS circuits of shape $(w,s)$, i.e., $\bar{\polarization}_{w,s}$ defined in Eq.~\eqref{eq:mean_process_polarization}. Executing the QUOPS benchmark therefore requires a method for estimating $\bar{\polarization}_{w,s}$. Here we state the method that we use, MCFE~\cite{Proctor2022-zs}.

The MCFE procedure given here estimates the process polarization of an individual QUOPS circuit. Because the QUOPS ensemble success metric, $\bar{\polarization}_{w,s}$, is the mean process polarization of randomly sampled QUOPS circuits of shape $(w,s)$, the method in this section provides a direct approach to estimating $\bar{\polarization}_{w,s}$. Namely, sample many QUOPS circuits of shape $(w,s)$, use MCFE to estimate the process polarization of each circuit, and average those estimates. Section~\ref{ssec:ensemble_mcfe}, however, introduces a more sample-efficient procedure that uses MCFE without first estimating the process polarization of each circuit individually. First we note, however, that QUOPS does not require the use of a specific procedure for estimating $\bar{\polarization}_{w,s}$---instead it is only necessary to use a polarization estimation procedure that is accurate under assumptions that are applicable to the system being benchmarked. MCFE is reliable under many physically-relevant circumstances (explained below). For any system covered by these circumstances, MCFE can be used in QUOPS without further justification. Otherwise, or if a procedure other than MCFE is used, arguments for the reliability of the technique should be provided.

Given a QUOPS circuit $c_{w,s}$, the quantity to be estimated is
\begin{equation*}
    \polarization(c_{w,s})
    \equiv
    \polarization(\mathcal{U}(c_{w,s}),\Lambda(g(c_{w,s}))),
\end{equation*}
which was defined in 
Eq.~\eqref{eq:circuit_process_polarization}. Here $\Lambda(g(c_{w,s}))$ is the CPTP map representing the noisy execution of the compiled circuit $g(c_{w,s})$. $\polarization(c_{w,s})$ is reduced by noise in the implementation of $g(c_{w,s})$ and by any approximation errors introduced during compilation, i.e., whenever $\mathcal{U}(g(c_{w,s})) \neq \mathcal{U}(c_{w,s})$.

\subsubsection{MCFE circuit ensembles}\label{ssec:circuit-classes}

MCFE requires a \emph{reference compiler} $g_{\textrm{ref}}$ that creates a compilation of a QUOPS circuit that, in expectation, experiences only stochastic errors. Specifically, $g_{\textrm{ref}}$ should compile a QUOPS circuit $c_{w,s}$ into a (typically randomized) circuit $g_{\textrm{ref}}(c_{w,s})$ such that 
\begin{align}
   \mathbb{E}\{\Lambda(g_{\textrm{ref}}(c_{w,s}))\} &= \mathcal{E}_{\textrm{ref}}(c_{w,s})\mathcal{U}(c_{w,s}) 
   \label{eq:reference-compiler-condition-1}\\
      \mathbb{E}\{\Lambda(g_{\textrm{ref}}(c_{w,s}^{\dagger}))\} &= \mathcal{E}_{\textrm{ref}}(c_{w,s}^{\dagger})\mathcal{U}(c_{w,s}^{\dagger})
         \label{eq:reference-compiler-condition-2}
\end{align} 
where $c_{w,s}^{\dagger}$ is the inverse of $c_{w,s}$, obtained by reversing the order of its layers and replacing each gate by its inverse. The channels $\mathcal{E}_{\textrm{ref}}(c_{w,s})$ and $\mathcal{E}_{\textrm{ref}}(c_{w,s}^{\dagger})$ should be approximately stochastic error channels with approximately equal process polarizations (to the identity) \cite{Proctor2022-zs}. The conditions of  Eqs.~\eqref{eq:reference-compiler-condition-1}--\eqref{eq:reference-compiler-condition-2} are a joint property of a compiler and the noise impacting the quantum computing system being tested. It is permissible for $g_{\textrm{ref}} = g$ if the compiler $g$ satisfies the necessary conditions of $g_{\textrm{ref}}$, and this approach reduces the sample complexity of the experiments. Reference compilation approaches suitable for physical-qubit architectures and many FTQC architectures are included in Section~\ref{sec:reference_compiler}.

MCFE estimates $\polarization(c_{w,s})$ using three mirror circuit ensembles, illustrated in Fig.~\ref{fig:mcfe}:
\begin{itemize}
    \item $M_1(g(c_{w,s}), g_{\textrm{ref}}(c_{w,s}))$,
    \item $M_2(g_{\textrm{ref}}(c_{w,s}))$,
    \item $M_3(w)$.
\end{itemize}
These are the three ensembles of mirror circuits defined in Ref.~\cite{Proctor2022-zs} but specialized to QUOPS circuits and adapted to the case where a circuit can be approximately compiled.

The $M_1$ ensemble contains both $g(c_{w,s})$ and $g_{\textrm{ref}}(c_{w,s})$ and has the form:
\begin{equation}
M_1(g(c_{w,s}), g_{\textrm{ref}}(c_{w,s}))  = L'\, g_{\mathrm{ref}}(c_{w,s}^{\dagger})\, g(c_{w,s})\, L,
\end{equation}
where $L$ consists of independent random single-qubit Clifford gates on each qubit, and $L'$ is distributed such that $U(L'L)$ corresponds to an independent and uniformly random Pauli operator on each qubit. Equivalently, $L'$ implements the inverse of $L$ multiplied by a uniformly random Pauli operator. In the idealized case where all parts of the $M_1$ circuit are error free except $g(c_{w,s})$, this ensemble alone would be sufficient to estimate $\polarization(c_{w,s})$~\cite{Proctor2022-zs}. The role of the two additional ensembles, $M_2$ and $M_3$, is to separate the error in $g(c_{w,s})$ from the error in the rest of the $M_1$ circuit.

The $M_2$ ensemble contains only the reference compilations and has the form:
\begin{equation}
M_2(g_{\textrm{ref}}(c_{w,s})) = L'\, g_{\mathrm{ref}}(c_{w,s}^{\dagger})\, g_{\mathrm{ref}}(c_{w,s})\, L.
\label{eqn:2xreference-ensemble-definition}
\end{equation}
Note that if $g=g_{\mathrm{ref}}$, then $M_1$ and $M_2$ are identically distributed. In this case it is unnecessary to sample and execute $M_2$ circuits separately. The data analysis in this case is simpler, and the sample complexity lower, as described below.

The $M_3$ ensemble consists of state preparation and measurement (SPAM) reference circuits of the form:
\begin{equation}
 M_3(w) = L'L.
\end{equation}
The $M_3$ circuit ensemble enables estimation of the error in $L$, $L'$ and the native SPAM operations. The combination of data from $M_3$ and $M_2$ circuits enables estimation of the error in the reference compilation $g_{\mathrm{ref}}(c_{w,s}^{\dagger})$, and then estimation of $\polarization(c_{w,s})$ using data from $M_1$ circuits.

\subsubsection{MCFE single-circuit data analysis}

All circuits in the $M_i$ mirror circuit ensembles are \emph{definite outcome} circuits. A definite outcome circuit is a circuit $c$ that, in the absence of errors, always outputs a particular bit string $x_c$, which is called its \emph{target} bit string. The MCFE data analysis uses a quantity called the \emph{effective polarization} \cite{Proctor2022-zs}, which is closely related to a definite outcome circuit's success probability, i.e., the probability it returns $x_c$. For an $n$-qubit definite outcome circuit $c$, the effective polarization is defined by
\begin{equation}
    \epolarization(c) = \frac{4^n}{4^n-1} \left( \sum_{k=0}^n \left(-\frac{1}{2}\right)^k h_k(c) \right) - \frac{1}{4^n - 1}, 
\end{equation}
where $h_k(c)$ is the probability that the output bit string $y$ has Hamming distance $k$ from the circuit's target bit string $x_c$, i.e.,
\begin{equation}
    h_k(c)=\sum_{y\in D_k} P_y(c), \qquad
    D_k=\{y \mid h(x_c,y)=k\},
\end{equation}
and $P_y(c)$ is the probability that $c$ outputs $y$. From $\kappa$ executions of $c$, this is estimated by
\begin{equation}
    \hat{\epolarization}(c,N) = \frac{4^n}{4^n-1} \left( \sum_{k=0}^n \left(-\frac{1}{2}\right)^k \hat{h}_k(c,\kappa) \right) - \frac{1}{4^n - 1}, \label{eq:effective_polarization_estimator}
\end{equation}
where $\hat{h}_{k}(c,\kappa)$ is the empirical frequency of $h_k(c)$ in the $\kappa$ circuit executions.

MCFE estimates $\polarization(c_{w,s})$ by executing mirror circuits from each of the $M_1$, $M_2$ and $M_3$ ensembles. The mean effective polarization of each kind of mirror circuit is used in the MCFE procedure to estimate the process polarization of the circuit $c_{w,s}$. In particular, the theory of MCFE shows that 
\begin{equation}
    \polarization(c_{w,s}) \approx  \frac{\mathbb{E}_{M_1}\left\{\epolarization(M_1(c_{w,s}))\right\}}{\sqrt{\mathbb{E}_{M_3}\left\{ \epolarization(M_3(w))\right\} }\sqrt{\mathbb{E}_{M_2}\left\{\epolarization(M_2(c_{w,s}))\right\}}}, \label{eq:mcfe_equation}
\end{equation}
with small approximation error. The assumptions and approximations encompassed in this approximate equality are described in Ref.~\cite{Proctor2022-zs}. When $g_\textrm{ref} = g$, this simplifies to
\begin{equation}
    \polarization(c_{w,s}) \approx  \frac{\sqrt{\mathbb{E}_{M_1}\left\{\epolarization(M_1(c_{w,s}))\right\}}}{\sqrt{\mathbb{E}_{M_3}\left\{ \epolarization(M_3(w))\right\} }}.
    \label{eq:mcfe_equation-simplified}
\end{equation}

MCFE estimates $\polarization(c_{w,s})$ by sampling $K_i \geq 1$ instances of the $M_i$ circuits, denoted $m_{i,k}$ with $k=1,2,\dots,K_i$ and with $i=1,2,3$, executing each circuit $N$ times to estimate its effective polarization, and then applying Eq.~\eqref{eq:mcfe_equation} (or Eq.~\eqref{eq:mcfe_equation-simplified} if no $M_2$ circuits are required). The estimator is
\begin{equation}
    \hat{\polarization}(c_{w,s}) =  \frac{ \frac{1}{K_1}\sum_{k=1}^{K_1}\left\{\hat{\epolarization}(m_{1,k}(c_{w,s}), \kappa)\right\}}{\sqrt{ \frac{1}{K_3}\sum_{k=1}^{K_3}\left\{ \hat{\epolarization}(m_{3,k}(w), \kappa)\right\} }\sqrt{ \frac{1}{K_2}\sum_{k=1}^{K_2}\left\{\hat{\epolarization}(m_{2,k}(c_{w,s}),\kappa)\right\}}}. \label{eq:mcfe_estimator}
\end{equation}

The sample complexity of $ \hat{\polarization}(c_{w,s})$ is provided in Ref.~\cite{Proctor2022-zs} and it does not increase with circuit size or width, if the mirror circuit effective polarizations are $\mathcal{O}(1)$. In QUOPS, this is the case because it is only necessary to check if the mean polarization of QUOPS circuits of a given shape is above $1/\sqrt{e}$, corresponding to mirror circuit effective polarizations of approximately $1/e$.

\subsection{Estimating QUOPS circuit success: reference compilations}\label{sec:reference_compiler}

MCFE requires a reference compiler $g_{\mathrm{ref}}$ whose output circuits experience, to a good approximation, only stochastic errors [Eqs.~\eqref{eq:reference-compiler-condition-1}--\eqref{eq:reference-compiler-condition-2}]. This section describes three constructions of $g_{\mathrm{ref}}$, two of which are used in the experiments presented in this paper. Each rests on different assumptions and is most applicable to a different setting. The \emph{standard randomized compiler} (Section~\ref{sssec:standard_rc}) applies when QUOPS circuits can be compiled exactly, as they can on physical qubits; it was used for all of the physical-qubit experiments. The other two constructions apply when the $R_P(\theta)$ rotations must be synthesized approximately from a discrete gate set, as in FTQC architectures, and they differ in how they make MCFE sensitive to the resulting synthesis error. The \emph{weighted synthesis-randomization compiler} (Section~\ref{sssec:weighted_synthesis}) randomizes the synthesis itself, so that synthesis errors become stochastic. \emph{Depolarizing channel insertion} (Section~\ref{sssec:depolarizing_insertion}) instead synthesizes each rotation deterministically and reproduces its known infidelity by inserting a random logical Pauli gate after it, with a probability given by this infidelity. This last construction requires logical Pauli gates to be essentially error free, which is true in many FTQC architectures, and in exchange it bounds $\bar{\polarization}_{w,s}$ from below using a single circuit ensemble. This third approach was used for the logical-qubit experiments on \helios.

\subsubsection{Standard randomized compilation}\label{sssec:standard_rc}
The standard randomized compiler is the construction used in the original MCFE work~\cite{Proctor2022-zs}. It compiles $c_{w,s}$ into a gate set $G_{\mathrm{arch}}'$ that need not equal the gate set $G_{\mathrm{arch}}$ of the compilation under test, but that must contain arbitrary single-qubit gates and two-qubit Clifford gates such as CNOT, possibly with restricted connectivity. It proceeds in two stages:

\begin{enumerate}
\item $c_{w,s}$ is compiled exactly into a circuit $c'_{w,s}$ over $G_{\mathrm{arch}}'$, so that $U(c'_{w,s}) = U(c_{w,s})$, with $c'_{w,s}$ arranged as alternating layers of single-qubit and two-qubit gates. For example, a CNOT gate between qubits that are not connected is replaced by a chain of CNOT gates between connected qubits.
 
\item Pauli frame randomization~\cite{Knill2005-xm, Wallman2016-rd, Hashim2021-my, Hashim2023-qk} is applied to $c'_{w,s}$: an independent, uniformly random Pauli layer is inserted before each two-qubit gate layer, the compensating Pauli layer is inserted after it, and both are absorbed into the neighboring single-qubit layers, which is possible because $G_{\mathrm{arch}}'$ contains arbitrary single-qubit gates.
\end{enumerate}

Averaged over this randomization, the effective error in $g_{\mathrm{ref}}(c_{w,s})$ is approximately stochastic whenever any one of the following conditions holds~\cite{Wallman2016-rd}: (1) single-qubit gate errors are small compared with two-qubit gate errors; (2) single-qubit gate errors are approximately gate independent; or (3) single-qubit gate errors are predominantly stochastic. These conditions are commonly met on physical qubits, as randomized compilation experiments have demonstrated~\cite{Hashim2021-my, Hashim2023-qk}.

\subsubsection{Weighted synthesis randomization}\label{sssec:weighted_synthesis}
In FTQC architectures the $R_P(\theta)$ rotations must be synthesized from a discrete gate set, so synthesis error is unavoidable and can be significant. A naive mirror circuit, in which $g(c_{w,s})$ is followed by its exact inverse, would coherently cancel these synthesis errors, resulting in unrealible estimates of circuit polarization. The weighted synthesis-randomization compiler prevents this cancellation by making the synthesis error of the reference circuit stochastic, using the method of Campbell~\cite{Campbell_2017}.

Campbell's method converts coherent synthesis error into incoherent error by sampling each synthesized gate from a weighted ensemble. Because QUOPS circuits contain only Pauli-axis rotations, the axial-rotation construction of Ref.~\cite{Campbell_2017} (Sec.~V therein) suffices. For each $R_P(\theta)$ gate, two approximations $U_1$ and $U_2$, each within $\epsilon$ of the target, are produced by any synthesis algorithm with a guaranteed precision, such as \texttt{GridSynth}~\cite{selinger2012efficient, ross2016optimal}, together with their conjugates $U_3 = ZU_1Z$ and $U_4 = ZU_2Z$. Sampling from $\{U_1,U_2,U_3,U_4\}$ with the weights prescribed in Ref.~\cite{Campbell_2017} yields an approximation to $R_P(\theta)$ whose synthesis error is incoherent.

The reference circuit, over the gate set $\{\mathrm{CNOT},H,T,S,X,Y,Z\}$, is then built in three steps:
\begin{enumerate}
\item Each $R_P(\theta)$ gate in $c_{w,s}$ is replaced by a randomly synthesized approximation as above, with the precision $\epsilon$ chosen to balance synthesis error against noise in the implemented gates. \item Random Pauli layers and their compensating layers are inserted around each CNOT layer, exactly as in the standard randomized compiler.
\item CNOT gates are routed onto the available connectivity where necessary.
\end{enumerate}

The result satisfies the requirements of MCFE provided that the errors on the single-qubit gates $H$, $T$, $S$, $X$, $Y$ and $Z$ are approximately stochastic. This is expected in FTQC architectures, because error correction renders logical errors approximately stochastic.

\subsubsection{Depolarizing channel insertion}\label{sssec:depolarizing_insertion}
The third construction, used for the logical-qubit experiments on \helios, also targets FTQC architectures. Rather than randomizing the synthesis, it inserts after each synthesized rotation $R_P(\theta)$ a logical depolarizing channel whose error probability equals the process infidelity of that rotation's approximate compilation. Because the inserted channel is realized by sampling logical Pauli gates, the mirror circuits remain directly executable, and the mean process polarization $\bar{\polarization}_{w,s}$, including synthesis error, can be bounded from below using a single circuit ensemble, which substantially reduces the sample complexity. Below we state the assumptions under which this bound is accurate.

Given an arbitrary-angle axial rotation, $R_Z(\theta)$, and its associated unitary channel $\mathcal{R}_\theta = R_Z(\theta) ( \, \cdot \, ) R_Z(\theta)^\dagger$, the process fidelity $F$ between $\mathcal{R}_\theta$ and another arbitrary-angle rotation $\mathcal{R}_\varphi$ is given by
\begin{equation}
    F(\mathcal{R}_\theta, \mathcal{R}_\varphi) = {\cos\left( \frac{\theta - \varphi}{2} \right)}^2 
    \eqqcolon 1 - r(\theta, \varphi),
    \label{eqn:axial-fidelity-expression}
\end{equation}
where we use the convention $R_Z(\theta) = e^{-i(\theta/2)Z}$ for axial rotations.
Let $\mathcal{D}_p(\rho) = (1-p)\rho + p (X \rho X + Y \rho Y + Z \rho Z) / 3$ denote a single-qubit depolarizing channel with error probability $p$ (i.e., with depolarizing parameter $4p/3$). By unitary invariance of the process fidelity,
\begin{equation}
    F(\mathcal{R}_\varphi, \mathcal{D}_p \mathcal{R}_\varphi) =
    F(\mathcal{I}, \mathcal{D}_p) = 1 - p.
    \label{eqn:depolarizing-channel-fidelity}
\end{equation}
Suppose that $\mathcal{R}_\theta$ represents the target unitary channel and synthesis of $\mathcal{R}_\theta$ results in the unitary channel $\mathcal{R}_\varphi \approx \mathcal{R}_\theta$.
We can enforce that the process fidelity between the target and $\mathcal{R}_\varphi$ is captured by applying the approximate rotation $\mathcal{R}_\varphi$ followed by the depolarizing channel $\mathcal{D}_p$, i.e., $F(\mathcal{R}_\theta, \mathcal{R}_\varphi) = F(\mathcal{R}_\varphi, \mathcal{D}_p \mathcal{R}_\varphi)$, by equating $p = r(\theta, \varphi)$. 
The depolarizing channel can then be achieved in expectation by applying $P=I$ with probability $1-p$, otherwise $P$ is chosen uniformly from the set of non-identity Paulis $\{X, Y, Z \}$. Note that the expression~\eqref{eqn:axial-fidelity-expression} requires that both the target and synthesized unitaries are axial rotations.

Let $g(c)$ denote approximate and deterministic compilation of the logical circuit $c$. We assume that $g(c)$ approximates arbitrary-angle Pauli-axis rotations with easy-to-implement Pauli-axis rotations and that no further compilation takes place.
For example, given a circuit $c_\theta$ consisting of a single $R_z(\theta)$ gate, we have $\mathcal{U}[g(c_\theta)] = \mathcal{R}_\varphi$.
We additionally introduce a {nondeterministic} logical \emph{reference compiler} $g_\text{ref}(\, \cdot \,)$.
Specifically, $g_\text{ref} \sim \mathcal{G}$ with $\mathcal{G}$ a probability distribution over deterministic compilations. A randomized reference compilation is then realized by sampling $g_\text{ref} \sim \mathcal{G}$ and outputting $c' = g_\text{ref}(c)$. The reference compiler is chosen such that, in expectation, it creates an approximate compilation of a QUOPS circuit that experiences only stochastic errors:
\begin{subequations}
\begin{align}
    \mathbb{E}_\mathcal{G} \{\Lambda(g_{\textrm{ref}}(c))\} 
    &= 
    \mathcal{U}(g(c)) \mathcal{E}_{\textrm{ref}}(g(c)) ,
    \label{eq:logical-reference-compiler-condition-1} \\
    \mathbb{E}_\mathcal{G} \{\Lambda(g_{\textrm{ref}}(c^{\dagger}))\} 
    &= 
    \mathcal{E}_{\textrm{ref}}(g(c^{\dagger}))\mathcal{U}(g(c^{\dagger})) .
    \label{eq:logical-reference-compiler-condition-2}
\end{align}
\end{subequations}
As before, the channels $\mathcal{E}_\text{ref}(g(c))$ and $\mathcal{E}_\text{ref}(g(c^\dagger))$ should be approximately stochastic error channels with approximately equal process polarizations.
In practice, we implement either logical Pauli frame twirling ~\cite{beale2023logicalRC,mclaren2025benchmarkingquantuminstruments}, or physical twirling of 2Q gates~\cite{Wallman2016-rd}. 

In the following we set $g = g_\text{ref}$ to minimize the number of experiments that need to be performed, as in Eq.~\eqref{eq:mcfe_equation-simplified}. Our goal is therefore to find a MCFE procedure that estimates $\mathbb{E}[\polarization(\mathcal{U}(c_{w,s}), \Lambda[g_\text{ref}(c_{w,s})])]$ with access only to noisy discrete-angle rotations. To incorporate synthesis error, we introduce a third {nondeterministic} compiler, $f_\text{ref} \sim \mathcal{F}$, which randomly inserts logical Pauli operators before or after synthesized rotations into the circuits $g_\text{ref}(c)$ with error probability equal to the process infidelity of the synthesized rotation, as in Eq.~\eqref{eqn:depolarizing-channel-fidelity}.

We write a QUOPS circuit as $c_{w, s}= e_d l_d \cdots e_2 l_2 e_1 l_1$, where layers of single-qubit (two-qubit) gates are denoted by $l_i$ ($e_i$).
Note that the following discussion uses the convention in which rotations $R_P(\theta)$ are applied \emph{before} the layer of CNOT gates.
Then, for a fixed circuit $c_{w, s}$, 
\begin{equation}
    \mathbb{E}_{\mathcal{F}} [\Lambda(f_{\text{ref}}(c_{w,s}))]
    =
    \prod_{k=1}^d \mathcal{U} (e_k) \mathcal{E}(e_k) \mathcal{U}(g(l_k)) \mathcal{D}(l_k) \mathcal{E}(g(l_k)) ,
    \label{eqn:depolarizing-reference-compiler-expectation}
\end{equation}
where $\mathcal{E}(e_k)$ and $\mathcal{E}(g(l_k))$ are stochastic Pauli channels and $\mathcal{D}(l_k)$ is the composition of depolarizing channels for each $R_P(\theta)$ gate in layer $l_k$. Note that Eq.~\eqref{eqn:depolarizing-reference-compiler-expectation} can be rewritten as
\begin{gather}
    \mathbb{E}_{\mathcal{F}} [\Lambda(f_{\text{ref}}(c_{w,s}))]
    = \mathcal{U}(g(c)) \tilde{\mathcal{E}}_{\textrm{ref}}(g(c)) \\
    \tilde{\mathcal{E}}_{\textrm{ref}}(g(c)) 
    = 
    \prod_{k=1}^d \mathcal{U}^\dagger_k \mathcal{E}(e_k) \mathcal{U}_k \, \mathcal{U}^\dagger_{k-1/2} \mathcal{D}(l_k) \mathcal{E}(g(l_k)) \mathcal{U}_{k-1/2}
\end{gather}
where $\mathcal{U}_{k} = \mathcal{U}( e_kl_k \dots e_1l_1)$ and $\mathcal{U}_{k-1/2} = \mathcal{U}(l_k \dots e_1l_1)$ only includes half of the final double layer. Since we assume the channels $\mathcal{E}(e_k)$ and $\mathcal{E}(g(l_k))$ to be Pauli stochastic,
under the assumptions of MCFE~\cite{Proctor2022-zs}, the process polarization approximately factorizes as
\begin{equation}
    \polarization(\tilde{\mathcal{E}}_{\textrm{ref}}(g(c))) \approx \prod_{k=1}^d \polarization(\mathcal{E}(e_k)) \polarization(\mathcal{D}(l_k)) \polarization(\mathcal{E}(g(l_k))) .
\end{equation}
To estimate $\mathbb{E}[\polarization(\mathcal{U}(c_{w,s}), \Lambda[g_\text{ref}(c_{w,s})])]$, we introduce a single circuit ensemble
\begin{equation}
    M = L' f_\text{ref}(c_{w,s}^\dagger) f_\text{ref}(c_{w,s}) L
    \label{eqn:logical-circuit-ensemble}
\end{equation}
where $L$ consists of independent, random single-qubit Clifford gates, and $L'$ is distributed such that $U(L'L)$ corresponds to an independent and uniformly random Pauli operator.  This ensemble is analogous to the $M_2$ ensemble~\eqref{eqn:2xreference-ensemble-definition}, except that $f_\text{ref}$ additionally inserts stochastic Pauli errors. We assume, as in MCFE~\cite{Proctor2022-zs}, that $\Lambda(L) = \mathcal{E}_{\text{sp}}\mathcal{L}$, where $\mathcal{L} = \mathcal{U}(L)$. Similarly, $\mathcal{L} = \mathcal{L}' \mathcal{P}$ with $\mathcal{P} = \mathcal{U}(P)$ for a uniformly random Pauli operator $P$.
Averaging over the circuit ensemble 
\begin{align}
    \mathbb{E} [ \hat{\polarization}({M}) ] &= 
    \mathbb{E}_{c_{w,s}} \left[
        \polarization(
            \tilde{\mathcal{E}}_\text{ref}(g(c_{w,s})) \mathcal{E}_\text{sp} ,
            \tilde{\mathcal{E}}_\text{ref}(g(c_{w,s})) \mathcal{E}_\text{sp} 
        ) \label{eqn:averaged-logical-polarization-estimator}
    \right] \\
    &\approx \mathbb{E}_{c_{w,s}} \left[  
        \polarization(\mathcal{E}_\text{ref}(g(c_{w,s})))^2 \, \polarization(\mathcal{E}_\text{sp})^2 \, \prod_k\polarization(\mathcal{D}(l_k))^2  
    \right] \notag
\end{align}
where we use the shorthand notation $\polarization(\mathcal{E}) = \polarization(\mathcal{I}, \mathcal{E})$.

As required, the resulting expression accounts for synthesis error through the product of polarizations $\polarization(\mathcal{D}(l_k))^2$ for each layer of 1Q rotations.
Since $\polarization(\mathcal{E}_\text{sp}) < 1$, we are able to bound the quantity of interest by writing
\begin{equation}
    \bar{\polarization}_{w,s} \gtrsim \sqrt{ \mathbb{E} [ \hat{\polarization}({M}) ] }
    .
    \label{eqn:single-circuit-estimator}
\end{equation}
This allows us to bound the circuit-averaged polarization by running only one ensemble of circuits. Alternatively, we could use an expression analogous to Eq.~\eqref{eq:mcfe_equation-simplified} to divide off the SPAM contribution.

In order for Eq.~\eqref{eqn:averaged-logical-polarization-estimator} to hold quantitatively, the only additional assumptions that are required beyond standard MCFE~\cite{Proctor2022-zs} are that (i) logical Paulis are error free, (ii) locally reproducing the process polarization gives an approximately unbiased estimator of the true, global process polarization, and (iii) the root-mean-square (RMS) polarization is close to the mean polarization. Points (ii) and (iii) are justified below sequentially, whereas point (i) needs to be shown on a case-by-case basis. See Sec.~\ref{sec:helios-synthesis} for a justification of point (i) in the context of the $\qeccode{7}{1}{3}$ code experiments performed on \helios.

\begin{figure}[t]
    \centering
    \includegraphics[width=0.75\linewidth]{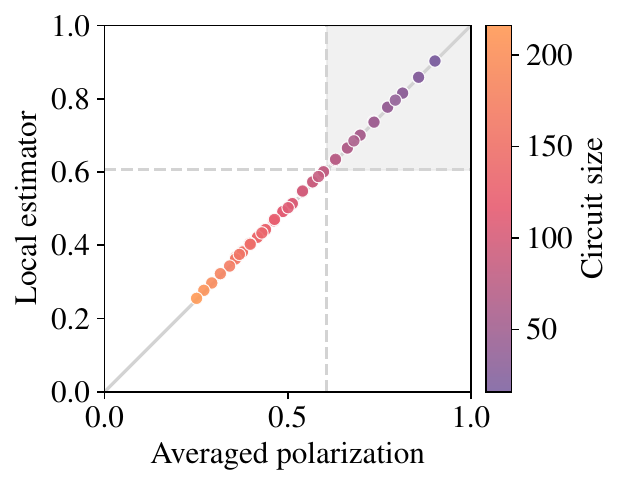}
    \caption{\textbf{Accuracy of depolarizing channel insertion MCFE method.} Approximate vs exact estimators for the process fidelity averaged over circuits using the depolarization channel insertion method of Section~\ref{sssec:depolarizing_insertion}. The exact polarization estimator is constructed by sampling $10^5$ QUOPS circuits and computing the process polarization $\polarization(\mathcal{U}(c_{w,s}), \Lambda[g(c_{w,s})])$ averaged over these randomly chosen circuits for various shapes and sizes. The approximate compilation $g(\,\cdot\,)$ uses three bits of precision (as used in the experiments on \helios). The approximate local estimator is constructed by running noiseless state-vector simulations of the logical circuits drawn from the ensemble Eq.~\eqref{eqn:logical-circuit-ensemble} and evaluating Eq.~\eqref{eqn:single-circuit-estimator}, in which synthesis error is accounted for by inserting stochastic Pauli operators with probability equal to the local process infidelity. The gray dashed lines indicate the $1/\sqrt{e}$ threshold used to determine the score in the QUOPS benchmark.}
    \label{fig:estimated-fidelity}
\end{figure}

To provide evidence that the strategy of \emph{locally} reproducing the process fidelity of synthesized unitaries gives rise to an accurate estimator of the \emph{global} process fidelity---even in the absence of noise---we perform noiseless numerical simulations of QUOPS circuits. The simulations additionally show that RMS and mean polarizations approximately coincide. We plot the results of evaluating $\mathbb{E}[\hat{\polarization}(M)]$ (using noiseless simulation and $10^5$ shots) and taking its square root, against an unbiased estimator for the mean process polarization (evaluating the empirical mean of the exact process polarization over $10^5$ random QUOPS circuits). We plot these two estimators against one another for a variety of QUOPS circuit shapes and sizes in Fig.~\ref{fig:estimated-fidelity}, which illustrates that the estimator Eq.~\eqref{eqn:single-circuit-estimator} accurately incorporates the degradation in process polarization due to synthesis error.

\subsection{Estimating QUOPS circuit ensemble success}\label{ssec:ensemble_mcfe}
This section specifies the estimator for the QUOPS circuit ensemble success metric, $\bar{\polarization}_{w,s}$. The QUOPS experiment at fixed $(w,s)$ consists of sampling $K \geq 1$ QUOPS circuits of shape $(w,s)$, denoted $c_{w,s,k}$ with $k=1,2,\dots,K$. Each QUOPS circuit is converted into 
\begin{enumerate}
\item $L_1 \geq 1$ type-1 mirror circuits $m_{1,l}(c_{w,s,k})$ with $l=1,2,\dots,L_1$, each sampled independently from the distribution of $M_1(c_{w,s,k})$, and \item $L_2 \geq 1$ type-2 mirror circuits $m_{2,l}(c_{w,s,k})$ with $l=1,2,\dots,L_2$, each sampled independently from the distribution of $M_2(c_{w,s,k})$ (unless, $M_1$ and $M_2$ are identically distributed, in which case separate type-2 mirror circuits are not needed).
\end{enumerate}
In addition, $L_3 \geq 1$ type-3 mirror circuits are sampled (which can be shared across all width-$w$ shapes), $m_{3,l}(w)$ with $l=1,2,\dots,L_3$ which are independently sampled from the distribution of $M_3(w)$. In each case, each circuit is executed $\kappa \geq 1$ times (different numbers of circuit executions are possible but we do not include this in our notation), to estimate its effective polarization using the estimator $\hat{\epolarization}(c,\kappa)$ given in Eq.~\eqref{eq:effective_polarization_estimator}, which is applicable to any definite-outcome circuit.

The effective polarization estimates for individual circuits are combined to estimate $\bar{\polarization}_{w,s}$ as follows. When the $M_1$ and $M_2$ circuit ensembles are not identically distributed (which is the standard assumption), we use the following estimator: 
\begin{equation}
    \hat{\bar{\polarization}}_{w,s} =
\frac{1}{\sqrt{\frac{1}{L_3}\sum_{l=1}^{L_3}\hat{\epolarization}(m_{3,l}(w),\kappa)}} \,
    \frac{\frac{1}{L_1 K}\sum_{k=1}^{K}\sum_{l=1}^{L_1} \hat{\epolarization}(m_{1,l}(c_{w,s,k}),\kappa)}{\sqrt{\frac{1}{L_2 K}\sum_{k=1}^{K}\sum_{l=1}^{L_2} \hat{\epolarization}(m_{2,l}(c_{w,s,k}),\kappa)}}.
    \label{eq:meanpol-estimator-1}
 \end{equation}
When the $M_1$ and $M_2$ circuit ensembles are identically distributed, and so separate $m_2$ circuits are not run, we use the simpler estimator
\begin{equation}
    \hat{\bar{\polarization}}_{w,s} =
   \sqrt{\frac{\frac{1}{L_1 K}\sum_{k=1}^{K}\sum_{l=1}^{L_1} \hat{\epolarization}(m_{1,l}(c_{w,s,k}),\kappa)}{\frac{1}{L_3}\sum_{l=1}^{L_3}\hat{\epolarization}(m_{3,l}(w),\kappa)}}. \label{eq:meanpol-estimator-2}
 \end{equation}
Equation~\eqref{eq:meanpol-estimator-2} follows directly from the simplified MCFE equation, Eq.~\eqref{eq:mcfe_equation-simplified}, by replacing each expectation value with the corresponding sample mean over all sampled circuits and executions, and then averaging the resulting squared polarizations over the $K$ QUOPS circuits before taking the square root. Equation~\eqref{eq:meanpol-estimator-1} is obtained in the same way from an approximation to the general MCFE equation, as we now explain

The estimator in Eq.~\eqref{eq:meanpol-estimator-1} comes from applying the estimator for effective polarization from Eq.~\eqref{eq:mcfe_equation-simplified} and the standard sample mean estimator of the mean to the following equation:
\begin{equation}
    \bar{\polarization}_{w,s}' =  \frac{1}{\sqrt{\mathbb{E}_{M_3}\left\{ \epolarization(M_3(w))\right\} }}   \frac{\mathbb{E}_{C_{w,s},M_1}\left\{\epolarization(M_1(C_{w,s}))\right\}}{\sqrt{\mathbb{E}_{C_{w,s}}\mathbb{E}_{M_2}\left\{\epolarization(M_2(C_{w,s}))\right\}}}.
    \label{eq:approximate_estimator}
\end{equation}
This is obtained by taking an approximation to the MCFE estimator for an individual QUOPS circuit, given in Section~\ref{ssec:mcfe}. We now explain this relationship and why we make this approximation. 

Applying the MCFE formula for a circuit's polarization, given in Eq.~\eqref{eq:mcfe_equation}, and the defining formula for the QUOPS success metric  $\bar{\polarization}_{w,s}$ in Eq.~\eqref{eq:mean_process_polarization} gives
\begin{align}
    \bar{\polarization}_{w,s} &=  \mathbb{E}_{C_{w,s}}\{\polarization(C_{w,s})\}\\
    & \approx
    \frac{1}{\sqrt{\mathbb{E}_{M_3}\left\{ \epolarization(M_3(w))\right\} }} \, \mathbb{E}_{C_{w,s}}\left\{\frac{\mathbb{E}_{M_1}\left\{\epolarization(M_1(C_{w,s}))\right\}}{\sqrt{\mathbb{E}_{M_2}\left\{\epolarization(M_2(C_{w,s}))\right\}}}\right\}. \label{eq:standard_mcfe_equation}
\end{align}
This can be rewritten as follows:
\begin{equation}
    \bar{\polarization}_{w,s} \approx \frac{1}{\sqrt{\mathbb{E}_{M_3}\left\{ \epolarization(M_3(w))\right\} }} \Bigg( \,  \frac{\mathbb{E}_{C_{w,s},M_1}\left\{\epolarization(M_1(C_{w,s}))\right\}}{\mathbb{E}_{C_{w,s}}\sqrt{\mathbb{E}_{M_2}\left\{\epolarization(M_2(C_{w,s}))\right\}}}  + 
     \mathbb{COV}\Bigg).
    \label{eq:expanded_mcfe_equation}
\end{equation}
where
\begin{multline}
     \mathbb{COV} = 
    \mathbb{E}_{C_{w,s}}\left(\frac{ \mathbb{E}_{M_1}\left\{\epolarization(M_1(C_{w,s}))\right\}}{\sqrt{\mathbb{E}_{M_2}\left\{\epolarization(M_2(C_{w,s}))\right\}}}  \right)
     -  \frac{\mathbb{E}_{C_{w,s}} \mathbb{E}_{M_1}\left\{\epolarization(M_1(C_{w,s}))\right\}}{\mathbb{E}_{C_{w,s}}\sqrt{\mathbb{E}_{M_2}\left\{\epolarization(M_2(C_{w,s}))\right\}}}.
\end{multline}
We obtain Eq.~\eqref{eq:approximate_estimator} under the assumption that $\mathbb{COV} = 0$. This covariance term is typically non-zero and negative, but is also typically small in practice. This means that the estimator in Eq.~\eqref{eq:meanpol-estimator-1} is typically biased towards slightly over-optimistic (too large) estimates of $\bar{\polarization}_{w,s}$. However, the advantage of the estimator of Eq.~\eqref{eq:meanpol-estimator-1} is that it has lower variance.

\emph{Uncertainty in the estimate.} The statistical uncertainty in $\hat{\bar{\polarization}}_{w,s}$ has two sources: the finite number $K$ of sampled QUOPS circuits (and of sampled mirror circuits per QUOPS circuit), and the finite number $\kappa$ of executions of each circuit. We quantify this uncertainty with a non-parametric bootstrap over circuits and circuit executions: the $K$ sampled QUOPS circuits are resampled with replacement together with all of their mirror circuits and the shot noise in their executions, and the $L_3$ SPAM reference circuits are resampled with replacement in the same way. The estimator of Eq.~\eqref{eq:meanpol-estimator-1} [or Eq.~\eqref{eq:meanpol-estimator-2}] is recomputed on each resample. Resampling in this way captures the circuit-to-circuit variation in polarization as well as the shot noise, and it is the procedure used for all experiments in this paper. The bootstrap distribution provides the standard deviation $\sigma_{w,s}$ of $\hat{\bar{\polarization}}_{w,s}$ and a one-sided $95\%$ lower confidence bound on $\bar{\polarization}_{w,s}$, which we compute as $\hat{\bar{\polarization}}_{w,s} - \Phi^{-1}(0.95)\,\sigma_{w,s}$ using a normal approximation (or the $5$th percentile of the bootstrap distribution can be used instead). Section~\ref{ssec:estimating_regions_scores} specifies how these per-shape estimates and uncertainties are used to test whether a shape is in the capability region.

\subsection{Estimating QUOPS capability regions and scores}
\label{ssec:estimating_regions_scores}

This section specifies how QUOPS capability regions and scores are estimated. Sections~\ref{ssec:mcfe}--\ref{ssec:ensemble_mcfe} specify how to estimate the QUOPS ensemble success metric $\bar{\polarization}_{w,s}$ at a single circuit shape. This section specifies how those per-shape estimates are turned into an estimated QUOPS capability region and an estimated QUOPS score, with statistical guarantees. The capability region and score are defined in Sections~\ref{ssec:capability_regions} and \ref{ssec:quops_score} using the true and \emph{unknown} values of $\bar{\polarization}_{w,s}$. The estimate $\hat{\bar{\polarization}}_{w,s}$ at each circuit shape has statistical uncertainty and a typical QUOPS experiment is likely to test many circuit shapes (a few to hundreds). So, declaring a shape ``successful'' whenever $\hat{\bar{\polarization}}_{w,s} \geq \alpha$ would systematically over-estimate the QUOPS capability region and score (for example, if $\bar{\polarization}_{w,s}=\alpha - \delta$ for some tiny $\delta$ at every tested shape, we would declare success roughly half of the time even though the polarization is below the threshold at every shape). QUOPS therefore requires that the QUOPS capability region and score are estimated with stated confidence:
\begin{itemize}
    \item The QUOPS score $Q$ must be estimated with at least 95\% confidence, meaning that the procedure produces an estimate $\hat{Q}$ satisfying $\hat{Q} \leq Q$ with probability at least 95\%.
    \item The QUOPS capability region must be estimated with at least 90\% confidence, meaning that the estimated region contains one or more shapes that are outside the true capability region with probability at most 10\%.
\end{itemize}
Any statistical procedure that meets these criteria is permitted, and the confidence levels must be stated alongside the results if they differ from these defaults.

The procedures used for all results in this paper are stated below. They are constructed from a hypothesis test implemented at each circuit shape (Section~\ref{sssec:quops_test}), a gated sequence of tests that estimates the QUOPS score (Section~\ref{sssec:gated_score}), and a multiple-comparisons procedure applied to the remaining shapes that completes the QUOPS capability region estimate (Section~\ref{sssec:region_estimation}). Section~\ref{sssec:wagering} then describes a more flexible \emph{significance wagering} procedure for estimating the QUOPS score, of which the gated procedure is a special case. Throughout, $\alpha$ is the polarization threshold ($\alpha = 1/\sqrt{e}$ for the standard capability region and score, and $\alpha < 1/\sqrt{e}$ for error-mitigated regions and scores; see Section~\ref{ssec:conditional_quops}). The procedures are run separately, on the same data, for each value of $\alpha$ of interest.

\subsubsection{The QUOPS test at a single circuit shape}\label{sssec:quops_test}
At each tested shape $(w,s)$, the mirror circuit data are used to compute a point estimate $\hat{\bar{\polarization}}_{w,s}$ of the mean process polarization (using Eq.~\eqref{eq:meanpol-estimator-1}, or Eq.~\eqref{eq:meanpol-estimator-2}) together with an estimate $\sigma_{w,s}$ of the standard deviation of $\hat{\bar{\polarization}}_{w,s}$. We estimate $\sigma_{w,s}$ using a non-parametric bootstrap in which the sampled circuits within each mirror circuit ensemble are resampled with replacement. This bootstrap accounts for both the finite number of sampled circuits and the finite number of executions of each circuit, which are the two sources of statistical uncertainty in $\hat{\bar{\polarization}}_{w,s}$.

The \emph{QUOPS test} at shape $(w,s)$ with threshold $\alpha$ is a one-sided hypothesis test of
\begin{equation}
    H_{w,s}:\ \bar{\polarization}_{w,s} < \alpha \qquad \textrm{against} \qquad \bar{H}_{w,s}:\ \bar{\polarization}_{w,s} \geq \alpha .
\end{equation}
The null hypothesis $H_{w,s}$ is that the shape is \emph{outside} the capability region, and the shape is declared successful only if the null hypothesis is rejected. A false rejection (called a type I error) causes an over-estimate of the QUOPS capability region and/or score. We use the normal approximation to the sampling distribution of $\hat{\bar{\polarization}}_{w,s}$,  so our test statistic and one-sided $p$-value are
\begin{equation}
    z_{w,s} = \frac{\hat{\bar{\polarization}}_{w,s} - \alpha}{\sigma_{w,s}}, \qquad
    p_{w,s} = 1 - \Phi(z_{w,s}), \label{eq:quops_test_pvalue}
\end{equation}
where $\Phi$ is the cumulative distribution function of the standard normal distribution. A test at significance level $\delta$ rejects $H_{w,s}$---i.e., the shape \emph{passes the QUOPS test at level $\delta$}---if and only if $p_{w,s} \leq \delta$. Equivalently, the shape passes if and only if the one-sided lower confidence bound
\begin{equation}
    \hat{\polarization}_{\mathrm{lower},\beta,w,s} = \hat{\bar{\polarization}}_{w,s} - \Phi^{-1}(\beta)\,\sigma_{w,s}, \qquad \beta = 1 - \delta, \label{eq:lower_bound}
\end{equation}
satisfies $\hat{\polarization}_{\mathrm{lower},\beta,w,s} \geq \alpha$, which is the form in which the test is stated in the Methods.

A single QUOPS test controls the probability of a false QUOPS test pass at one shape. The estimated capability region and score are, however, functions of the outcomes at all tested shapes, and so the procedures that produce them must control the probability of \emph{any} false QUOPS test pass, among the shapes that contribute to them. This is a multiple-hypothesis test problem. We address this using different approaches for the QUOPS score and capability region, which are designed so that taking data at more circuit shapes does not dilute the statistical significance allocated to estimating the QUOPS score, which would suppress the estimated QUOPS score and discourage ``mapping out'' of the capability region. The QUOPS score is estimated using a sequential (gated) procedure that requires no correction for multiple comparisons, while the capability region is completed using a family-wise error rate (FWER) controlling procedure.

\subsubsection{Estimating the QUOPS score with gated tests}\label{sssec:gated_score}

The QUOPS score is estimated using a gated testing procedure \cite{Mascha2012-vg}. Before the tests are performed, an ordered sequence of circuit shapes within the admissible cone of Eq.~\eqref{eq:cone},
\begin{equation}
    \mathcal{S} = \big((w_1,s_1), (w_2,s_2), \dots, (w_n,s_n)\big),
\end{equation}
such that 
\begin{equation}
w_i^2 \leq s_i \leq w_i^3, \qquad s_1 < s_2 < \dots < s_n,
\end{equation}
is chosen. The shapes are tested in this order, each at level $\delta = 0.05$, and the procedure stops at the first shape that fails. Let $m$ be the number of shapes that passed before the procedure stopped ($m = n$ if every shape in $\mathcal{S}$ passed). The estimated QUOPS score is
\begin{equation}
    \hat{Q} = s_m, \label{eq:gated_score}
\end{equation}
the size of the last shape that passed, with $\hat{Q} = 0$ if $m = 0$. $\hat{Q}$ is achieved at width $\hat{w}_Q = w_m$. Shapes in $\mathcal{S}$ that come after the first failure are not tested (and, if the experiment is run sequentially, their data need not be collected). The shape at which the procedure stopped, and all shapes after it in $\mathcal{S}$, are also not included in the estimated capability region, as explained in Section~\ref{sssec:region_estimation}.

This procedure estimates $Q$ with 95\% confidence, i.e., 
\begin{equation}
\Pr(\hat{Q} > Q) \leq 0.05,
\end{equation}
 without any correction for the number of shapes tested and without any assumption about how $\bar{\polarization}_{w,s}$ varies with $w$ and $s$. To see this, note that $\hat{Q} > Q$ requires that some shape in $\mathcal{S}$ with size larger than $Q$ passed, and every such shape is outside the capability region (by the definition of $Q$ as the largest size of any in-cone shape in the capability region), so its null hypothesis is true. Consider the first shape in $\mathcal{S}$ whose null hypothesis is true. If it is reached, it passes with probability at most $0.05$. If it fails, the procedure stops before any other true-null shape is tested, so no false pass can occur. So the probability of one or more false passes, and therefore of $\hat{Q} > Q$, is at most $0.05$. The same argument shows that all shapes that passed are simultaneously in the true capability region with probability at least $0.95$, which is used in Section~\ref{sssec:region_estimation}.

The guarantee relies on two conditions. First, the sequence $\mathcal{S}$ must be fixed independently of the data used in the tests: it may be chosen before any data are collected, or it may be generated by an online rule in which the choice of $(w_{i+1}, s_{i+1})$ depends only on the outcomes of the tests at $(w_1,s_1),\dots,(w_i,s_i)$ and on any data not used in the tests (e.g., preliminary experiments or simulations). Choosing or reordering $\mathcal{S}$ after examining the estimates $\hat{\bar{\polarization}}_{w,s}$ at the candidate shapes invalidates the guarantee. Second, the data used to test different shapes must be independent, which holds when each shape's test uses only the mirror circuits sampled for that shape (the width-$w$ SPAM reference circuits from the $M_3$ ensemble may be shared between shapes of the same width, which introduces a weak positive dependence between the tests at those shapes that does not affect the validity of the gated procedure).

The sequence $\mathcal{S}$ should be chosen to contain the shapes at which the QUOPS score is expected to be maximized, ordered by increasing size. For the results in this paper, $\mathcal{S}$ consisted of all tested in-cone shapes at a single width, ordered by increasing size, with the width chosen as that at which the score was expected to be maximized based on preliminary simulations and experiments (width 6 for \willow and \boston, width 16 for \helios, and width 4 for the logical-qubit experiments on \helios). For \htwoone, $\mathcal{S}$ contained shapes at two widths, $\mathcal{S} = ((8,512), (12,1320), (12,1392), (12,1488))$. The experimental sections of this SI describe the shapes run on each system.

The gated procedure is simple and powerful when the tester is confident in their prior knowledge of how polarization varies with circuit shape (e.g., that narrower circuits have higher polarizations at the same circuit size). However, the entire 5\% significance budget is spent on the first failure. If, for example, a failure occurs earlier than expected in the sequence, the tester cannot then test different shapes to potentially increase the estimate of the QUOPS score. Section~\ref{sssec:wagering} describes a generalization that removes this restriction, at the cost of not implementing each test at 5\% significance.

\subsubsection{Estimating the QUOPS capability region}\label{sssec:region_estimation}

The capability region is estimated from the outcomes of two families of tests. The first family consists of the gated tests of Section~\ref{sssec:gated_score} (or their generalization in Section~\ref{sssec:wagering}). The second family consists of tests at all tested shapes that are \emph{not} in $\mathcal{S}$,  denoted by $\mathcal{R}$. These could be, for example, shapes outside the admissible cone. Because the shapes in $\mathcal{R}$ are not tested sequentially, and the aim is to include as many of them as the data justify, a correction for multiple comparisons is required. We control the FWER of the second family at 5\%, i.e., the probability that one or more shapes in $\mathcal{R}$ falsely pass is at most 5\%. The simplest FWER-controlling procedure is the Bonferroni correction, which tests each of the $\zeta = |\mathcal{R}|$ shapes at level $0.05/\zeta$ (equivalently, at confidence $\beta = 1 - 0.05/\zeta$). Our analysis instead uses Hochberg's step-up procedure \cite{Hochberg1988}, which is uniformly more powerful than Bonferroni and is valid when the tests are independent or positively dependent \cite{Simes1986, Sarkar1997}, as they are here. Hochberg's procedure orders the $p$-values of the shapes in $\mathcal{R}$ as $p_{(1)} \leq p_{(2)} \leq \dots \leq p_{(\zeta)}$, finds the largest $k$ such that
\begin{equation}
    p_{(k)} \leq \frac{0.05}{\zeta - k + 1}, \label{eq:hochberg}
\end{equation}
and declares the shapes with the $k$ smallest $p$-values successful (if no such $k$ exists, no shape in $\mathcal{R}$ is declared successful). Equivalently, the procedure computes Hochberg-adjusted $p$-values $\tilde{p}_{(i)} = \min_{j \geq i} \min\{(\zeta - j + 1)\, p_{(j)},\, 1\}$ and declares successful all shapes with $\tilde{p} \leq 0.05$.

Denote by $\mathcal{D}$  the set of \emph{declared-successful} shapes: the shapes in $\mathcal{S}$ that passed the gated tests together with the shapes in $\mathcal{R}$ that were declared successful by Hochberg's procedure. The estimated capability region is the downward closure of $\mathcal{D}$,
\begin{equation}
    \hat{Q}_{\textrm{region},\alpha} = \left\{ (w,s) \;\mid\; w \leq w' \textrm{ and } s \leq s' \textrm{ for some } (w',s') \in \mathcal{D} \right\}, \label{eq:region_closure}
\end{equation}

The estimated region has 90\% confidence. The probability that $\mathcal{D}$ contains a shape outside the true capability region is at most the probability of a false QUOPS test pass in the gated tests, which is at most 5\% (Section~\ref{sssec:gated_score}) added to the probability of a false QUOPS test pass among the shapes in $\mathcal{R}$, which is at most 5\% by FWER control, giving at most 10\% in total by a union bound. The downward closure in Eq.~\eqref{eq:region_closure} adds shapes that were not tested, and it is justified by the assumption that $\bar{\polarization}_{w,s}$ is non-increasing in both $w$ and $s$, i.e., that the true capability region is itself downward closed. Under this assumption the closure adds no shapes outside the true region, and so $\hat{Q}_{\textrm{region},\alpha}$ contains any shape outside the true region with probability at most 10\%. This monotonicity is expected to typically hold in practice but it can be violated in some circumstances such as, for example, for a system that drifts between the experiments at different shapes. Without this assumption, the 90\% guarantee applies to $\mathcal{D}$ but not to its closure.

\subsubsection{Significance wagering: a flexible procedure for estimating the QUOPS score}\label{sssec:wagering}
The gated procedure of Section~\ref{sssec:gated_score} is the simplest member of a family of sequential procedures that estimate the QUOPS score with 95\% confidence. In the general procedure, which we call \emph{significance wagering}, the tester holds a significance budget of $\delta_{\textrm{tot}} = 0.05$ and wagers a part of it on each test. A wager is lost if the test fails and is returned if the test passes. This allows a tester who is uncertain about which width (or which sizes) will maximize the score to hedge, continuing to test after a failure at the cost of testing each shape at a smaller significance level.

The procedure is as follows. Initialize the spent significance to $\beta_{\textrm{spent}} = 0$ and the set of passed shapes to $\mathcal{P} = \emptyset$. Then repeat the following steps until the budget is exhausted ($\beta_{\textrm{spent}} = \delta_{\textrm{tot}}$), or until the experimenter chooses to stop.
\begin{enumerate}
    \item Choose a circuit shape $(w,s)$ in the admissible cone and a significance level $\delta$ for the test at that shape, satisfying
    \begin{equation}
        0 < \delta \leq \delta_{\textrm{tot}} - \beta_{\textrm{spent}}. \label{eq:wager_constraint}
    \end{equation}
    Both choices may depend on the outcomes of all previous tests, on any data not used in the tests, and on prior knowledge, but not on the data that will be used in this test. Any user-specified algorithm for making these choices is permitted.
    \item Collect the mirror circuit data at $(w,s)$ (or use previously collected data at $(w,s)$ that has not been used in any earlier test) and perform the QUOPS test of Section~\ref{sssec:quops_test} at level $\delta$.
    \item If the test passes, add $(w,s)$ to $\mathcal{P}$; $\beta_{\textrm{spent}}$ is unchanged. If the test fails, update $\beta_{\textrm{spent}} \to \beta_{\textrm{spent}} + \delta$.
\end{enumerate}
The estimated QUOPS score is the largest size of any passed shape,
\begin{equation}
    \hat{Q} = \max\{ s \mid (w,s) \in \mathcal{P} \}, \label{eq:wagered_score}
\end{equation}
with $\hat{Q} = 0$ if $\mathcal{P}$ is empty. The gated procedure of Section~\ref{sssec:gated_score} is the special case in which every wager is the entire remaining budget ($\delta = \delta_{\textrm{tot}}$ on every test), so that the first failure uses up the entire budget, and in which the shapes are chosen from a pre-specified list in increasing size.

The significance wagering procedure controls the probability of one or more false passes at $\delta_{\textrm{tot}}$, and therefore satisfies $\Pr(\hat{Q} > Q) \leq \delta_{\textrm{tot}} = 0.05$, for any adaptive choice of shapes and wagers and with no assumption about the dependence of $\bar{\polarization}_{w,s}$ on $w$ and $s$. The argument is a generalization of that for the gated procedure. Consider the sequence of tests performed, and let the $i$th test be at level $\delta_i$. Conditioned on the outcomes of the first $i-1$ tests, the $i$th test is a valid level-$\delta_i$ test (its shape and level are fixed given those outcomes, and its data are independent of them), so if its null hypothesis is true it passes with probability at most $\delta_i$. Let $T$ be the index of the first test at which a false pass occurs ($T = \infty$ if there is none). Every true-null test before $T$ failed, and so its wager was added to $\beta_{\textrm{spent}}$; together with Eq.~\eqref{eq:wager_constraint} for the $T$th wager this implies that, on every possible run of the procedure,
\begin{equation}
    \sum_{i \leq T,\ H_i\ \textrm{true}} \delta_i \;\leq\; \beta_{\textrm{spent},T} + \delta_T \;\leq\; \delta_{\textrm{tot}},
\end{equation}
where $\beta_{\textrm{spent},T}$ is the spent significance immediately before the $T$th test. Summing the conditional probabilities of a first false pass at each step therefore gives $\Pr(T < \infty) \leq \delta_{\textrm{tot}}$. As in Section~\ref{sssec:gated_score}, $\hat{Q} > Q$ requires a false pass, so $\Pr(\hat{Q} > Q) \leq \delta_{\textrm{tot}}$. This procedure is an instance of alpha-spending with recycling of the significance of rejected hypotheses, closely related to the fallback procedure \cite{Wiens2003} and to online FWER-controlling procedures \cite{Tian2021}; the independence of the tests at different shapes is what permits the fully adaptive choice of shapes and wagers.

Two rules are essential to the validity of this procedure. First, the shape and the wager for each test must be fixed before the data for that test are examined. Second, each test must use data that have not been used in any previous test. A shape that failed may be tested again, with a new wager, but only using newly collected data; the earlier data at that shape must not be pooled with the new data.

\subsection{Estimating QUOPS rate}\label{ssec:operational_quops_per_second}

The foundational definition of the QUOPS rate (Sec.~\ref{ssec:foundational_quops_per_s}) is not directly measurable in a QUOPS experiment, both because the QUOPS experiments execute mirror circuits rather than QUOPS circuits and because various time overheads must be accounted for. Here we provide an operational definition of the QUOPS rate. The (operational) QUOPS rate $\Omega(w,s)$ at circuit shape $(w,s)$ is defined by
\begin{equation}
    \Omega(w, s)
    =
    \frac{2 s \, \hat{\polarization}_{w, s}^2 }{\tau_{\text{wall}}(w, s)} \left( N_\text{total}(w, s) N_\text{kept}(w, s) \right)^{\frac{1}{2}}
    \label{eq:omega_operational}
\end{equation}
where $\hat{\polarization}_{w,s}$ is the estimated mean process polarization for shape $(w,s)$, $\tau_{\mathrm{wall}}(w,s)$ is the measured wall-clock time elapsed while executing the mirror circuits (of types $M_1$ and $M_2$) for shape $(w,s)$ QUOPS circuits, and $N_{\mathrm{kept}}(w,s)$ and $N_{\mathrm{reject}}(w,s)$ are the total numbers of executions of these mirror circuits that are accepted or rejected by any error detection procedure, respectively, and 
\begin{equation}
    N_\text{total}(w, s) = N_{\mathrm{kept}}(w,s) +N_{\mathrm{reject}}(w,s).
\end{equation}
The QUOPS rate is intended to be reported alongside the QUOPS score $Q$ at the circuit shape at which it was obtained, i.e., $Q$ is intended to be reported alongside $\Omega=\Omega(w_Q,Q)$.

We now explain why we have chosen this definition for $\Omega(w,s)$. Equation~\eqref{eq:omega_operational} is an operational version of the idealized definition of $\Omega$ given in Eq.~\eqref{eq:omega_foundational} (see Sec.~\ref{ssec:foundational_quops_per_s}). There, $\Omega(w,s)$ is defined as
\[
\Omega(w,s)=\frac{s\,\polarization_{w,s}^2}{\tau(w,s)},
\]
where $\tau(w,s)$ is the execution time per sampled QUOPS circuit of shape $(w,s)$. In the foundational definition,
\[
\tau(w,s)=\mathrm{Median}\{\tau_{\mathrm{circ}}(g(C_{w,s}))\},
\]
where $\tau_{\mathrm{circ}}(g(C_{w,s}))$ is the time required to generate one sample from the compiled QUOPS circuit $g(C_{w,s})$. However, this quantity does not account for several operational overheads in quantum computing systems, and is not defined with respect to the circuit actually executed in a QUOPS experiment (mirror circuits). Equation~\eqref{eq:omega_operational} is designed to address these practical complications while preserving the intended meaning of $\Omega$ as a metric for the useful operations executed per second.

The need for an operational definition of $\Omega$ arises from several features of the QUOPS experiments and real quantum computing systems:
\begin{enumerate}
    \item The time required to gather $N$ samples from $\kappa$ QUOPS circuits of shape $(w,s)$ is often not simply proportional to $\kappa N$, because there is typically an overhead associated with starting execution of a new circuit, for example loading it into control hardware or performing on-the-fly compilation. Generating $N_{\mathrm{total}}$ samples from a single QUOPS circuit may therefore take substantially less time than generating $N_{\mathrm{total}}/\kappa$ samples from each of $\kappa \gg 1$ distinct circuits.

    \item Additional tasks may be performed during the execution of a set of circuits, such as recalibration. Some of these are not cleanly separable from circuit execution as they are necessary for circuit executions to occur. For example, replacing lost ions in a trapped-ion device may be necessary before the next circuit can be executed, but this task is not usually regarded as part of the circuit execution.

    \item QUOPS is implemented using proxy circuits rather than direct executions of the compiled QUOPS circuits. Some of these proxy circuits contain the QUOPS circuits embedded within deeper mirror circuits, and are therefore slower to execute than the corresponding compiled QUOPS circuits.

    \item QUOPS permits error detection and postselection. Some complete circuit executions are therefore discarded, and error detection may even terminate an execution before the circuit has completed.
\end{enumerate}

The time $\tau_{\mathrm{wall}}(w,s)$ in Eq.~\eqref{eq:omega_operational} is defined as the total time required to gather all data from the mirror circuits associated with the QUOPS circuits of shape $(w,s)$. Operationally, this is the elapsed time between the start of the first execution and the end of the final execution of the mirror circuits used for the shape-$(w,s)$ experiment. This time need not include the execution of the width-$w$ SPAM-reference circuits, from the $M_3$ ensemble (and note that those circuits are independent of $s$ and may be reused for all circuit shapes with width $w$). For cloud-access experiments, in which precise circuit start and stop times may not be available, the start and end times of submitted circuit batches or jobs may be used to estimate $\tau_{\mathrm{wall}}(w,s)$. This should not include additional time overheads such as queue time or data transfer time, when these times can be excluded.

The parameters $N_{\mathrm{kept}}(w,s)$ and $N_{\mathrm{reject}}(w,s)$ in Eq.~\eqref{eq:omega_operational} account for the number of circuit executions while accounting for the time cost of any error detection and post-selection. We use 
\begin{equation}
N_{\mathrm{kept},\textrm{QUOPS}}(w,s) = \left( N_\text{total}(w, s) N_\text{kept}(w, s) \right)^{\frac{1}{2}}
\end{equation}
as a proxy for the total number of shots that would have been retained in a QUOPS circuit of shape $(w,s)$, which assumes that a rejection is caused by an error in the first or second half of the mirror circuits with equal probability. This is a heuristic that may not be appropriate for all error detection schemes, and it can be replaced with a more accurate heuristic where one exists. 

The factor of $2$ in Eq.~\eqref{eq:omega_operational} accounts for the fact that each mirror circuit is approximately twice as deep as a shape $(w, s)$ circuit, and so corresponds to circuits that are approximately twice as large.

\section{QUOPS targets from challenge problems}
\label{ssec:challenge_problem_conversion}

The main text compares QUOPS scores and rates with the requirements of two representative challenge problems: factoring RSA-2048 with Shor's algorithm~\cite{Shor1994-zh,Gidney2025-hg} and estimating the ground-state energy of FeMoco~\cite{Low2025-nc} (Fig.~\ref{fig:main:capability:quops-vs-rate}). Resource estimates for these problems are stated as a number of logical data qubits (i.e., not including logical qubits used for architecture-specific tasks like $T$-state routing) and a number of Toffoli gates. This section describes how we convert these resources estimates into a \emph{QUOPS target}: the QUOPS score that a fault-tolerant quantum computer would have if it were just capable of executing the challenge-problem circuit. The conversion is a \emph{non-Clifford resource-matching} heuristic. Both the algorithm circuit and the QUOPS circuits are assumed to be compiled into Clifford$+T$ gates, only the non-Clifford resources are assumed to be costly or noisy, and the QUOPS target is the largest QUOPS circuit whose non-Clifford resources fit within the same error budget as the algorithm.

\noindent
\emph{Inputs.}
Table~\ref{tab:challenge_inputs} lists the resource estimates that we use. For RSA-2048 we use the estimate of Gidney~\cite{Gidney2025-hg}, in which a single execution (``shot'') of the period-finding circuit uses $1399$ logical qubits and takes approximately $12$ hours. On average $9.2$ shots are needed, for an expected total of $6.5\times 10^{9}$ Toffoli gates and a total runtime of a little under five days. The Toffoli count per shot is therefore $N_{\textrm{Tof}} = 6.5\times 10^{9}/9.2 = 7.1\times 10^{8}$. For FeMoco we use the estimate of Low \emph{et al.}~\cite{Low2025-nc} for the 76-orbital active space: $1459$ logical qubits and $N_{\textrm{Tof}} = 9.99\times 10^{8}$ Toffoli gates for a single run of phase estimation. No runtime is given in Ref.~\cite{Low2025-nc}, so we use the same five-day target as for RSA-2048.

\begin{table*}[t]
\centering
\label{tab:challenge_inputs}
\begin{ruledtabular}
\begin{tabular}{lrrrrrrrr}
Problem & $w$ & $N_{\textrm{Tof}}$ & $N_T$ & $p_T$ & $n_T^{\star}$ & $Q_{\textrm{target}}$ & $\tau$ & $\Omega_{\textrm{target}}$ (QUOPS/s) \\
\midrule
RSA-2048~\cite{Gidney2025-hg} & 1399 & $7.1\times 10^{8}$ & $2.8\times 10^{9}$ & $1.8\times 10^{-10}$ & 22.2 & $2.5\times 10^{8}$ & 12.07 h & $5.7\times 10^{3}$ \\
FeMoco~\cite{Low2025-nc}      & 1459 & $1.0\times 10^{9}$ & $4.0\times 10^{9}$ & $1.3\times 10^{-10}$ & 22.5 & $3.4\times 10^{8}$ & 5 days  & $8.0\times 10^{2}$ \\
\end{tabular}
\end{ruledtabular}
\caption{\textbf{QUOPS targets for challenge problems.} The resources estimates for challenge problems and the estimated QUOPS score and rate targets. Here $w$ is the number of logical data qubits, $N_{\textrm{Tof}}$ the Toffoli count of one circuit execution, $N_T = 4N_{\textrm{Tof}}$ the corresponding $T$ count used in our analysis, $p_T = 1/(2N_T)$ the logical $T$-gate error rate at which one execution has polarization $1/\sqrt{e}$, $n_T^{\star}$ the optimal number of $T$ gates per synthesized $R_P(\theta)$ rotation [Eq.~\eqref{eq:challenge_optimal_nT}], $Q_{\textrm{target}}$ the QUOPS target [Eq.~\eqref{eq:challenge_target}], $\tau$ the time budget per execution, and $\Omega_{\textrm{target}} = Q_{\textrm{target}}/\tau$ the required QUOPS rate.}
\end{table*}

\noindent
\emph{Step 1: Toffoli count to $T$ count.}
We convert each Toffoli count into a $T$ count using $N_T = 4N_{\textrm{Tof}}$. A Toffoli gate can be synthesized exactly with seven $T$ gates, but most of the Toffoli gates in these algorithms compute a temporary logical AND that is later uncomputed, and a compute--uncompute pair costs four $T$ gates when the uncomputation is implemented by measurement~\cite{gidney2018halving,jones2013low}. The factor of four is therefore a good approximation to the $T$ count of an implementation that synthesizes Toffoli gates from $T$ gates. Fault-tolerant architectures could instead prepare Toffoli (or $\ket{CCZ}$) magic states directly~\cite{eastin2013distilling,jones2013low,Gidney2025-hg}, in which case a Toffoli gate is the natural non-Clifford resource. This changes the target by a constant factor (see below) but not the conclusions of the main text.

\noindent
\emph{Step 2: The logical error budget of the algorithm.}
We model each logical $T$ gate as having process infidelity $p_T$ and we neglect the errors of logical Clifford gates. The polarization of the algorithm circuit is then $\polarization_{\textrm{alg}} = (1-p_T)^{N_T}$. The processor is defined to be just capable of the algorithm if this equals the QUOPS success threshold, $\polarization_{\textrm{alg}} = 1/\sqrt{e}$, which gives
\begin{equation}
    p_T = 1 - e^{-1/(2N_T)} \approx \frac{1}{2N_T}.
    \label{eq:challenge_pT}
\end{equation}
Alternative polarization thresholds that were tailored to the noise tolerance of a particular algorithm could be used instead. The resulting $T$-gate error rates are $1.8\times 10^{-10}$ (RSA-2048) and $1.3\times 10^{-10}$ (FeMoco). The threshold $1/\sqrt{e}$ is the same one used to define QUOPS capability regions, so a machine that meets Eq.~\eqref{eq:challenge_pT} executes the algorithm circuit ``successfully'' in the same sense in which it executes QUOPS circuits successfully.

\noindent
\emph{Step 3: The QUOPS circuits with the same error budget.} Consider QUOPS circuits of width $w$ and size $s$ executed on the same processor. Half of the size of a dense QUOPS circuit is contributed by $R_P(\theta)$ rotations and half by CNOT gates, so the circuit contains $s/2$ rotations (Section~\ref{ssec:quops_circuits}). In a Clifford$+T$ architecture each rotation is approximately synthesized into a sequence containing, on average, $n_T$ $T$ gates, and the inexact synthesis also introduces error. The CNOT gates and the Clifford gates within the synthesized sequences are again assumed to be error free. The mean polarization of the QUOPS circuits is then
\begin{equation}
    \bar{\polarization}_{w,s} \approx (1-p_T)^{n_T s/2}\,\big(1-\epsilon_{\textrm{synth}}(n_T)\big)^{s/2}
    \approx \exp\!\left[-\frac{s}{2}\big(n_T\,p_T + \epsilon_{\textrm{synth}}(n_T)\big)\right],
    \label{eq:challenge_quops_polarization}
\end{equation}
where $\epsilon_{\textrm{synth}}(n_T)$ is the process infidelity per synthesized rotation. Throughout, errors are assumed to be stochastic and independent, so that the polarizations of individual gates multiply.

For the synthesis cost we use the mixed-fallback method of Kliuchnikov \emph{et al.}~\cite{kliuchnikov2023shorter}, which approximates an arbitrary single-qubit rotation to diamond-norm accuracy $\epsilon$ using an average $T$ count of
\begin{equation}
    n_T(\epsilon) = 0.53\log_2(1/\epsilon) + 4.86 ,
\end{equation}
so
\begin{equation}
    \epsilon_{\textrm{synth}}(n_T) = 2^{-(n_T - 4.86)/0.53}.
    \label{eq:challenge_synthesis_cost}
\end{equation}
Mixed-fallback approximations are probabilistic mixtures of exactly implementable unitaries, whose error channel is stochastic rather than coherent. The diamond-norm error and the process infidelity of such a channel are equal, so we identify $\epsilon$ with the polarization loss per rotation. (For a single-qubit Pauli channel the polarization loss is $4/3$ of the diamond-norm error and process infidelity, but this factor shifts $n_T^{\star}$ by $0.2$ and the targets by about $1\%$, and we neglect it.) The stochastic character of the residual error also justifies combining it multiplicatively with the $T$-gate errors. For deterministic unitary approximations, such as those of Ross and Selinger~\cite{ross2016optimal}, the process infidelity is instead of order $\epsilon^{2}$, but the $T$ count is roughly six times larger at fixed $\epsilon$. So a compiler maximizes circuit polarization by using the mixed method.

The QUOPS target is the largest size at which Eq.~\eqref{eq:challenge_quops_polarization} meets the threshold $1/\sqrt{e}$, with $n_T$ chosen to maximize that size:
\begin{equation}
    Q_{\textrm{target}}
    = \max_{n_T}\;\frac{1}{n_T\,p_T + \epsilon_{\textrm{synth}}(n_T)}
    = \frac{2N_T}{\min_{n_T}\big[n_T + 2N_T\,\epsilon_{\textrm{synth}}(n_T)\big]} ,
    \label{eq:challenge_target_def}
\end{equation}
where the second equality uses Eq.~\eqref{eq:challenge_pT}. The minimization balances the two error sources: adding a $T$ gate to each rotation costs $p_T$ in polarization per rotation and reduces the synthesis error by a factor of $2^{1/0.53}\approx 3.7$. With the cost model of Eq.~\eqref{eq:challenge_synthesis_cost} the optimum is attained at a synthesis error of $\epsilon^{\star} = 0.53/(2N_T\ln 2)$, that is, at
\begin{equation}
    n_T^{\star} = 0.53\log_2\!\left(\frac{2N_T\ln 2}{0.53}\right) + 4.86 .
    \label{eq:challenge_optimal_nT}
\end{equation}
This gives
\begin{equation}
    Q_{\textrm{target}} = \frac{2N_T}{n_T^{\star} + 0.53/\ln 2} \approx \frac{2N_T}{n_T^{\star} + 0.76} .
    \label{eq:challenge_target}
\end{equation}
For both problems $n_T^{\star}\approx 22$ (Table~\ref{tab:challenge_inputs}), so 
\begin{equation}
Q_{\textrm{target}} \approx N_T/11.5 \approx 0.35\,N_{\textrm{Tof}}.
\end{equation}
Equivalently, the target QUOPS circuit contains $n_T^{\star} Q_{\textrm{target}}/2 \approx 0.97\,N_T$ $T$ gates. So, to within a few percent, the QUOPS target is simply the QUOPS circuit that is compiled into the same number of $T$ gates as the algorithm's $T$ count. The residual $3\%$ of the error budget is spent on synthesis error. The resulting targets are
\begin{equation}
    Q_{\textrm{target}}^{\textrm{RSA-2048}} = 2.5\times 10^{8},
    \qquad
    Q_{\textrm{target}}^{\textrm{FeMoco}} = 3.4\times 10^{8} .
\end{equation}

We report each target at the logical width of the corresponding algorithm, i.e., at the shapes $(1399, 2.5\times 10^{8})$ and $(1459, 3.4\times 10^{8})$. Both shapes lie inside the admissible cone of Eq.~\eqref{eq:cone}, so a processor whose capability region contains the target shape has a QUOPS score of at least $Q_{\textrm{target}}$. Conversely, a processor whose QUOPS score is below $Q_{\textrm{target}}$, at whatever width that score is achieved, cannot have the target shape in its capability region, and so cannot execute the algorithm circuit under the assumptions of this model. This is the sense in which QUOPS scores and QUOPS targets are directly comparable in Fig.~\ref{fig:main:capability:quops-vs-rate}.

\noindent
\emph{Step 4: Required QUOPS rates.} A QUOPS rate target follows from a time budget $\tau$ for one execution of the algorithm circuit, $\Omega_{\textrm{target}} = Q_{\textrm{target}}/\tau$. For RSA-2048 we take $\tau = 12.07$ hours, the per-shot time of Gidney's estimate~\cite{Gidney2025-hg}, so that the expected $9.2$ shots complete in $4.6$ days (giving 0.4 days of spare time with a five-day target). This gives $\Omega_{\textrm{target}} = 5.7\times 10^{3}$ QUOPS/s. For FeMoco we take $\tau = 5$ days for its single execution, giving $\Omega_{\textrm{target}} = 8.0\times 10^{2}$ QUOPS/s. These QUOPS rate targets are \emph{raw} rates (size divided by time). The QUOPS rate $\Omega$ of Section~\ref{ssec:foundational_quops_per_s} additionally attenuates the raw rate by $\bar{\polarization}^{2}$, so comparing $\Omega_{\textrm{target}}$ with a machine's reported $\Omega$, as in Fig.~\ref{fig:main:capability:quops-vs-rate}, is conservative by a factor of at most $e$.

\noindent
\emph{Sensitivity and limitations.}
Equation~\eqref{eq:challenge_target} makes the dependence of the QUOPS targets, in our model, on the resource estimate inputs explicit: $Q_{\textrm{target}}$ is proportional to $N_T$ up to a slowly varying logarithm, so a factor-of-two change in the Toffoli count, or in the Toffoli-to-$T$ conversion, changes the target by a factor of two. For example, using seven $T$ gates per Toffoli gives $4.2\times 10^{8}$ and $5.9\times 10^{8}$. If Toffoli gates are instead the elementary non-Clifford resource of the architecture, the target depends on the relative error rates of its Toffoli and $T$ gates, and if they are similar it is still within a factor of about two of the values above. The conversion neglects all Clifford errors in both the algorithm and the QUOPS circuits, neglects idling errors, and uses a specific synthesis cost model. It also ignores the very different structure of algorithm and random circuits. Because the targets are compared with QUOPS scores that are currently five orders of magnitude smaller, these approximations do not affect the conclusions of the main text.

\section{Projecting QUOPS for FTQC architectures}
\label{app:projecting-quops}
This section details how the projected QUOPS capability regions and scores were computed, for two families of hypothetical early FTQC systems. Section~\ref{app:flasq} details the calculations for the FTQC architecture based on the surface code and a non-reconfigurable architecture with 2D-planar-grid connectivity (results shown in Fig.~\ref{fig:main:projections:surface_code}). Section~\ref{app:fully_connected} details the calculations for FTQC architectures with all-to-all physical connectivity, based on either the surface code or a concatenated symplectic double code (results shown in Fig.~\ref{fig:main:projections:fully_connected}).

\subsection{Surface code with 2D-planar-grid connectivity}\label{app:flasq}

We consider a family of devices based on the same architectural model used in Ref.~\cite{huggins2025fluidallocationsurfacecode}: each device consists of a monolithic two-dimensional square lattice of locally connected physical qubits implementing the two-dimensional (rotated) surface code. Each device is assumed to support the same logical primitive gate set,
\[
\{H, S, S^\dagger, T, \mathrm{CNOT}\}.
\]
We assume logical CNOT gates performed by lattice surgery, and we assume that the computation logical qubits are arranged in a line when determining the ancilla volume cost of CNOTs. As in Ref.~\cite{huggins2025fluidallocationsurfacecode}, we assume H gates are performed via physical single-qubit gates plus patch rotation, $S$/$S^{\dagger}$ gates are performed via teleportation and Y-basis measurements, and non-Clifford operations are supplied by cultivated \(T\)-states. In this architecture, the single-qubit Clifford gates \(H\), \(S\), and \(S^\dagger\) incur no magic-state cost, and all non-Clifford cost is represented through \(T\)-state consumption. We consider devices with total physical-qubit budgets
\[
N_{\mathrm{pq}} \in \{5000, 10000, 25000, 50000\},
\]
while fixing the syndrome-extraction cycle time to \(t_{\mathrm{cyc}}=1\,\mu\mathrm{s}\), the classical reaction time to \(t_{\mathrm{react}}=10\,\mu\mathrm{s}\), and the physical gate noise model to uniform depolarizing noise with error rate $p_{\mathrm{phys}}=10^{-3}$. These assumptions define the device family for which we forecast QUOPS.

Our goal is to estimate, for each device and each benchmark point $(w,s)$, the average probability that a QUOPS circuit of width $w$ and size $s$ executes successfully, and from this to infer the predicted QUOPS of that device. We accomplish this task by using a simplified version of the FLASQ resource model~\cite{huggins2025fluidallocationsurfacecode}, together with a simple error-budgeting rule for synthesized rotations, logical Clifford operations, and $T$-state infidelity.

\subsubsection{Estimating circuit polarizations using FLASQ}

To estimate the success probability of a benchmark circuit on a fault-tolerant device, we use a simplified version of the FLASQ cost model. FLASQ was developed as a coarse-grained resource model for early fault-tolerant quantum computers based on the two-dimensional lattice surface code architecture introduced above. FLASQ goes beyond simple proxies such as circuit depth or non-Clifford gate count as they do not adequately capture the true implementation cost of a computation on such devices, because they neglect overheads associated with routing, Clifford operations, ancilla management, and magic-state preparation. FLASQ instead provides a heuristic estimate of the spacetime resources required to execute a circuit on a two-dimensional surface-code architecture. 

In the form relevant here, FLASQ takes as input: (i) a circuit $C$ expressed in logical primitive operations, (ii) $N_{\mathrm{tot}}$, the total number of logical qubits available on the device, including ancillas, (iii) $t_{\mathrm{react}}$, the processor's reaction time in units of logical timesteps, and (iv) a dictionary assigning each primitive gate an ancilla volume. We use the same per-gate ancilla-volume assignments as Ref.~\cite{huggins2025fluidallocationsurfacecode}. A logical timestep is the time required for $d$ rounds of syndrome extraction, where $d$ is the code distance.

The FLASQ model outputs a set of coarse-grained resource estimates for the circuit, including the total number of logical timesteps $L$, the total spacetime volume $S$, and the number $M$ of consumed magic states. The FLASQ model also computes several useful quantities as intermediary steps, including $V$, the total ancilla volume of $C$, as well as the Clifford spacetime volume
\begin{equation}
    V_{\mathrm{cliff}} = V - v_{\mathrm{cult}} M,
\end{equation}
where $v_{\mathrm{cult}}$ is the spacetime volume associated with the cultivation of a single $T$-state. Under our model $M$ is equal to the logical $T$-count $N_T$ of the compiled circuit.

We convert these FLASQ outputs into a heuristic estimate of circuit success probability using 
\begin{equation}
    P_{\mathrm{success}} = (1-\epsilon_{\textrm{rot}}^2)^{N_{\textrm{rot}}}(1-p_{\mathrm{cyc}})^{dV_{\mathrm{cliff}}}(1-p_{\mathrm{mag}})^M,
    \label{app:flasq:eq:psuccess}
\end{equation}
where $\epsilon_{\textrm{rot}}$ is the per-rotation synthesis precision in diamond distance, $N_{\textrm{rot}}$ is the number of rotations in the QUOPS circuit, $p_{\mathrm{cyc}}$ is the logical error rate per logical cycle, and $p_{\mathrm{mag}}$ is the infidelity of each consumed $T$-state. This model is the same coarse-grained model used in Eq.~D7 of Ref.~\cite{huggins2025fluidallocationsurfacecode} updated to include variable synthesis error. In our projected QUOPS analysis, we use this no-fault probability as a proxy for the circuit-success metric and compare it directly to the QUOPS success threshold. 

\begin{table*}[t]
\begin{ruledtabular}
\begin{tabular}{cccc}
Physical qubits & Predicted QUOPS & Circuit width ($w$) & T-gates per rotation (average)\\
\hline
5,000   & 1950 & 13 & $20.40 \pm 1.54$ \\
10,000  & 9240 & 21 & $23.98 \pm 2.09$ \\
25,000  & 68470 & 41 & $28.56 \pm 2.96$ \\
50,000  & 273632 & 68 & $32.56 \pm 2.67$ \\
\end{tabular}
\end{ruledtabular}
\caption{\textbf{Projected QUOPS for future fault-tolerant devices.}
For each hypothetical device, we report the total physical-qubit count, the predicted QUOPS, the benchmark point \((w,s)\) at which that score is attained, and the average number of T gates per rotation at the benchmark point. Rough estimates for the number of T-gates each device can perform---the typical resource considered in resource estimates---can be calculated by multiplying each device's score by the mean number of T-gates per rotation synthesis all divided by two, which we report here from samples of $100$ rotations per device with $1\sigma$ error bars. The increase in the mean number of T-gates per rotation comes from synthesizing rotations to greater precisions in larger circuits in order to maintain the same overall synthesis error.}
\label{app:flasq:tab:ftqc-predictions}
\end{table*}

\subsubsection{Compilation model and error budget}

To apply the FLASQ model to QUOPS circuits, we map each benchmark circuit to the logical primitive gate set $\{H, S, S^\dagger, T, \mathrm{CNOT}\}$. Logical CNOT gates are implemented by lattice surgery, where we account for the distance between the target qubits in determining the required ancilla volume. Arbitrary single-qubit QUOPS rotations $R_P(\theta)$, with $P \in \{X, Y, Z\}$, are compiled first by changing basis, when necessary, to convert the rotation to a $Z$-axis rotation, and then synthesizing that rotation into Clifford+$T$.

For these projections, we assigned each benchmark circuit a total allowable error budget equal to 
\begin{equation}
    \epsilon_{\mathrm{tot}} = 1 - \frac{1}{\sqrt{e}},
\end{equation}
and split this budget evenly between three error sources: logical Clifford-gate infidelity (estimated using the error rate of the syndrome extraction cycle), $T$-state infidelity, and single-qubit rotation synthesis error.

A size-$s$ QUOPS circuit contains $s/2$ single-qubit rotations, so each such rotation was synthesized to diamond-distance precision
\begin{equation}
    \epsilon_{\mathrm{synth}} = \sqrt{\frac{1-1/\sqrt{e}}{3(s/2)}}.
    \label{app:flasq:eq:synth-budget}
\end{equation}
We used \texttt{pygridsynth}~\cite{yamamoto2026pygridsynthfastnumericaltool}, a Python implementation of the \texttt{GridSynth} algorithm~\cite{selinger2012efficient, ross2016optimal}, to synthesize these rotations. The exterior square root comes from assuming that coherent synthesis errors combine incoherently, leading to quadratic contributions to a circuit's infidelity. If the synthesized implementation of a benchmark circuit contained $N_T$ total $T$ gates, then we set
\begin{equation}
    M = N_T
\end{equation}
and assigned each consumed $T$-state the infidelity
\begin{equation}
    p_{\mathrm{mag}} = \frac{1-1/\sqrt{e}}{3N_T}.
    \label{app:flasq:eq:pmag}
\end{equation}
The remaining one-third of the total budget was allocated to Clifford gates. This equal three-way split is a simple heuristic chosen to make the resource forecasts explicit, simple, and uniform across circuit sizes. It is possible that higher QUOPS could be achieved using optimized error budgets. 

\subsubsection{Simplifying assumptions}

We make two simplifying assumptions relative to the full FLASQ model. First, we assume that all computations are spacetime-limited rather than reaction-limited. In FLASQ notation, this means
\begin{equation}
    L = V/A,
    \label{app:flasq:eq:ancilla-limited}
\end{equation}
where $L$ is the number of logical timesteps, $V$ is the total ancilla volume required by the circuit, and $A$ is the number of ancilla qubits available to support the computation. Second, we fix the number of non-ancilla logical qubits used by the circuit to be equal to its benchmark width. In \emph{FLASQ notation}, $Q = w$,
where $w$ is the width of the QUOPS circuit (here, and only here, $Q$ denotes the FLASQ data-qubit count of Ref.~\cite{huggins2025fluidallocationsurfacecode} rather than the QUOPS score). This fixing replaces the more detailed active-qubit accounting used in the full FLASQ model with a simpler approximation that is sufficient for our coarse-grained forecasts.

Under these assumptions, the cost estimate for a benchmark circuit is determined by a small set of aggregate resource counts, rather than by a detailed schedule of every logical operation.

\subsubsection{Predicted QUOPS}

A direct application of the above procedure would require explicitly generating each benchmark circuit at each point $(w,s)$, synthesizing every Pauli-axis rotation in that circuit into Clifford+$T$, and then evaluating the corresponding FLASQ cost. This end-to-end computation becomes computationally expensive for large circuits. A size-$s$ QUOPS circuit contains $s/2$ random single-qubit rotations, so explicit synthesis of every rotation in circuits with $s \gtrsim 10^6$ is costly. 

To make the forecasts tractable, we used representative gate-count statistics rather than exact compilations of complete benchmark circuits. For each $(w,s)$ pair, we generated $10$ QUOPS circuits. The number of CNOT gates in these circuits is fixed by the circuit shape, and because QUOPS samples CNOTS uniformly over qubit pairs, the expected count of CNOTS between any ordered qubit pair $(q_1, q_2)$ is determined directly by the total number of CNOTs and the number of allowed control-target pairs. 

To estimate the cost of the single-qubit rotations, we sampled $100$ representative Pauli-axis rotations for each circuit, synthesized them using \texttt{pygridsynth} at the precision given by Eq.~\eqref{app:flasq:eq:synth-budget}, and used these samples to estimate the number of $H$, $S$, $S^\dagger$, and $T$ gates required per QUOPS rotation in the circuit. We multiply these average synthesis statistics by the number of single-qubit rotations in a benchmark circuit to estimate the primitive logical gate counts of that circuit.

Given these estimated primitive-gate counts, we evaluated the simplified FLASQ model to obtain \(V\), \(V_{\mathrm{cliff}}\), and \(M=N_T\), and then computed the success probability using Eq.~\eqref{app:flasq:eq:psuccess}. For each tuple of device, benchmark point \((w,s)\), and circuit, we scanned odd code distances
\begin{equation}
d \in \{3,5,\dots,21\}
\label{eq:appendix-code-distance-scan}
\end{equation}
in increasing order and identified the smallest code distance for which the predicted circuit-success probability exceeded the QUOPS threshold \(1/\sqrt{e}\). When such a code distance existed, we terminated the scan and recorded the corresponding success probability. When no code distance in the scan range crossed threshold, we recorded the maximum success probability attained. We truncated the scan at \(d=21\), which is already large relative to the code distances typically considered in early fault-tolerant resource estimates. Furthermore, in the cultivation data we use for our estimates \cite{huggins2025fluidallocationsurfacecode} the error rate of $T$ states for a physical gate error rate of $0.1\%$ is lower bounded by $2.6 \times 10^{-6}$. We are therefore limited to probing circuit sizes of $O(10^{6})$ in our resource estimates. We note that this is not a fundamental limitation of these hypothetical processors---it is possible to remove this limitation by allowing magic state distillation on top of cultivation or using a variation of cultivation that reaches a lower error rate.

Finally, for each $(w,s)$ (and a given device), we averaged these predicted success probabilities over the 10 sampled benchmark circuits. Denoting this average by $\overline{P}_{\mathrm{success}}(w,s)$, we claimed that a device passed at $(w,s)$ if 
\begin{equation}
    \overline{P}_{\mathrm{success}}(w,s) > \frac{1}{\sqrt{e}}.
    \label{app:flasq:eq:pass-threshold}
\end{equation}
Applying this criterion across the admissible cone yields the predicted QUOPS capability region for the device, from which the projected QUOPS reported in the main text is extracted.

We also estimate the QUOPS rate for each hypothetical processor at the circuit shape $(w,s)$ maximizing QUOPS. Because we have assumed that computations are spacetime-limited rather than reaction-limited, the execution time of a QUOPS circuit is given by $t_{\textrm{cyc}}L = t_{\textrm{cyc}}V/A$. We use estimates of $V$ and $A$ from our resource estimates to estimate the time required for each QUOPS circuit. 

\subsubsection{Projected QUOPS for hypothetical fault-tolerant devices}
Applying the procedure above yields projected QUOPS capability regions and scalar QUOPS estimates for each of the four hypothetical fault-tolerant devices. As expected, the predicted QUOPS increases strongly with total physical-qubit budget: larger devices can support wider logical circuits and larger code distances, and therefore successfully execute larger benchmark circuits.

For the device family considered here, we presently predict that the \(N_{\mathrm{pq}}=5000\) device achieves a QUOPS of approximately \(1950\) at width \(w=13\), the \(N_{\mathrm{pq}}=\num{10000}\) device achieves a QUOPS of \(\num{9240}\) at \(w=21\), the \(N_{\mathrm{pq}}=\num{25000}\) device achieves a QUOPS of \(\num{68470}\) at \(w=41\), and the \(N_{\mathrm{pq}}=\num{50000}\) device achieves a QUOPS of \(\num{273632}\) at \(w=68\). These estimates are shown in Fig.~\ref{fig:main:projections:surface_code} in the main text and summarized in Table~\ref{app:flasq:tab:ftqc-predictions}.

Outside of the cone, the capability regions are capped by device constraints. Pushing to higher-width circuits, while still passing the QUOPS threshold, requires increasingly large code distances. Ultimately, devices run out of physical qubits and are unable to support high-width circuits. At low widths, pushing to larger sizes requires increasingly higher fidelity magic state generation. Existing cultivation protocols exhibit asymptotic floors in the logical error rates they can achieve at fixed physical error rates. As a result, the performance of cultivation ultimately caps QUOPS sizes achievable by each device. This illustrates that, even within the simplified model used here, the dominant bottleneck can shift from logical spacetime overhead to non-Clifford resource quality as device scale increases.

\subsection{Fully connected FTQC architectures}\label{app:fully_connected}

This section describes the projected QUOPS capability regions and scores for hypothetical FTQC systems built on a monolithic, all-to-all connected physical architecture, such as a large-scale trapped-ion or neutral-atom system, encoded in either the surface code or a concatenated symplectic double code. Unlike the coarse-grained model of Section~\ref{app:flasq}, these projections are obtained by explicitly compiling and scheduling QUOPS circuits with a logical resource-estimation framework, and then fitting the resulting execution times and logical error rates to an approximate cost model that is extrapolated over circuit shape. Section~\ref{sssec:fc_assumptions} states the hardware and logical-architecture assumptions, Section~\ref{sssec:fc_methodology} the scheduling framework, Section~\ref{sssec:fc_cost_model} the approximate cost model and its fit, and Section~\ref{sssec:fc_frontiers} the resulting QUOPS capability regions and scores. Throughout this section, $d$ denotes the number of \emph{benchmark layers} in a QUOPS circuit (each consisting of a CNOT layer and an $R_P(\theta)$ layer), so that a dense QUOPS circuit of even width $w$ has size $s = 2wd$; this differs from the depth convention of Section~\ref{ssec:circuits_definitions}, in which $d$ counts individual layers. The code distance is also denoted $d$ where indicated.

\subsubsection{Architectural assumptions}\label{sssec:fc_assumptions}
We consider a system with $n_q$ physical qubits with
\begin{equation}
n_q \in \{ \num{1000}, \num{5000}, \num{10000} \},
\end{equation}
and all-to-all connectivity. We assume physical two-qubit gate and SPAM error rates of $p_{\mathrm{2Q}} = p_{\mathrm{SPAM}} = 1\times 10^{-4}$, and physical single-qubit and memory error rates of $p_{\mathrm{1Q}} = p_{\mathrm{mem}} = 1\times 10^{-5}$ per depth-1 cycle. We define the depth-1 cycle as the worst-case time required to arbitrarily pair all of the physical qubits on the device and run a circuit layer composed of physical two-qubit gates sandwiched by single-qubit gates and possibly followed by measurement and reset. Although we expect the transport time required by a depth-1 cycle to scale as $\sqrt{n_q}$, we set $t_{\mathrm{d1}} = 10\ \mathrm{ms}$
irrespective of $n_q$, as we expect the time spent gating and cooling physical qubits to dominate the budget. 

We compare two different instantiations of a logical (fault-tolerant) quantum computing architecture based on the above hardware model. The first, based on the \textit{surface code}, is scalable according to the code distance $d$ and relies on transversal logical CNOT gates ($\tau_{\mathrm{CNOT}} = \ t_{\mathrm{d1}}$), fold-transversal mid-cycle logical phase gates ($\tau_{\mathrm{S}} = 5 \ t_{\mathrm{d1}}$)~\cite{chen2026transversal}, SWAP-transversal logical Hadamard gates ($\tau_{\mathrm{H}} = 0$)~\cite{chen2026transversal}, quantum error correction via bare-ancilla syndrome extraction ($\tau_{\mathrm{EC}} = 4 \ t_{\mathrm{d1}}$)~\cite{fowler2012surface}, and standard T injection via a logical CNOT between the gate target and a magic state ($\tau_{\mathrm{T}} = \ t_{\mathrm{d1}}$). We assume magic states are produced according to the fold-transversal protocol of Ref.~\cite{sahay2026fold}. A second logical architecture, based on the $\qeccode{20}{2}{6}$ instantiation of the \textit{concatenated symplectic double} (CSD) code construction~\cite{berthusen2025simple}, assumes all intra-block logical Clifford gates are implemented as layers of SWAP-transversal gates  and global S gates, with maximum three global S gates per Clifford gate. The SWAP-transversal layers can be implemented as physical single-qubit Clifford gates and qubit relabeling, resulting in an effectively depth-0 sub-circuit, and global S gates can be implemented using a $\tau_{\mathrm{GlobalS}} = 4 \ t_{\mathrm{d1}}$ sub-circuit~\cite{Berthusen2026experimental}. Recent work has discovered how to directly inject surface code magic states into $\qeccode{20}{2}{6}$ code blocks using depth-2 chain maps~\cite{benhemou2026automated}. Therefore, the CSD logical architecture also relies on fold-transversal surface code magic state cultivation. Global CNOT gates are implemented transversally, while targeted inter-block CNOT gates are assumed to be implemented via the depth-2 chain maps discovered in Ref.~\cite{benhemou2026automated}. Modified Shor-style syndrome extraction in the $\qeccode{20}{2}{6}$ code can be accomplished with $\tau_{\mathrm{EC}} = 16 \ t_{\mathrm{d1}}$ using an overhead of sixteen ancilla qubits~\cite{Berthusen2026experimental}. In both logical architectures, we assume that scalable correlated decoding allows logical gadgets to be followed by only a single round of syndrome extraction, instead of $O(d)$ rounds~\cite{cain2025fast}.

For more detailed information about these architectural assumptions and the resource estimation methodology to follow, see Ref.~\cite{LeBlond2026InPrep}. 

\subsubsection{Methodology}\label{sssec:fc_methodology}

In this work, we calculate shot times and total logical error rates for various $(w, s)$ QUOPS circuits applied to the above-mentioned architectures. To accomplish this, we make use of a to-be-published resource estimation framework that relies on a logical compilation and dynamic logical scheduling pipeline as its backbone~\cite{LeBlond2026InPrep}. This framework includes the explicit tracking of all data and resource code-block lifetimes during a simulated logical execution trace, responds to real-time resource state demand with state preparations according to a given policy, handles stochastic state preparation failures and Clifford corrections, and dynamically manages error correction schedules for idling code blocks (inserting error correction cycles as needed according to simulations of logical error rate vs.\ per-block accumulated memory error). Crucially, this framework enables us to realistically estimate the achievable logical-layer parallelism for a given $(w, s)$ QUOPS circuit at a given $n_q$. 

The total logical error rate of a scheduled quantum circuit is calculated within the resource scheduler as 
\begin{equation}
\label{eq:p_succ_retrace}
    \epsilon_{\mathrm{sched}} = 1 - (1 - \epsilon_{\mathrm{mag}})^{n_T}\prod_{i}^{n_{\mathrm{EC}}}(1 - \epsilon_{\mathrm{EC}}(m_i)),
\end{equation} where $n_{\mathrm{EC}}$ is the total number of error correction cycles (including both cycles inserted due to excessive idling and cycles inserted as follow-ups to logical operations), $\epsilon_{\mathrm{EC}}(m_i)$ is the per-cycle logical error rate parametrized by $m_i$ depth-1 cycles of logical memory preceding error correction cycle $i$, and $\epsilon_{\mathrm{mag}}$ is the logical error rate of $n_T$ prepared magic states. Arbitrary single-qubit rotation gates $R_P(\theta)$ are synthesized using \texttt{pygridsynth} with identical $\epsilon_{\mathrm{synth}}$ to what is described in Sec.~\ref{app:flasq}. Rotation synthesis is likewise given one third of the total allowed error budget of $\epsilon_{\mathrm{tot}} = 1 - \frac{1}{\sqrt{e}}$. Unlike the methodology described in Sec.~\ref{app:flasq}, however, this scheduler allows the flexible distribution of the remaining error budget between error correction and consumed magic states; the exact proportion will be circuit-dependent.

\subsubsection{Approximate model}\label{sssec:fc_cost_model}

Because the brute-force compilation and scheduling of QUOPS circuits over a large $(w, s)$ grid is very computationally expensive, we chose to run simulations for a relatively small number of circuit shapes and fit our results to an approximate cost model. Also, instead of directly fitting to \eqref{eq:p_succ_retrace}, which relies on a large number of schedule-dependent parameters, we employed a simpler logical error model based on an effective logical error rate per logical qubit per depth-1 cycle, $\epsilon_{\mathrm{eff, EC}}$, and an effective logical error rate per magic state injection, $\epsilon_{\mathrm{eff, mag}}$ (which includes contributions from injection itself, error correction, and a potential Clifford correction). This logical error model,

\begin{align}
\tilde{\epsilon}_{\mathrm{sched}}(w, d) = 1 - (1 - \epsilon_{\mathrm{eff, mag}})^{n_T}(1 - \epsilon_{\mathrm{eff, EC}})^{\tilde{\tau}_{\mathrm{idle}}(w, d)}, \label{eq:sched_error} 
\end{align}

is designed to capture the freedom that the scheduler has to trade between logical error accumulated from the insertion of error correction cycles and logical error accumulated from magic state injection. In \eqref{eq:sched_error}, we model the number of T gates as 
\begin{equation}
    n_T = w \times d \times l_T(w,d), \label{eq:num_T}
\end{equation}
where the approximate T-depth of a rotation layer, $l_T(w,d)$, is related to the expected gridsynth scaling~\cite{ross2016optimal}
\begin{equation}
l_T(w, d) = a_T (\log(wd) + b_T),
\end{equation}
where $a_T$ and $b_T$ are fit parameters.

In Eq.~\eqref{eq:sched_error}, logical error trades between logical qubit idling and magic state injection through $\tilde{\tau}_{\mathrm{idle}}$, which we define as
\begin{equation}
       \tilde{\tau}_{\mathrm{idle}} = \max\left(0, \lceil w/k \rceil \tilde{\tau}_{\mathrm{tot}} - \tau_{\mathrm{injec}} n_T / k \right). \label{eq:time_idling}
\end{equation}

In Eq.~\eqref{eq:time_idling}, $\tilde{\tau}_{\mathrm{tot}}(w, d)$ is the approximate total number of scheduled time steps for a $(w, 2wd)$ QUOPS circuit, $\tau_{\mathrm{injec}}$ is the magic state injection depth, and $k$ is the number of logical qubits encoded in a single code block. As such, $\lceil w/k \rceil$ is the number of data code blocks present in the computation, $\lceil w/k\rceil \tilde{\tau}_{\mathrm{tot}}$ is the total number of data code-block-cycles, and $\tau_{\mathrm{injec}}n_T/k$ is the total number of data code-block-cycles taken up by T injection. Hence, $\tilde{\tau}_{\mathrm{idle}}$ represents the sum of data code-block-cycles during which code blocks are mostly idling. Importantly, all of the $(w, d)$ dependence of $\tilde{\epsilon}_\mathrm{sched}$ comes in through the trade-off we are describing, whereas $\epsilon_{\mathrm{eff, EC}}$ and $\epsilon_{\mathrm{eff, mag}}$ are treated as static quantities.

To model the total number of scheduled time steps for a QUOPS circuit, we treat $\tilde{\tau}_{\mathrm{tot}}$ as a piecewise function describing three regimes: a T-count-limited regime (serial magic state production), a T-depth-limited regime where magic state production is not a bottleneck, and a regime which interpolates between the two, e.g., where there is space on the processor to parallelize some but not all T gates in a given layer. The following equations specify $\tilde{\tau}_{\mathrm{tot}}(w, d)$:
\begin{align} 
 \tilde{\tau}_{\mathrm{tot}}(w, d) &= d \times\tilde{\tau}_{\mathrm{layer}}(w, d), \label{eq:cost1} \\    \tilde{\tau}_{\mathrm{layer}}(w, d) &= l_T(w, d) \times \max\left[ \tau_{\mathrm{injec}}, w \tau_{\mathrm{prep}} / n_{\mathrm{prep}}(w) \right] \label{eq:cost2} \\
    n_{\mathrm{prep}}(w) &= \max\left( 1, \frac{n_q - \tilde{N}_{\mathrm{cb}} \left\lceil \frac{w}{k} \right\rceil}{N_{\mathrm{prep}}}\right) \label{eq:cost4}
\end{align} In Eq.~\eqref{eq:cost2}, $\tilde{\tau}_{\mathrm{layer}}(w, d)$ represents the approximate number of scheduled time steps for a single QUOPS layer. It is modeled as the product of $l_T(w,d)$ and either $\tau_{\mathrm{injec}}$, the depth of $T$-state injection, or $w \tau_{\mathrm{prep}} / n_{\mathrm{prep}}(w)$, the effective magic state preparation depth, whichever is greater. The former case captures the T-depth-limited regime, and the latter case captures the magic-state-preparation-limited regimes in which not enough magic state preparation instructions $n_{\mathrm{prep}}(w)$ can be parallelized to support a full T layer at a given $w$ without waiting. Equation~\eqref{eq:cost4} discriminates between two magic-state-preparation limited regimes, where in one regime there is only space for a single magic state preparation instruction to be scheduled, and in the other regime there is space to parallelize a greater number of these preparation instructions. In Eq.~\eqref{eq:cost4}, $\tilde{N}_{\mathrm{cb}}$ is the effective number of physical qubits consumed by each code block, and incorporates both the number of physical qubits required by the data block itself as well as a circuit-and-schedule-dependent effective error correction overhead.

\begin{figure*}
    \centering
    \includegraphics[width=\linewidth]{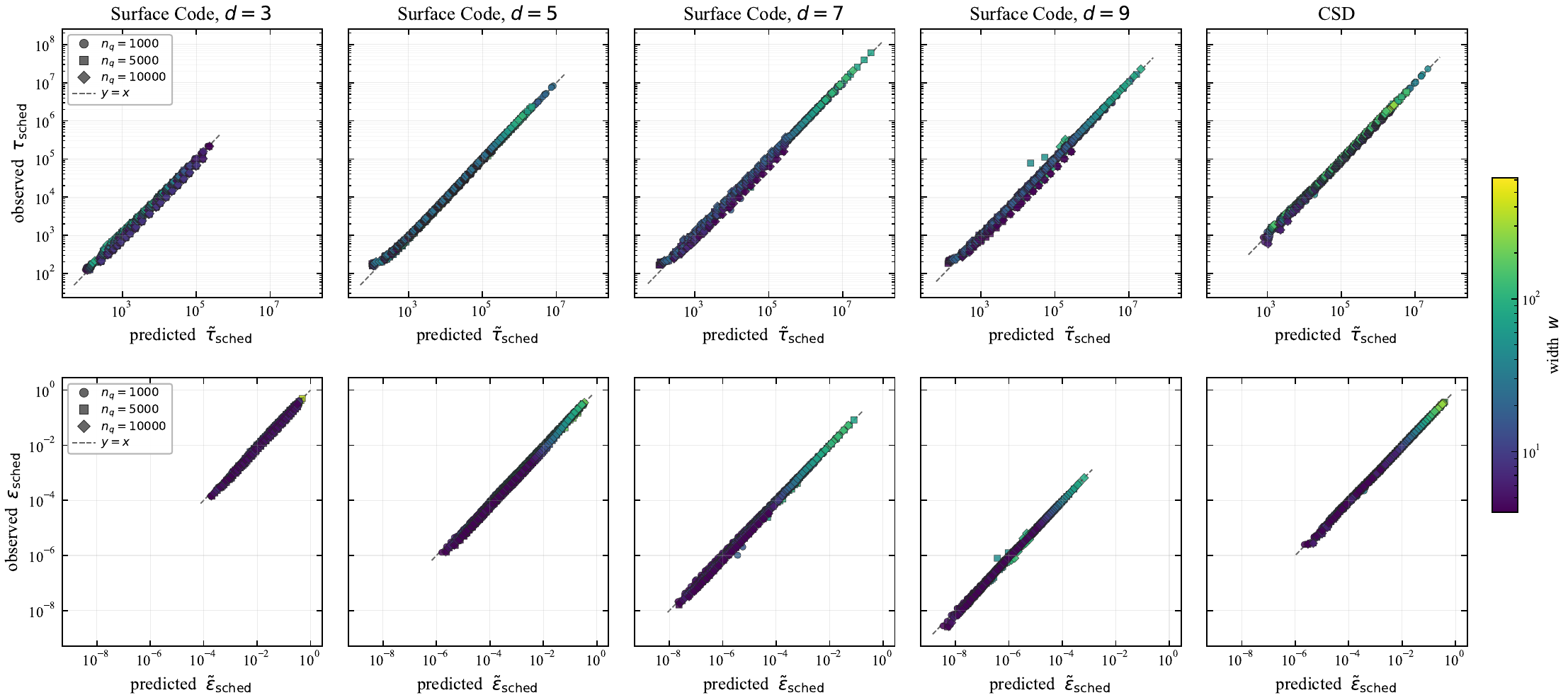}   \caption{\textbf{Cost model comparison.} Comparison between our fitted approximate scheduler cost models $\tilde{\tau}_{\mathrm{sched}}$ (denoted $\tilde{\tau}_{\mathrm{tot}}$ in the text) [see Eq.~\eqref{eq:cost1}] (top row) and $\tilde{\epsilon}_{\mathrm{sched}}$ [see Eq.~\eqref{eq:sched_error}] (bottom row) and observed data points $\tau_{\mathrm{sched}}$ and $\epsilon_{\mathrm{sched}}$ for the two logical architectures under consideration, and for a range of code distances where applicable (see column titles). It is evident from these plots that our fitted cost model agrees well with our observed data.}
    \label{fig:collapse_quality}
\end{figure*}

At a given $(w,s)$, Eq.~\eqref{eq:cost1} contains six parameters, $\tau_{\mathrm{injec}}$, $\tau_{\mathrm{prep}}$,  $\tilde{N}_{\mathrm{cb}}$, $N_{\mathrm{prep}}$, $a_T$, and $b_T$, some but not all of which have values that can be known \textit{a priori}. That said, the model above is not able to capture all the complexities of our scheduler exactly (for example, it only incorporates the costs of T-state preparation and injection and neglects the costs of other logical operations). Therefore, in fitting to this model, we have chosen to let all parameters vary so as to make the model less rigid and able to absorb the effects of details not explicitly modeled. In Fig.~\ref{fig:collapse_quality}, we show the quality of our fitted models relative to our observed points. It is evident from these plots, as well as from the agreement between observed [pass, fail] brackets and our modeled frontiers (see Figs.~\ref{fig:frontiers_qntm} and~\ref{fig:frontiers_fom_qntm}), that our fitted cost model agrees well with our observed data.

\subsubsection{QUOPS capability regions}\label{sssec:fc_frontiers}
\begin{figure*}
    \centering
    \includegraphics[width=\linewidth]{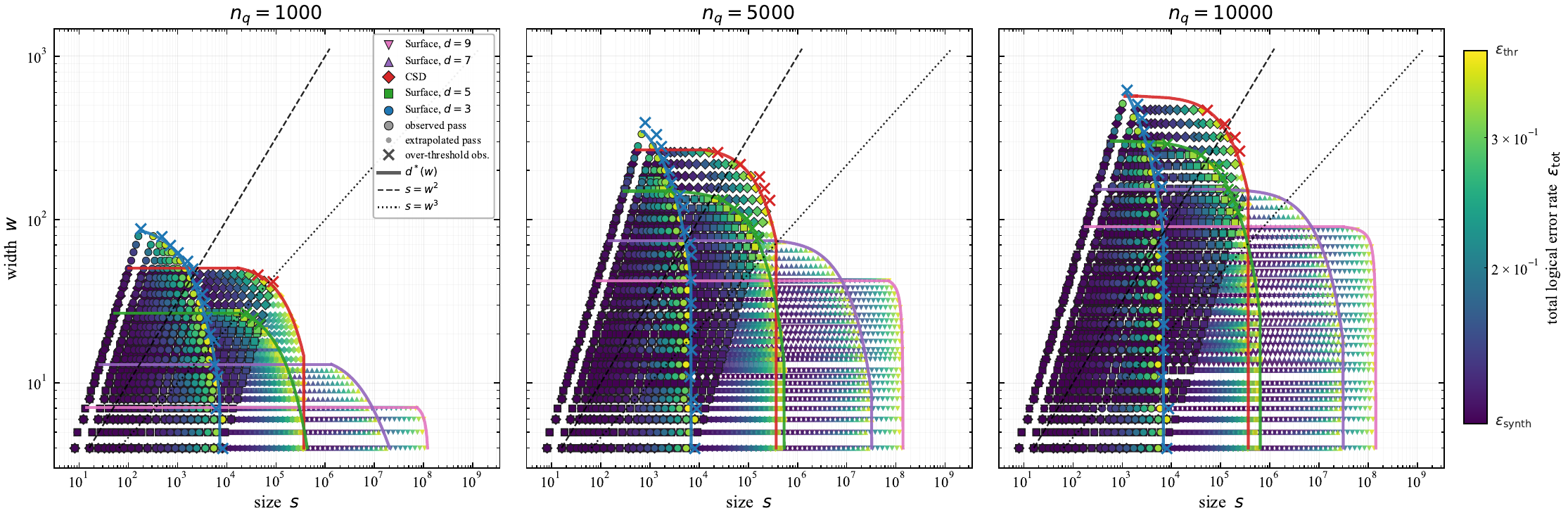}
    \caption{\textbf{Overlaid QUOPS frontiers for various logical architectures.} For $n_q = 1000$ (left), $n_q  = 5000$ (middle), and $n_q = 10000$, (right) observed scheduler results are given by large, bold icons, with a different shape for each logical architecture. Extrapolated scheduler results are given by smaller icons of the same shape for each logical architecture. Both observed data points and extrapolated data points are filled according to the total logical error rate $\epsilon_{\mathrm{tot}}$. The colored solid lines, $d^{*}(w)$, are the maximum size circuit at each width able to run within the infidelity threshold. The widths available to each logical architecture are bounded by a threshold $w_{\mathrm{max}}$ (see text for definition), here represented as a horizontal line continuous with $d^{*}(w)$. Observed data points that were over-threshold are marked with bold X's. Good agreement is apparent between the $d^{*}(w)$ curves and the location of the bold X's, providing further validation for our approximate cost model.}
    \label{fig:frontiers_qntm}
\end{figure*}

\begin{figure*}
    \centering
    \includegraphics[width=\linewidth]{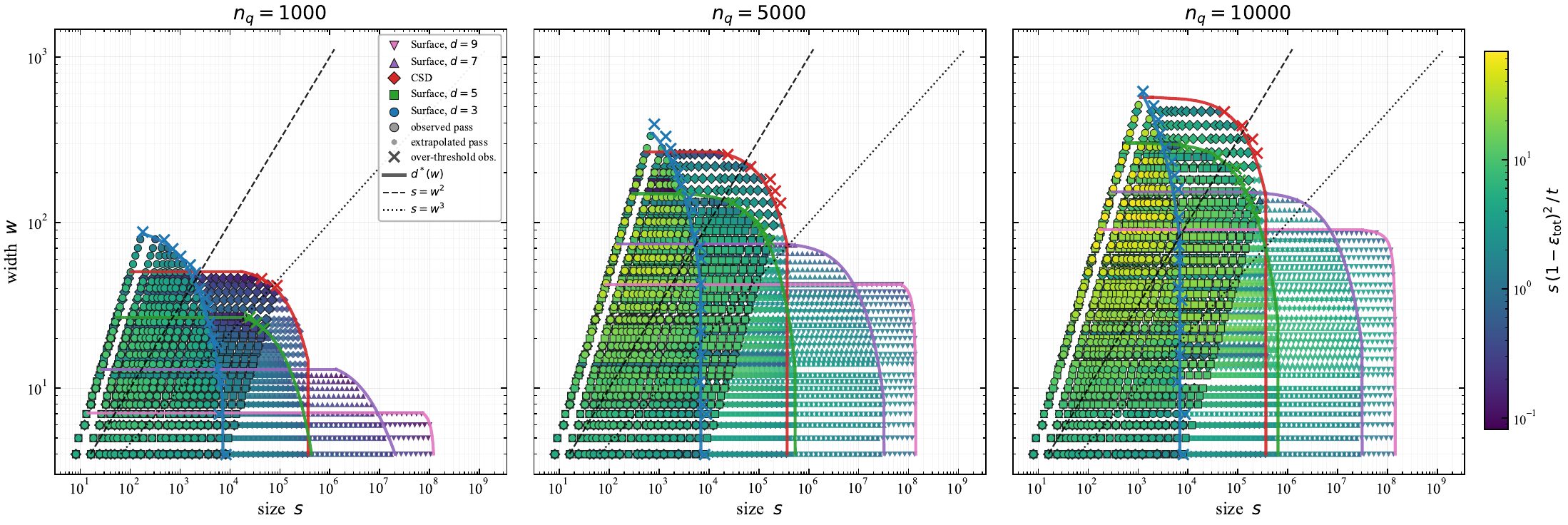}
    \caption{\textbf{QUOPS rate heatmap.} This plots is the same as Fig.~\ref{fig:frontiers_qntm} except that it shows the scaled QUOPS rate on the heatmap instead of logical error rate.}
    \label{fig:frontiers_fom_qntm}
\end{figure*}

Fig.~\ref{fig:frontiers_qntm} and~\ref{fig:frontiers_fom_qntm} show both our observed and extrapolated scheduler results overlaid for various logical architectures, with different sub-panels representing different $n_q$. In Fig.~\ref{fig:frontiers_qntm}, data is colored according to the total logical error rate, while in Fig.~\ref{fig:frontiers_fom_qntm} data is colored according to the scaled QUOPS throughput. For each logical architecture, a distinct frontier curve $d^*(w)$ is shown that is calculated according to our fitted models. To create these frontier curves, we evaluate the maximum passing circuit size at every given width beneath a threshold width of $w_{\mathrm{max}}$. The fitted formula \eqref{eq:cost4}, which differentiates between two different magic-state-production-limited regimes, also allows us to determine the maximum runnable width $w_{\mathrm{max}}$ for every logical architecture at every $n_q$. We expect, and our observations substantiate, that at values of $w$ for which room on the processor exists for only a single magic state preparation instruction to be scheduled at a time, the $n_q$ will also not be able to support greater numbers of code blocks beyond $\lceil w / k \rceil$.  We have seen exceptions to this rule only where the interaction graph of the QUOPS circuit is not fully connected, allowing it to be decomposed into sub-circuits that run sequentially. We do not consider this scenario to be realistic at relevant circuit depths, and it is also forbidden by the QUOPS benchmark, which does not permit data qubit reuse. Therefore, setting $n_\mathrm{prep} = 1$, we obtain $w_{\mathrm{max}} = k(n_q - N_{\mathrm{prep}})/\tilde{N}_{\mathrm{cb}}$, and combine the horizontal $w=w_\mathrm{max}$ with our extrapolated $d^*(w)$ curves to form the frontier curves in Figs.~\ref{fig:frontiers_qntm} and \ref{fig:frontiers_fom_qntm}.

\section{Quantinuum experiments}
Quantinuum's tests were run on \helios, a 98-qubit trapped-ion QCCD quantum computer~\cite{Ransford2026-ny}, and \htwoone, an older generation 56-qubit trapped-ion QCCD quantum computer \cite{Moses_2023}. Timed QUOPS tests were run at the physical level and within a $\llbracket 7, 1, 3 \rrbracket$ quantum error correcting code. At the physical level, additional untimed experiments were performed, as shown in Table~\ref{tab:physical-quops-h2-1}, in order to fill out the capability region shown in Fig.~\ref{fig:Helios-physical-capability}.

In this section we first detail general features of---and methods for programming---\helios (Section~\ref{ssec:helios_system}) and \htwoone (Section~\ref{ssec:h2_system}), followed by a comprehensive explanation of design decisions made for the logical-qubit experiments on \helios (Section~\ref{ssec:helios_logical}) and the physical-qubit experiments on both systems (Section~\ref{ssec:quantinuum_physical}.

\subsection{\helios}
\label{ssec:helios_system}
\helios features all-to-all qubit connectivity via ion transport and rearrangements. \helios also enables real-time classical compute and qubit routing from mid-circuit measurements. These features make \helios ideal for exploring near-term fault-tolerant implementations of different QEC codes.

\subsubsection{Hardware}
\helios supports a native gate set consisting of single-qubit (1Q) gates
$R_{\phi}(\theta) = e^{-i(\cos(\phi)X+\sin(\phi)Y)\theta/2}$
and 
$R_z(\theta)=e^{-iZ\theta/2}$,
and a two-qubit (2Q) gate 
$R_{zz}(\theta)=e^{-iZZ\theta/2}$.
The $R_{\phi}$ gate is performed by a laser-driven Raman transition,
the $R_z$ gate is implemented in software by phase-tracking,
and the $R_{zz}$ gate is performed by a laser-driven M\o lmer-S\o renson interaction~\cite{molmer_2000, Lee_2005}.

A comprehensive set of benchmarks was presented in Ref.~\cite{Ransford2026-ny}, which quantified \helios's performance at component- and system-level operations. The dominant sources of errors on \helios are from two-qubit gates (typical errors around $8\times10^{-4}$) and ion transport (typical errors around $5\times10^{-4}$ per qubit per dense layer of two-qubit gates on random qubit pairs of all 98 qubits), both of which are measured with Clifford randomized benchmarking. A large fraction of these errors cause leakage that incoherently moves population from the qubit subspace to other hyperfine levels of the ion. These errors can be detected at the end of the circuit with a ternary measurement that works by shelving qubit population to other long-lived atomic sublevels to measure both qubit populations and leakage population. Leakage errors can also be mitigated by a leakage-repump scheme, similar to that of Ref.~\cite{Hayes2020_leakage_repump}, which can be enabled as a compiler option.

\subsubsection{Real-time engine control, Guppy, and Selene}\label{sec: Real-time engine control}
\helios operates using a real-time control engine (the Helios runtime)~\cite{Ransford2026-ny}, which allows for classical co-compute and conditional ion transport. The primary role of the runtime is to efficiently map virtual qubits to physical qubits on the device, map gate and measurement requests on virtual qubits into low-level operations on physical qubits, and to dispatch commands to transport batches of qubits from storage into quantum operation zones. 
Crucially, all these decisions are made on-the-fly during program execution and can be dependent on the outcomes of mid-circuit-measurements or classical pseudo-random numbers generated at runtime.

\helios can be programmed using Guppy~\cite{Koch2024guppy}, a pythonic hybrid-compute quantum programming language, which introduces the high-level constructs required to exploit the Helios runtime's capabilities. These constructs allow for dynamic qubit allocation, classical control flow (including if-then-else statements, for loops, and while loops), and early termination of programs based on arbitrary classical logic that can, e.g., depend on mid-circuit measurements.

Emulation of \helios is performed using \texttt{Selene}~\cite{selene}, an open-source framework to run quantum programs written in Guppy. It can be configured to emulate \helios using either state-vector or stabilizer simulators. Notably, \texttt{Selene} can be integrated with the Helios runtime and can therefore take into account errors associated with idling and (conditional) ion transport. Stabilizer simulation was used for the emulation results presented in Fig.~\ref{fig:Helios-physical-capability} and was validated to give similar performance to state-vector simulation up to 24 qubits.

\subsubsection{\helios compiler options}

\helios jobs can be submitted with a number of compiler options. The options that we made use of in the following experiments are:
\begin{enumerate}[label=(\roman*)]
\item \emph{leakage repump}~\cite{Hayes2020_leakage_repump}, which selectively pumps leaked population back into the computational subspace. For qubits based on barium-137 ions, this is implemented by driving a narrow-band quadrupole transition out of the leaked hyperfine states, followed by a dipole transition up to a short-lived excited state, from which the ion decays back toward either the computational subspace or the leaked hyperfine states—similar to Ref.~\cite{An_2022}. Because each cycle removes a fixed fraction of the leaked population, leakage decays exponentially with cycle number.

\item \emph{dynamical decoupling}~\cite{biercuk2009DD, Watkins26}, which {reduces memory error through the insertion of canceling $X$ gates}.
The real-time compiler estimates idling times on qubits and schedules two $X$ gates at regular intervals whenever the idling time exceeds a threshold. This threshold  can either be chosen automatically or specified explicitly.

\item \emph{single-qubit gate squashing}, in which {the real-time compiler attempts to} combine multiple single-qubit gates on a given qubit into a single $R_{\phi}$ gate. Since the compiler runs in real-time, the computation of the combined single-qubit gate must occur with low latency during a look-ahead window before the gate is executed.
\end{enumerate}

A summary of which options were enabled for each logical experiment is detailed in Table~\ref{tab:[[7,1,3]]-helios} (all physical experiments enabled only single-qubit gate squashing). To satisfy the assumptions of MCFE---for all experiments---we additionally insert compilation barriers in the middle of the circuit between ``forward'' and ``reverse'' circuits, as well as after (respectively, before) random 1Q Cliffords at the beginning (respectively, end) of the circuit. These compilation barriers prevent 1Q gate squashing from occurring between gates on different sides of the barrier and are necessary to satisfy the assumptions of MCFE.

\subsubsection{Timing and QUOPS rate}
Reported QUOPS rates should be indicative of the average runtime of a QUOPS circuit including any overhead, such as ion loading and calibrations, encountered during normal operation. To capture these effects we performed each timed experiment (i.e., each data point $(w, s)$)
within a reserved time window, unless otherwise stated. Timing for each such experiment was then simply calculated as the time from the start of the first shot until the end of the last shot. 

Data points without timing information were submitted as part of Helios's normal mode of operation through \texttt{Quantinuum Nexus} where jobs across different projects are allowed to be interleaved in accordance with the fair-queue system~\cite{nexus}.

\subsection{\htwoone}
\label{ssec:h2_system}
\htwoone is a 56-qubit trapped-ion QCCD quantum computer \cite{Moses_2023}. Qubits are encoded in the hyperfine clock states of $^{171}\mathrm{Yb}^{+}$ ions, each paired with a $^{138}\mathrm{Ba}^{+}$ ion used for sympathetic cooling. \htwoone uses a race-track trap in which ions circulate between storage locations and four active gate zones. \htwoone features all-to-all qubit connectivity via ion transport and rearrangements, and supports the same native gates as \helios, namely $\{ R_{\phi}(\theta) , R_z(\theta), R_{zz}(\theta) \}$. The dominant sources of errors on \htwoone are from two-qubit gates (with typical errors around $1.1\times10^{-3}$) and ion transport (typical errors around $3\times10^{-4}$ per qubit per dense layer of two-qubit gates on random qubit pairs of all 56 qubits) \cite{h2-1-error-rates-docs}. A full set of benchmarks can be found in \textcite{Moses_2023}. Like \helios, \htwoone utilizes reservations and queues which we use to gather timed and untimed results respectively.

\htwoone allows for real-time classical operations on bit registers and classically conditioned gate operations on qubits. Mid-circuit measurement and reset operations, or classical pseudo-random numbers generated at runtime, enable per-shot conditional gate operations during job execution. The primary difference between the run-time capabilities of \htwoone and \helios is that while \helios calculates ion transport in real time, \htwoone calculates all ion transport before the circuit is run. This means that on \htwoone all control flow constructs must be unrolled at compile time. 

We use pytket~\cite{Sivarajah_2020}, a software platform for the development and execution of gate-level quantum computation, to construct circuits for this experiment. Classical control flow, including if-then-else statements and for loops, are supported by pytket. While loops and early termination of programs are not supported by pytket, in contrast to programming \helios with Guppy. Guppy can also be used to program \htwoone via intermediate conversion to QIR, although the full command set of Guppy is not supported by that workflow.

On \htwoone, spontaneous scattering during two-qubit gates can transfer a \(^{171}\mathrm{Yb}^{+}\) ion from the computational subspace into other hyperfine states. Subsequent gates then act incorrectly, while measurement typically misidentifies the leaked state as $\left| 1 \right\rangle$. We use a leakage-detection gadget introduced and benchmarked in \textcite{Moses_2023} which couples each qubit in the circuit to an ancilla before the final measurement. The ancilla flags leakage, allowing flagged shots to be discarded. We only apply this technique to QUOPS circuits with $w<56$ to avoid serialization of terminal measurements. As with \helios, \htwoone also supports leakage repump and dynamical decoupling, although we do not use this functionality in the results reported. 

\subsection{Logical experiments on \helios}\label{ssec:helios_logical}
This section describes the QUOPS experiments in which each computational qubit was encoded in a Steane code logical qubit on \helios. Section~\ref{sssec:helios_logical_summary} summarizes the experiments and their results. Sections~\ref{sec:helios-synthesis} and \ref{sec:helios-randomness} describe two ingredients common to all of them: the synthesis of the $R_P(\theta)$ rotations, and the per-shot runtime randomization. Sections~\ref{sec:helios-baseline-implementation}--\ref{sec:helios-fully-ft} describe the three successive implementations of the Steane code that were run. Section~\ref{sec:helios-T-QUOPS} describes how the QUOPS score was computed, and Section~\ref{sec:logical-capability-region} includes an emulated logical capability region.

\subsubsection{Summary}\label{sssec:helios_logical_summary}
Our logical QUOPS experiments were run using the Steane code, a $\qeccode{7}{1}{3}$ quantum error correction code~\cite{Steane1996}. We ran QUOPS circuits with up to eight Steane code logical qubits, and we achieved a maximum logical QUOPS of $Q=40$ at width $w=4$.
At this size, the mirrored logical circuits that we use to estimate the polarization contain an average of 20 CNOT gates,  20 $T$ gates, and QEC gadgets involving real-time error correction. 

To achieve this score, each $R_P(\theta)$ gate was independently synthesized within a specified target precision (Sec.~\ref{sec:helios-synthesis}). Synthesis error was accounted for by inserting logical depolarizing channels after each $R_P(\theta)$ gate, with error probability equal to the process infidelity of the synthesized gate. This enables estimation of a lower bound on the process fidelity using an MCFE procedure involving a single circuit ensemble (as opposed to the three introduced in Sec.~\ref{ssec:circuit-classes}), thereby reducing the sample complexity of the estimation procedure [see Sec.~\ref{sec:reference_compiler} and Eq.~\eqref{eqn:single-circuit-estimator} for further details]. 
Beyond synthesis, we perform no compilation of the logical circuit. To minimize the variance of our polarization estimator, we use Guppy's runtime randomness to perform per-shot randomization of the logical circuit, i.e., a random logical circuit is drawn from the appropriate ensemble independently for each shot (Sec.~\ref{sec:helios-randomness}).

We iteratively implemented three versions of the Steane code experiment, which differ at the level of their primitive logical components and QEC strategies. The baseline implementation achieved 24 QUOPS. By making use of better parallelization, higher fidelity magic states, per-shot randomization, and only performing QEC where required, we both increased the QUOPS rate and improved the QUOPS score to 40 QUOPS. In Secs.~\ref{sec:helios-baseline-implementation}--\ref{sec:helios-fully-ft} we describe the logical primitives used in each of the three implementations.

We remark that during logical circuit execution,
the physical qubits needed for magic state preparation are allocated on-the-fly whenever a logical $T$ gate is called for. This is enabled by the real-time control engine described in Sec.~\ref{sec: Real-time engine control}. As fault-tolerant architectures continue to scale, a reservoir of magic states prepared in parallel will likely be needed, along with a notion of concurrency between the logical circuit execution and magic state preparation.

\subsubsection{Synthesis of \texorpdfstring{$R_P(\theta)$}{R(theta)} rotations} \label{sec:helios-synthesis}
The synthesis of arbitrary-angle rotations is performed by rounding each angle in the logical QUOPS circuit to $B=3$ bits of precision. This means that the angle in each $R_P(\theta)$ gate is rounded to the nearest discrete angle belonging to $\{ 2\pi k /2^B \mid k \in [2^B] \}$. Note that this synthesis procedure approximates axial rotations by axial rotations, as opposed to a more general $\mathrm{SU}(2)$ gate. Given a rounded angle $\varphi$ with binary fraction representation $\varphi = b_0.b_1b_2$, the gate $R_Z(\varphi)$ has the decomposition $Z^{b_0} S^{b_1} T^{b_2}$, which includes at most one $T$ gate per rotation. In practice, we decompose $R_Z(b_0.b_1b_2)$ rotations using the gate sequences $\{ I, T, S, ST, Z, T^\dagger S^\dagger, S^\dagger, T^\dagger \}$, which include at most one $T$ or $T^\dagger$ gate per rotation. These sequences ensure that $g(c^\dagger) = g(c)^\dagger$ where the compiler $g(\, \cdot \,)$ implements the synthesis method described above. To apply this procedure to a general $R_P(\theta)$ gate, we first compile each Pauli-axis rotation into an $R_z$ rotation conjugated by a Clifford gate and then synthesize the $R_z$ rotation as above. No further compilation of the logical circuits is applied, which permits per-shot randomization with minimal classical compute overhead (see Sec.~\ref{sec:helios-randomness} below).

Using $B$ bits of synthesis precision leads to notches separated by $\Delta = 2\pi/2^B$, which gives a worst-case process infidelity of $r = 1-\cos(\Delta/4)^2 \approx \Delta^2 / 16$ per rotation. Averaging uniformly over target angles $\theta \in [0, 2\pi)$ gives an average infidelity of $1/2-\sin(\Delta/2)/\Delta \approx \Delta^2 / 48$ per Pauli-axis rotation. The choice to use $B=3$ bits of precision was made to approximately balance synthesis error and the remaining logical error budget at the circuit sizes of interest. For example, for QUOPS circuits with $(w, s) = (4, 40)$, using $B=3$ bits of precision gives a process polarization ${\approx}77.3\%$ in the absence of noise. Note that this is approximately equal to $e^{-1/4} \approx 77.9\%$, which is the point at which the error budget is divided equally between synthesis and other sources of logical errors, assuming that they combine multiplicatively. For reference, at this size, using $B=2$ or $B=4$ would give ${\approx}36\%$ and ${\approx}94\%$, respectively. The largest circuit with $w=4$ logical qubits that can be run with $B=3$ before the error budget is entirely used up by synthesis is $(w, s)=(4, 72)$.

Synthesis error was accounted for using the depolarizing channel insertion technique described in Sec.~\ref{sec:reference_compiler}. Recall that this technique requires logical Pauli operators to be approximately error free. This assumption is satisfied since logical Paulis correspond to transversal 1Q gates in the $\qeccode{7}{1}{3}$ color code and, furthermore, when compiling to the native gate set, physical 1Q gates are combined with other 1Q gates that occur before or after physical 2Q gates if 1Q gate squashing is active. This means that no additional transport of ions is required, and the same number of native, arbitrary-angle 1Q gates are applied.

\subsubsection{Runtime randomization} \label{sec:helios-randomness}
All randomization is performed at runtime and on a per-shot basis.
Randomizing components \ref{itm:randomize-circuits}--\ref{itm:randomize-physical-rc} below on every shot reduces the variance of the estimator for the polarization given a fixed total number of shots. Equivalently, this leads to tighter confidence intervals (on average) for a fixed number of shots.
Since the QUOPS benchmark involves passing a polarization threshold with a fixed level of confidence, obtaining tighter confidence intervals leads to a larger lower boundary $\hat{\polarization}_\text{lower}$, which implies that we can pass the QUOPS test at a particular circuit size using fewer shots.
Note, however, that we always prespecify the number of shots to be executed prior to running the experiment to avoid p-hacking. A fully randomized implementation consists of
\begin{enumerate}[label=(\alph*)]
    \item \label{itm:randomize-circuits} a randomized circuit structure, consisting of random CNOT pairings in every layer, random Pauli axes for every 1Q rotation, and random angles for every 1Q rotation,
    \item applying random 1Q Cliffords on every logical qubit after state preparation and their inverses prior to measurement (up to a uniformly random Pauli operator so that $U(L'L)\ket{0}$ is a uniformly random bitstring),
    \item applying stochastic logical Paulis for the unraveling of the depolarizing channels that account for synthesis error, and
    \item \label{itm:randomize-physical-rc} Pauli twirling of all \emph{physical} 2Q gates.
\end{enumerate}
As detailed in Table \ref{tab:[[7,1,3]]-helios}, the baseline (v1) Steane implementation used most of the above per-shot randomization with the exception of \ref{itm:randomize-circuits} where a fixed number of random circuit structures were sampled for a fixed number of shots, and \ref{itm:randomize-physical-rc} where Pauli twirling of 2Q gates was performed at the logical level, with an additional randomization over stabilizer-equivalent representatives of each logical Pauli.

For subsequent experiments (v2 and v3), \emph{all} the above variables were drawn independently for every shot that ran on the device. This degree of runtime randomization is made straightforward to implement by Guppy~\cite{Koch2024guppy}. A complete summary of the logical experiments and the level of runtime randomness used is given in Table~\ref{tab:[[7,1,3]]-helios}.

\subsubsection{Baseline Steane code implementation (v1)} \label{sec:helios-baseline-implementation}

We use a repeat-until-success (RUS) protocol for FT $\ket{{0}}$ state preparation. Specifically, we use Goto's circuit~\cite{goto2016minimizing} involving a non-FT encoding circuit followed by a logical $Z$ measurement using a single ancilla.
To apply logical $T$ gates, we first fault-tolerantly prepare a magic $\ket{T}=T\ket{+}$ state in an ancillary Steane block,
and then inject the state using a teleportation gadget:
\begin{equation}
    \begin{quantikz}[]
        \lstick{$\ket{\psi}$} & \targ{} & \meter{}\wire[d][1]{c}  \\
        \lstick{$\ket{T}$} & \ctrl{-1} & \gate{SX} & \rstick{$T\ket{\psi}$}
    \end{quantikz}
    \label{eqn:T-gate-teleportation}
\end{equation}
which swaps the data qubits with freshly initialized ancilla qubits, thereby helping to ameliorate the effects of leakage. The logical $\ket{T}$ state is prepared by non-fault-tolerant encoding followed by a flagged measurement of $(X+Y)/\sqrt{2}$ and flagged extraction of the six Steane code stabilizers, adapting the postselected FT magic-state preparation scheme of Ref.~\cite{Chamberland2019faulttolerantmagic} to the $\ket{T}$-state basis.

All logical Clifford operations in the Steane code are implemented by transversal gates.
Real-time QEC was performed using the flag-fault-tolerant scheme of Ref.~\cite{Reichardt_2021}, which was previously implemented experimentally in Ref.~\cite{RyanAndersonRealization}.
This scheme uses 3 ancilla qubits that simultaneously measure half of the stabilizer generators while flagging each other. The QEC cycle consists of up to 2 rounds of flagged stabilizer measurements to measure all 6 stabilizer generators,
followed by conditional unflagged syndrome extraction circuits that occur if either round of flagged measurements returns a non-trivial outcome.
QEC is performed after every single-qubit logical $R_P(\theta)$ gate whose synthesis contains a $T$ or $T^\dagger$ gate, and on all logical qubits after every CNOT layer.

\subsubsection{Optimized Steane code implementation (v2)}
\label{sec:helios-qec-gadgets}
Here we describe the second, optimized version of the Steane code experiment. The QEC policy is identical to the baseline implementation with QEC being performed after every single-qubit rotation requiring magic state injection and after every CNOT. 

\begin{figure}
    \centering
    \includegraphics[width=\linewidth]{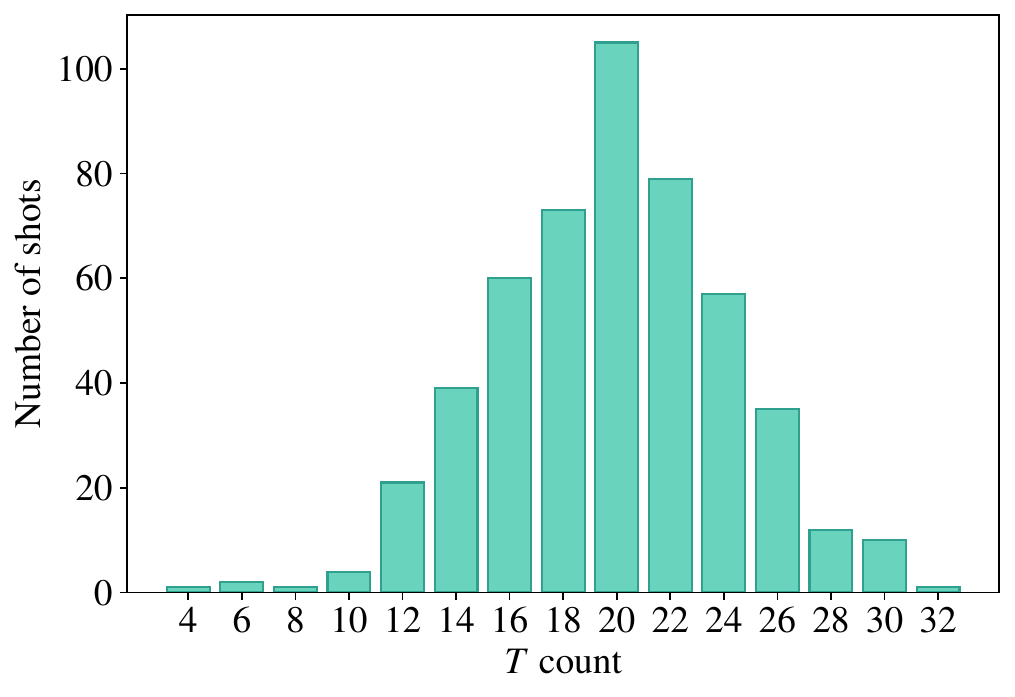}
    \caption{\textbf{Distribution of $\boldsymbol{T}$ count for size-40 logical experiment.} For circuits that ran during the $(w, s) = (4, 40)$ experiment using the fully fault-tolerant v3 Steane code implementation. The counts correspond to the mirrored circuits, so are equal to twice the number of $T$ gates in the forward QUOPS circuit. The median is equal to 20, leading to a T-QUOPS of $Q_T = 10$.}
    \label{fig:Helios-1-T-QUOPS}
\end{figure}

As described in Sec.~\ref{sec:helios-randomness}, as opposed to sampling a fixed number of circuits for a fixed number of shots, we randomize over the circuit on every shot. Due to the significant variance in $T$ count (see Fig.~\ref{fig:Helios-1-T-QUOPS}) between different random circuits, this reduces the variance of the final polarization estimator given a fixed number of shots. 

We additionally improve the parallelizability of the program by processing pairs of blocks in parallel after the layer of logical $R_P(\theta)$ rotations. As opposed to applying flagged syndrome extraction serially to each code block requiring QEC, we group the blocks requiring QEC into pairs. If the number of such blocks is odd, the residual block is left unpaired and processed separately. Processing the blocks in pairs means that the physical gates and measurements belonging to the QEC gadgets can occur in parallel, thereby reducing the overall circuit execution time and reducing memory error.

The $\ket{T}$ state is prepared by first preparing the $\ket{H}$ state using a RUS protocol from Ref.~\cite{perlin2026faulttolerantexecution},
which is a modification of the flagged FT magic state preparation circuit from Ref.~\cite{Chamberland2019faulttolerantmagic}.
This circuit involves a non-fault-tolerant $\ket{H}$ state encoding circuit,
followed by a flagged measurement of the logical $H$ operator along with additional flagging to make the circuit fault-tolerant.
The state is accepted if the logical $H$ measurement and all the flagged measurements yield the correct outcome.
All RUS state preparations use a RUS limit of 3, after which the shot is discarded. However, in our experiments that used this implementation, this limit was never reached and no postselection was necessary.
The $\ket{H}$ state is then rotated into a $\ket{T}$ state using $HS \ket{H} = \ket{T}$ and injected using the circuit~\eqref{eqn:T-gate-teleportation}. 

\subsubsection{Fully fault tolerant Steane code implementation (v3)} \label{sec:helios-fully-ft}
In the third implementation we used a different QEC policy that ensures the QUOPS circuits are fully fault-tolerant at distance three. The previous implementations made use of gadgets that were individually fault-tolerant, i.e., no fault within a gadget propagates to an error that is uncorrectable by the code. However, when these gadgets are composed, there can exist circuit locations at which an $O(p)$ fault can propagate to a logical error with the chosen QEC policy of applying QEC after $T$ and CNOT gates only.
In the third implementation, we modify the QEC policy to ensure that the circuits remain fault tolerant under composition of fault-tolerant gadgets and, furthermore, only insert QEC where needed to ensure that this property holds.
In the following, we allow all physical single-qubit and two-qubit gates, initializations, and measurements to fail with probability $p$.

\paragraph{Frame tracking}
The encoded computation is accompanied by a classical Pauli-frame label $\in \{ \{ Z, X \}, \{ Z, Y \}, \{X, Y\} \}$ specifying the error family associated with each logical block.
In the conventional CSS (i.e., $\{ Z, X \}$) setting, a single circuit fault may generate $X$-, $Z$-, or $\{ Z, X \}$-type errors with probability $O(p)$. Since $X$ and $Z$ errors can be treated separately, the $X$-weight and $Z$-weight must both at most $\lfloor (d-1)/2 \rfloor$ for an $\qeccode{n}{k}{d}$ code. Hence, for distance $d=3$ codes, $X \otimes Z$ errors can be generated with probability $O(p)$.
Under Clifford conjugation, the error frames are permuted and can efficiently be tracked classically through Clifford gates without performing error correction after every operation. 

We also make use of a type of ``correlated'' decoding, which refers to the joint decoding of the two Pauli components associated with the $\{X, Y\}$ or $\{ Z, Y \}$ frames using a correlated lookup table (as opposed to treating $X$- and $Z$-type errors independently).
Under conjugation by an $H$ gate, $X\leftrightarrow Z$, so $\{X, Y\} \leftrightarrow \{ Z, Y \}$. Under an $S$ gate, $X \leftrightarrow Y$ (up to phase), so $\{ Z, X \} \leftrightarrow \{ Z, Y \}$. When correlated QEC is performed (see below), it returns the logical block to the canonical $\{ Z, X \}$ frame. Together, frame tracking and correlated error correction ensure that the encoded QUOPS circuits remain fully fault tolerant at distance three, such that no single circuit-location fault propagates to an error uncorrectable by the code.

Correlated QEC is required at boundaries where the current frame could allow a correctable error to propagate into an uncorrectable one.
For a transversal CNOT, $X_c \otimes I_t \mapsto X_c \otimes X_t$ and $I_c \otimes Z_t \mapsto Z_c \otimes Z_t$, where $c$ and $t$ subscripts refer to the control and target qubits of the CNOT respectively.
Consequently, a block in the $\{X, Y\}$ frame must be decoded before being used as a control, since its $X$-type errors can propagate onto the target, whereas a block in the $\{ Z, Y \}$ frame must be decoded before being used as a target, since its $Z$-type errors can propagate onto the control.
Likewise, an $\{X, Y\}$-frame block must be decoded before a destructive $Z$-basis measurement, where its $X$-type errors could flip the measurement outcome.
A correlated decoding scheme that obeys the properties described above can be realized using Steane error correction based on logical teleportation~\cite{perlin2026faulttolerantexecution}:
\begin{equation}
\scalebox{1}{
\begin{quantikz}[row sep={0.8cm,between origins}]
    \lstick{$\ket{\psi_L}$}
        & \targ{}
        & \qw
        & \qw
        & \meter{}
        & \gate[2]{\text{LUT}} \setwiretype{c}
    \\
    \lstick{$\ket{+_L}_{\mathrm{strict}}$}
        & \ctrl{-1}
        & \ctrl{1}
        & \gate{H}
        & \meter{}
        & \wire[d]{c} \setwiretype{c}
    \\
    \lstick{$\ket{0_L}$}
        & \qw
        & \targ{}
        & \qw
        & \qw
        & \gate{P_L}
        & \rstick{$\ket{\psi_L}$}
\end{quantikz}
}
\label{eqn:correlated-Steane-QEC}
\end{equation}
The first teleportation uses a strictly fault-tolerant logical $|{+_L}\rangle$ resource and a destructive $Z$-basis measurement to obtain the syndrome sensitive to $X$-type errors; the second provides the complementary syndrome sensitive to $Z$-type errors.
The two syndrome strings are concatenated and decoded using a frame-specific lookup table (LUT), which we refer to as $L_{XY}$ or $L_{ZY}$.
These LUTs are constructed at compile time by exhaustively enumerating the relevant correlated error family over the Steane block. For example, the $\{X, Y\}$ table includes the relevant $X$-, $Y$-, and $\{X, Y\}$-type error patterns. 
The resulting LUT correction is then used to correct the state. Since the $\ket{0_L}$ state preparation can produce $X \otimes Z$ errors with probability $O(p)$, the frame at the output of the gadget~\eqref{eqn:correlated-Steane-QEC} is $\{ Z, X \}$.

\paragraph{Strict fault tolerance} 
The first Steane resource state must be prepared in a strict-FT manner. That is, a single preparation fault must not generate, with probability $O(p)$, a multi-qubit mixed-Pauli error, such as an $X\otimes Z$-type error. Absent the strict-FT state preparation, $X\otimes Z$-type errors can propagate to logical errors since they are incompatible with the assumed frame (note that \eqref{eqn:correlated-Steane-QEC} is only applied in the $\{ X, Y\}$ or $\{ Z, Y \}$ frames).
This strict-FT preparation ensures that such frame-incompatible errors require at least two independent physical faults and are therefore suppressed to $O(p^2)$.

\paragraph{Optimizing magic state preparation}
The logical $\ket{T}$ state is prepared using an optimized RUS protocol, obtained by modifying the flagged FT magic-state preparation circuit of Ref.~\cite{Chamberland2019faulttolerantmagic} to reduce the preparation overhead while preserving fault tolerance. 
In particular, each preparation attempt starts from a non-FT unitary encoding of $\ket{T_L}$, followed by a flagged measurement of the logical observable $T X T^\dagger=(X+Y)/\sqrt{2}$, for which $\ket{T_L}$ is the $+1$ eigenstate, and a final QED round consisting of three stabilizer measurements. The state is accepted only when all verification outcomes are trivial.
A further optimization is possible for magic-state injection, since the injection itself contains a teleportation step in which the data block participates in a transversal CNOT and is subsequently destructively measured in the $Z$ basis (see Eq.~\eqref{eqn:T-gate-teleportation}). When the incoming data block is in an $\{X, Y\}$ or $\{ Z, Y \}$ frame (and therefore requires correlated decoding before this teleportation), the second half of a separate Steane-QEC cycle can be merged with the injection:
\begin{equation}
\begin{quantikz}[row sep={0.8cm,between origins}]
    \lstick{$\ket{\psi_L}$}
        & \ctrl{1}
        & \gate{H}
        & \meter{}
        & \gate[2]{\text{LUT}} \setwiretype{c}
    \\
    \lstick{$\ket{0_L}_{\mathrm{strict}}$}
        & \targ{}
        & \targ{}
        & \meter{} 
        & \wire[d]{c} \setwiretype{c} 
    \\
    \lstick{$\ket{T_L}$}
        & \qw
        & \ctrl{-1}
        & \qw
        & \gate{U_L}
        & \rstick{$T\ket{\psi_L}$}
\end{quantikz}
\label{eqn:T-state-half-Steane}
\end{equation}
Instead of first performing a complete correlated Steane QEC round, only the complementary half is performed beforehand: a strict-fault-tolerant resource state $|0_L\rangle$ is coupled to the data and the appropriate $X$-basis syndrome is extracted destructively, providing one half of the joint syndrome. The complementary component is then obtained directly from the destructive $Z$-basis measurement already required by the magic-state-injection teleportation. The two are combined into a joint syndrome---as in the full correlated-QEC procedure~\eqref{eqn:correlated-Steane-QEC}---and decoded with the corresponding $L_{XY}$ or $L_{ZY}$ LUT.
Since the $\ket{T_L}$ state preparation can produce $X \otimes Z$ errors with probability $O(p)$, and the correction unitary $U_L$ can involve an $S$ gate, the frame on the output of the gadget~\eqref{eqn:T-state-half-Steane} is either $\{Z, X \}$ or $\{ Z, Y \}$ depending on the applied correction.

\begin{figure}
    \centering
    \includegraphics[width=0.8\linewidth]{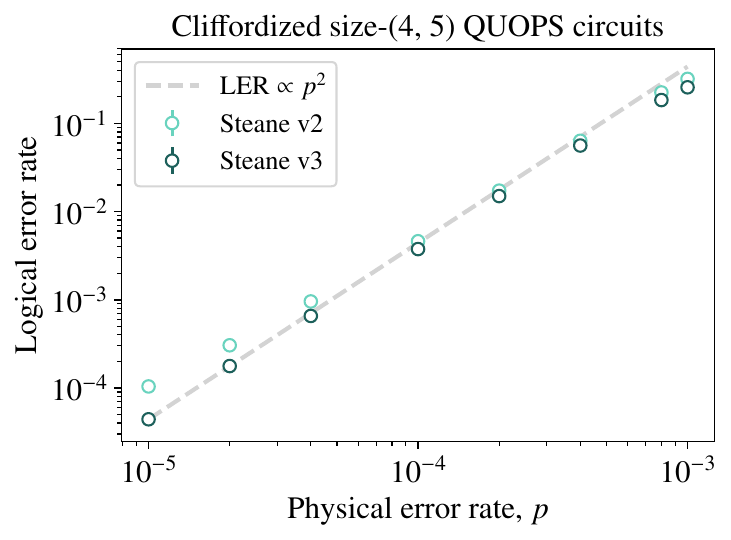}
    \caption{\textbf{Simulated scaling of logical error rate (LER) with physical error rate.} For mirrored random QUOPS circuits of shape $(w, s) = (4, 40)$ using the optimized (v2) and fully FT (v3) Steane code implementations. The data for the fully FT implementation are consistent with a LER scaling as $p^2$, whereas the v2 implementation departs from $p^2$ scaling at sufficiently small $p$. We use Cliffordized versions of the gadgets to permit efficient simulation and use a depolarizing error model with a uniform physical error rate $p$ for all 1Q and 2Q gates, initializations, and measurements. Note that the simulations are run without inserting stochastic Paulis to account of synthesis error so that $\text{LER}=0$ for $p=0$. One-sigma Wilson confidence intervals are included, but they are smaller than the markers for all data points.}
    \label{fig:error_rate_scaling}
\end{figure}

\paragraph{Improving parallelizability}
A further reduction in circuit execution time is obtained by using \emph{parallel state factories} whenever multiple $T$ gates are required in the same circuit layer. The corresponding encoded blocks are grouped into pairs, which are processed independently. If the number of $T$ gates is odd, the remaining block is processed individually. For each pair, the strictly fault-tolerant resource states required for the half-Steane QEC [see Eq.~\eqref{eqn:T-state-half-Steane}] are first prepared in parallel and immediately consumed by the two encoded blocks.
{Note that, if one of the state preparations fails during RUS, it is possible for the successfully prepared state to be used while the second, unsuccessful state is retried.}
The corresponding logical $T\ket{+}$ magic states are subsequently prepared in parallel and immediately used for the magic-state injection teleportations. The remaining teleportation measurements, correlated decoding, and logical feed-forward are then performed independently for each block. An analogous construction is used for performing the state preparations relevant to the correlated Steane QEC [see Eq.~\eqref{eqn:correlated-Steane-QEC}] that is performed before each CNOT layer in the QUOPS circuits.
Finally, for the initial $\ket{{0_L}}^{\otimes w}$ state preparation, we attempt the preparation of four $\ket{0_L}$ states in parallel.

\subsubsection{Estimating QUOPS and architecture-specific QUOPS} \label{sec:helios-T-QUOPS}
To determine whether a particular experiment passes the threshold, we must calculate a one-sided 95\% confidence interval. To do this, we compute the bootstrap distribution of the estimator~\eqref{eqn:single-circuit-estimator} using at least \num{9999} resamples of the experimental shots with replacement and find the associated confidence interval using scipy's \texttt{bootstrap} function~\cite{scipy}.

In addition to universal logical QUOPS, we also compute the T counts and the size of the low-level circuits associated with the largest logical circuits that we ran. The $T$ count at shape $(w,s)$ is computed as the median of the number of  $T$ gates in the compiled circuits as shape $(w,s)$. We find that the $(w, s) = (4, 40)$ experiment corresponds to a $T$ count of $Q_T = 10$ (see Fig.~\ref{fig:Helios-1-T-QUOPS} for the distribution of $T$ gates across shots). This coincides with the expected value since the average $T$ count per $R_P(\theta)$ rotation is $1/2$.
Due to the nondeterministic nature of programs submitted to \helios (e.g., as a result of RUS state preparation), we do not extract the physical gate counts from the classical output of the programs that ran. 
Instead, we use a local version of the \helios compiler to compute the expected distributions of native physical gates applied during program execution, which also allows us to place a tighter confidence interval on the estimate. Local emulation requires the use of Clifford simulations to determine the probabilities with which different branches in the program are taken.
We therefore use a Cliffordized version of magic-state injection that injects an $S$ gate. We ensure that the physical circuits used to implement $S$ state injection involve the same physical one-qubit and two-qubit gate locations with only their angles being modified to make the circuits efficiently simulable. It is therefore expected that transport and error rates for the $S$ and $T$ state injection circuits should be comparable, and indeed the polarizations obtained via Cliffordized simulations match well with \helios experimental data (see Sec.~\ref{sec:logical-capability-region}).

\begin{figure}[t]
    \centering
    \includegraphics[width=\linewidth]{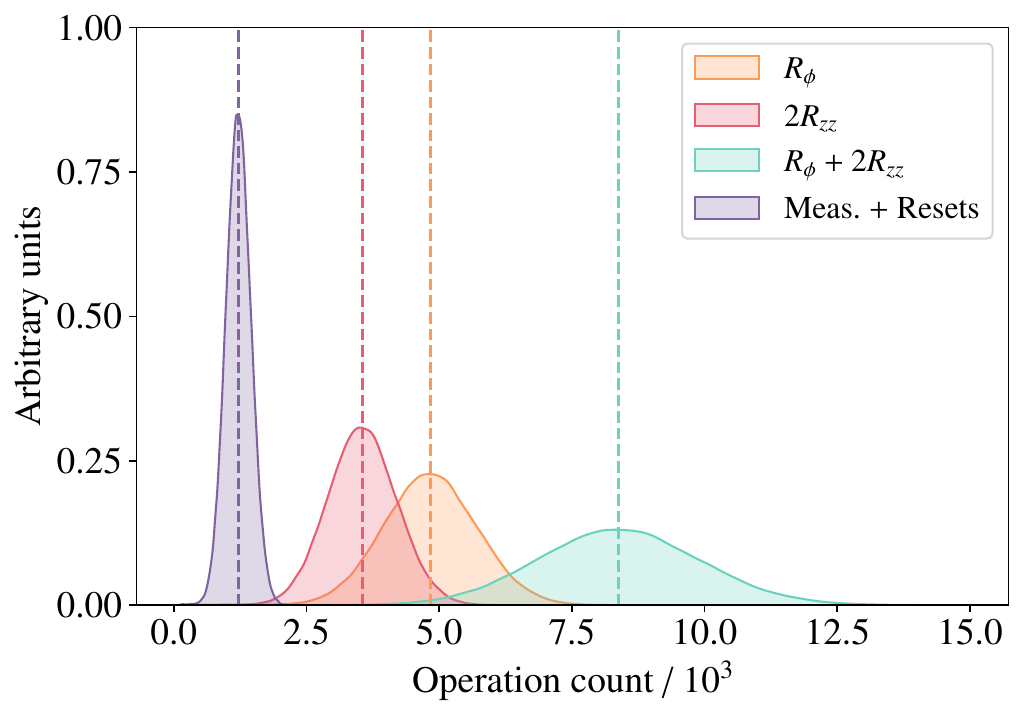}
    \caption{\textbf{Distributions of physical-level gate counts.} Empirical distribution of native single-qubit ($R_\phi$) and two-qubit ($R_{zz}$) gates, the associated size, and total number of physical measurement and reset operations applied during execution of the $(w, s)=(4, 5)$ logical QUOPS program with the $\qeccode{7}{1}{3}$ color code drawn from the circuit distribution~\eqref{eqn:logical-circuit-ensemble}. The empirical distributions are based on $10^5$ sampled random QUOPS circuits. The variance in the distribution comes both from the fluctuations in the number of $T$ gates in each approximately compiled random circuit, as well as from RUS state preparations. As in the experiments we perform, the repeat-until-success limit is set to 3. The dashed lines denote the medians of the individual empirical distributions.}
    \label{fig:size-distribution}
\end{figure}

Simulating $10^5$ circuits of shape $(w, s) = (4, 40)$ for the fully FT Steane code implementation and extracting their gate counts gives rise to the distributions shown in Fig.~\ref{fig:size-distribution}. To determine architecture-specific size, we use $\kappa(R_\phi)=1$, $\kappa(R_{zz})=2$, and $\kappa(R_z)=0$ in Eq.~\eqref{eq:circuit-kappa-size}. This gives an architecture-specific QUOPS of $Q_\text{arch}=\num{4189}(3)$ from the median of these distributions. The error is determined by bootstrapping the median estimator to obtain a $68\%$ confidence interval. 

\subsubsection{Full emulated capability region} \label{sec:logical-capability-region}
\begin{figure}[t]
    \centering
    \includegraphics[width=\linewidth]{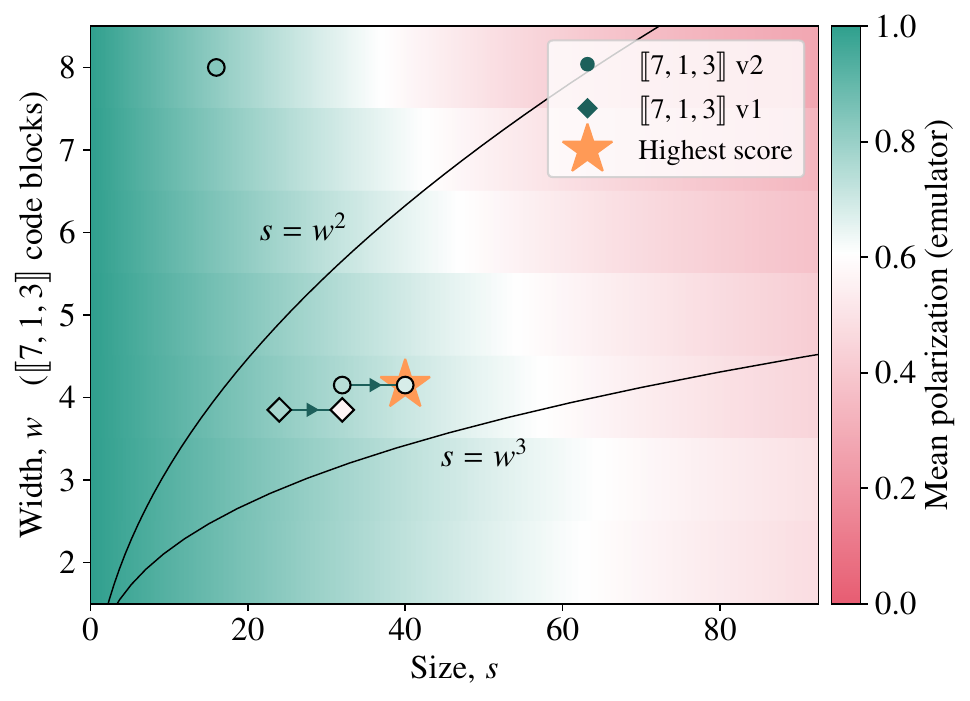}
    \caption{\textbf{Logical \helios dataset, with simulated capability region.} Uses $B=3$ bits of precision. Different Steane code experiment versions are offset for clarity at width $w=4$. Using $B=3$ sets an upper bound on achievable circuit sizes of $\lesssim 70$. The heatmap shows the estimated polarization based on Clifford simulations of $10^4$ circuits drawn from the ensemble in Eq.~\eqref{eqn:logical-circuit-ensemble} using the fully randomized (v2) Steane code implementation. The data are interpolated in the size direction. The v3 Steane code implementation achieved the same score $s=40$ with an improved QUOPS rate.}
    \label{fig:Helios-logical-capability}
\end{figure}

To estimate the full logical capability region of \helios using the procedure outlined in the previous sections, we perform Clifford simulations using \texttt{Selene}.
As explained in Sec.~\ref{sec:helios-T-QUOPS}, we make the compiled QUOPS circuits efficiently simulable by injecting $S$ and $S^\dagger$ gates in place of $T$ and $T^\dagger$ gates, respectively, using Cliffordized gadgets that are expected to lead to identical transport and have identical physical 1Q and 2Q gate counts.
The simulated capability region is illustrated in Fig.~\ref{fig:Helios-logical-capability} using $B=3$ bits of precision. Recall that $B=3$ bits of precision sets an upper limit of $s \sim 70$ on the circuit size that can be achieved in the absence of noise. As in the physical-level capability region shown in Fig.~\ref{fig:main:capability:Helios}, there is a weak falloff in the maximum achievable size with an increasing number of Steane code blocks due to ion transport phase errors. Note also that the largest achievable size for width $w=2$ is finite---this is because the QUOPS circuits are not compiled beyond synthesis of arbitrary angle rotations, since this enables maximal runtime randomization. 

\subsection{Physical experiments}\label{ssec:quantinuum_physical}
This section describes the physical-qubit QUOPS experiments on \helios and \htwoone, whose results are listed in Tables~\ref{tab:physical-quops-helios} and \ref{tab:physical-quops-h2-1} and shown in Figs.~\ref{fig:main:capability:Helios}, \ref{fig:main:capability:H2-1} and \ref{fig:Helios-physical-capability}. At the physical level we employed standard MCFE \cite{Proctor2022-zs}, and where needed employed Pauli frame randomization of the native entangling gates (the standard randomized compiler of Section~\ref{sssec:standard_rc}). Sections~\ref{sssec:helios_physical_circuits} and \ref{sssec:h2_physical_circuits} describe how the mirror circuit ensembles were constructed and sampled on each system, Section~\ref{sssec:helios_ternary} the ternary measurements used for optional leakage postselection on \helios, and Section~\ref{sec:physical-emulated-region} includes an emulated \helios capability region.

\subsubsection{\helios MCFE circuit class construction}\label{sssec:helios_physical_circuits}
We distinguish between compile-time randomness and runtime randomness enabled by Guppy. Circuits are drawn randomly at compile time, whereas random state preparation, and Pauli frame randomization of entangling gates and measurements are all performed at runtime and on a per-shot basis.

For each data-point $(w, s)$, $2K$ random circuits are generated $\{ c_{w,s} \}$ parameterized by $(w, s)$. Of these, $K$ are used for the $M_1$ ensemble and $K$ are used for the $M_2$ ensemble.
We exclusively consider even $w$, and choose $s$ such that the number of benchmarking layers $l$ is an integer---with the exception of $w=98$ for which $2l$ is an integer. While this is not a requirement of the QUOPS benchmark, we did this to maximize the ratio of 2Q to 1Q gates, thereby making the circuits as challenging as possible. In particular, if $w$ is even, $w$ and $w+1$ contain the same number of 2Q gates for the same number of benchmarking layers $l$.
Size $s$ can therefore be increased by $l$ using only 1Q gates by using odd $w$.
We fix $K=10$ circuits for all data points.

For each such circuit we use standard pytket compilation passes into the Quantinuum native gate set, $\{ R_z(\theta), R_{\phi}(\theta), R_{zz}(\theta) \}$, to obtain compiled circuits $g(c_{w,s})$. These passes are exact and permit compilation between benchmarking layers and permutations of qubits, as is natural on QCCD systems. Note, however, that when $g(c_{w,s})$ is combined with the reference circuit $g_\text{ref}(c_{w,s}^\dagger)$ below, a barrier is inserted between mirrored halves to prevent further compilation. 

For each circuit, $c_{w,s}$, we additionally construct a reference compilation, $g_\text{ref}(c_{w,s})$ where, in addition to applying standard pytket passes, $g_\text{ref}$ applies Pauli frame randomization (PFR) to all 2Q gates $R_{zz}(\theta)$ in the compiled circuit $g(c_{w,s})$. This PFR is performed at runtime on a per-shot basis. In general, $R_{zz}(\theta)$ is a non-Clifford gate; the commutation of a Pauli, $P$, through the gate requires the sign of the rotation angle to be flipped, $P R_{zz}(\theta) = R_{zz}(-\theta) P$, if $P$ anti-commutes with $Z \otimes Z$. As a random Pauli will anti-commute with $Z \otimes Z$ half of the time, we effectively twirl the set $\{ R_{zz}(\theta), R_{zz}(-\theta) \}$. In order for the assumptions underlying PFR to be satisfied, we require that the coherent elements of the error channels associated to the gates are uncorrelated with each other with respect to angle sign \cite{haghshenas2025digital}. A sufficient condition for this to hold would be if the two channels are the same.

The final circuit components required for MCFE are $L_w$ and $L_w'$. The former consists of iid random 1Q Clifford gates on each qubit, while the latter is its inverse with the addition of measurement randomized compiling (before measurement, a random Pauli $X$ gate is applied on each qubit with probability 50\%, which is corrected classically). Each 1Q Clifford or Pauli gate is drawn at random at runtime on a per shot basis. Putting these components together we can construct the three required circuit classes (using matrix multiplication order)
\begin{enumerate}[label=(\arabic*)]
    \item $L' \mid g_\text{ref}(c_\text{rev}) \mid g(c) \mid L$,

    \item $L' \mid g_\text{ref}(c_\text{rev}) \mid g_\text{ref}(c) \mid L$,

    \item $L' \mid  L$
\end{enumerate}
where $c_2 \mid c_1$ denotes circuit concatenation with a compilation barrier. The compilation barrier ensures that, when the program is compiled to lower-level instructions, all the operations before the barrier must have been performed before operations after the barrier commence, and that no compilation (including 1Q gate squashing) is performed across the barrier. 

For each data point on Fig.~\ref{fig:Helios-physical-capability}, circuit classes (1) and (2) were executed a total of $N_\text{shots}=2000$ times, with $\{c_{w,s}\}$ drawn independently for classes (1) and (2) a total of 10 times each. Circuit class (3) was also executed  $N_\text{shots}=2000$ times, and the data shared across all data points of the same width. Full details can be found in Table~\ref{tab:physical-quops-helios}.

To determine whether a particular experiment passes the threshold, we must calculate a one-sided 95\% confidence interval. To do this, we compute the bootstrap distribution of the estimator~\eqref{eq:meanpol-estimator-1} using 9999 resamples of the ensembles $M_1$ and $M_2$ with replacement and find the associated confidence interval using scipy's \texttt{bootstrap} function~\cite{scipy}.

\subsubsection{\htwoone MCFE circuit class construction}\label{sssec:h2_physical_circuits}

As in the case of the logical experiments on \helios, the experiment on \htwoone uses a MCFE procedure in which $g=g_\text{ref}$ (see Eq.~\eqref{eq:mcfe_equation-simplified}), i.e., in which both halves of the mirror circuit employ Pauli frame randomization. This approach reduces the sample complexity of the experiments since only circuits from the $M_2$ and $M_3$ ensembles need to be executed. Otherwise, circuits are constructed and optimized in the \htwoone experiments as they are in the \helios experiments. We sample 10 random QUOPS circuits. The 10 circuits are optimized with standard pytket optimization techniques, and rebased into the $\{ R_z(\theta), R_{\phi}(\theta), R_{zz}(\frac{\pi}{2}) = \text{ZZMax} \}$ gateset. These 10 circuits define 10 instances of the $M_2$ mirror-circuit ensemble. We sample from each such instance 200 times. For the SPAM measurement, there is a single ensemble for each width $w$. We sample from this ensemble 2000 times. Full details can be found in Table~\ref{tab:physical-quops-h2-1}. In the case of the size 448 circuit at width 56 we instead: sample 50 random QUOPS circuits, sample one randomly compiled instance of each of these circuits, and take 40 shots from each of these randomly compiled circuits.

\begin{table*}
    \bgroup
    \setlength{\tabcolsep}{0.35em}
    \scalebox{0.8}{
        \begin{tabular}{@{}llcc w{c}{1.2em} ccccc w{c}{1.2em} ccccc@{}}
            \toprule \toprule
            \multirow{2}{*}{Date (2026)}
            & \multirow{2}{*}{Runtime}
            & \multicolumn{2}{c}{\textbf{Pass}}
            & \qquad
            & \multicolumn{5}{c}{Circuit options}
            & \qquad
            & \multicolumn{5}{c}{Results}
            \\
            \cmidrule(r){3-4}
            \cmidrule(r){6-10}
            \cmidrule(l){12-16}
            & & Raw & PS
            &
            & Width & Size & Circuits & Shots/circuit & Total shots
            &
            & $\hat{\polarization}$ & 95\% CI & $\hat{\polarization}_{\mathrm{PS}}$
            & 95\% CI (PS) & Leakage discard
            \\
            \midrule

            August 19
            & ---
            & \cmark & \cmark
            &
            & 8 & 512 & 10 & 200 & 2000
            &
            & 0.8079 & $[0.7937,1]$ & 0.827 & $[0.8136,1]$ & 8.05\%
            \\

            August 29
            & \SI{99}{\minute} \SI{23}{\second}
            & \cmark & \cmark
            &
            & 12 & 1320 & 10 & 200 & 2000
            &
            & 0.6312 & $[0.6096,1]$ & 0.6733 & $[0.6545,1]$ & 19.7\%
            \\

            August 29
            & \SI{155}{\minute} \SI{19}{\second}
            & \xmark & \cmark
            &
            & 12 & 1392 & 10 & 200 & 2000
            &
            & 0.6119 & $[0.5983,1]$ & 0.6583 & $[0.646,1]$ & 20.7\%
            \\

            August 30
            & \SI{135}{\minute} \SI{5}{\second}
            & \xmark & \xmark
            &
            & 12 & 1488 & 10 & 200 & 2000
            &
            & 0.5784 & $[0.5652,1]$ & 0.6239 & $[0.6056,1]$ & 21.4\%
            \\

            September 3
            & ---
            & \xmark & \cmark
            &
            & 16 & 1024 & 10 & 200 & 2000
            &
            & 0.6081 & $[0.5799,1]$ & 0.6414 & $[0.6102,1]$ & 19.1\%
            \\

            September 2
            & ---
            & \xmark & \xmark
            &
            & 16 & 1184 & 10 & 200 & 2000
            &
            & 0.5774 & $[0.5484,1]$ & 0.6236 & $[0.5934,1]$ & 23.8\%
            \\

            August 21
            & ---
            & \xmark & \xmark
            &
            & 16 & 1344 & 10 & 200 & 2000
            &
            & 0.5206 & $[0.4821,1]$ & 0.5618 & $[0.5206,1]$ & 25.1\%
            \\

            August 26
            & ---
            & \xmark & \xmark
            &
            & 16 & 1504 & 10 & 200 & 2000
            &
            & 0.4775 & $[0.4539,1]$ & 0.5145 & $[0.4877,1]$ & 27.2\%
            \\

            August 27
            & ---
            & \xmark & \xmark
            &
            & 16 & 1664 & 10 & 200 & 2000
            &
            & 0.3986 & $[0.3691,1]$ & 0.4378 & $[0.4045,1]$ & 28.1\%
            \\

            August 27
            & ---
            & \cmark & \cmark
            &
            & 32 & 640 & 10 & 200 & 2000
            &
            & 0.6608 & $[0.6397,1]$ & 0.7018 & $[0.6782,1]$ & 26.1\%
            \\

            August 29
            & ---
            & \xmark & \xmark
            &
            & 32 & 960 & 10 & 200 & 2000
            &
            & 0.5364 & $[0.5226,1]$ & 0.5831 & $[0.5668,1]$ & 29.55\%
            \\

            August 30
            & ---
            & \cmark & ---
            &
            & 56 & 224 & 10 & 200 & 2000
            &
            & 0.8462 & $[0.828,1]$ & 0.8465 & [0.8301]] & 18.45\%
            \\

            September 4
            & ---
            & \cmark & ---
            &
            & 56 & 448 & 50 & 40 & 2000
            &
            & 0.7883 & $[0.7698,1]$ & --- & --- & ---
            \\
            
            \bottomrule\bottomrule
        \end{tabular}
    }
    \egroup
    \caption{
    \textbf{Physical experiments run on Quantinuum \htwoone.}
    PS denotes leakage postselection.
    A result passes when the lower endpoint of its 95\% confidence
    interval exceeds $1/\sqrt{e}$.
    Polarization estimates and confidence bounds are rounded to four
    decimal places. SPAM experiments were done for each point for an extra
    2000 shots.
    }
    \label{tab:physical-quops-h2-1}
\end{table*}

The \helios and \htwoone experiments differ in the means of sampling from the aforementioned ensembles, which is to say the means of achieving per-shot randomization. \htwoone supports populating classical registers with randomness which differs between shots. We can use this to perform randomized compiling, frame randomization, and measurement randomization. 
\begin{description}
    \item[Randomized compiling: ] For each ZZMax gate, populate 4 bits with random binary values drawn from the uniform distribution over $\{0, 1\}^4$. Use these bits to control $X$ and $Z$ gates acting on the two qubits immediately before the ZZMax gate. This has the effect of implementing a uniformly random two-qubit Pauli operator. Use a further 4 bits per ZZMax gate to control $X$ and $Z$ gates applied after the ZZMax gate. The values held by these additional 4 bits are chosen such that the resulting Pauli undoes the Pauli gates applied before the ZZMax gate. These values can be calculated at run-time.
    \item[Frame randomization: ] For each qubit generate an integer uniformly at random from the set $\{ 0, 1, 2, 3, 4, 5 \}$. Control $\{ I, X, H, HX, SH, SHX \}$ applied at the start of the circuit on this integer. This generates a random eigenstate of $X$ or $Y$ or $Z$. At the end of the circuit, control the appropriate inverse Clifford on the random integer.
    \item[Measurement randomization: ] For each qubit, generate a random bit. Control an $X$ gate on this bit immediately before the measurement. Apply a classical not gate on the measurement outcome, controlled on the random bit.
\end{description}
It is not possible to use run-time randomness to generate random rotations which must be generated at compile time.

Results from these experiments can be seen in Table~\ref{tab:physical-quops-h2-1}. As with \helios bootstrapping is used to calculate $95\%$ confidence intervals.

\subsubsection{Ternary measurements on \helios} \label{sssec:helios_ternary}
To enable optional post-selection of leakage on Helios, see Fig.~\ref{fig:main:capability:Helios}, we perform a ternary destructive measurement of each qubit at the end of the circuit, which returns $\{0,1,2\}$ where $2$ is indicative of leakage. This subroutine performs qubit state shelving in other long-lived atomic states and is detailed in Ref.~\cite{Ransford2026-ny}. The infidelity is comparable with that of the typical SPAM error.

Postselection based on leakage is straightforward, by discarding any shots where $2$ is measured for any qubit from the dataset. However to obtain results without postselection we must create the statistics that would have been obtained had a binary measurement \(\{0,1\}\) had been performed instead.  This is possible due to the following observations.

In general, when performing measurement randomized compiling for a binary measurement \(\{0,1\}\), the physical measurement outcomes must be flipped with 50\% probability. If leakage is state-independent---in the sense that the computational state of the qubit prior to leakage does not effect the Beta distribution describing the probability of obtaining 0 or 1 upon measurement of the leaked qubit---then the outcome obtained performing a binary measurement upon a leaked qubit after correction will be 0 or 1 with equal probability. 

\begin{figure}
    \centering
    \includegraphics[width=\linewidth]{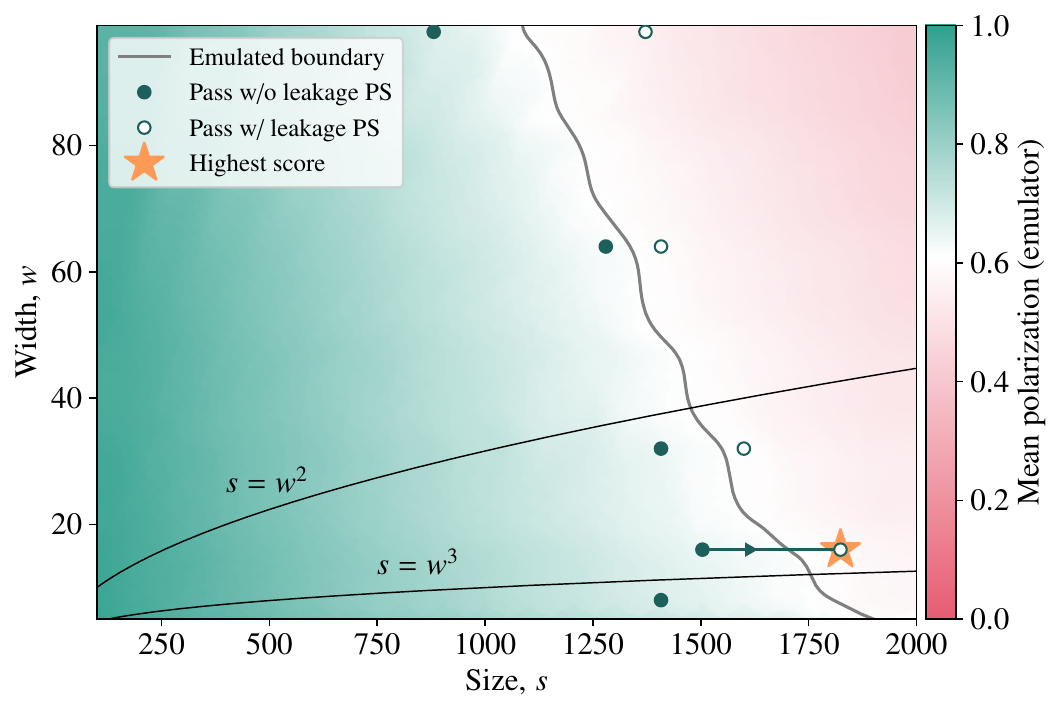}
    \caption{\textbf{Physical \helios dataset, with simulated capability region.} The heatmap shows the estimated polarization based on Clifford simulations, see Sec.~\ref{sec:physical-emulated-region}, and has been interpolated in both the width and size direction.}
    \label{fig:Helios-physical-capability}
\end{figure}

The above allows us to use results obtained using ternary measurements to be used to recreate binary measurement results by randomly substituting 0 or 1 for the obtained value 2 with equal probability. 

\subsubsection{Full emulated \helios capability region}\label{sec:physical-emulated-region}
Similar to the logical experiments, see Sec.~\ref{sec:logical-capability-region}, we employ emulation to fill out the full capability region substituting Clifford gates such that the simulation can be performed efficiently. For each compiled circuit $g(c_{w,s})$, in the Quantinuum native gate set $\{ R_z(\theta), R_{\phi}(\theta), R_{zz}(\theta) \}$, we replace each angle with a uniformly random setting from the discrete set of Clifford gate angles for each gate. The simulated capability region is shown in Fig.~\ref{fig:Helios-physical-capability}.
The heatmap shows the estimated polarization based on Clifford simulations with $10^5$ shots per circuit to effectively remove the effect of finite sampling.

\begin{table*}
\caption{ \textbf{Logical experiments run with the $\boldsymbol{\qeccode{7}{1}{3}}$ color code on Quantinuum \helios.} Optional features included real time randomness (RTR)---used for various randomization subroutines, as detailed in the main text---as well as leakage repump (LR), dynamical decoupling (DD), and single-qubit gate squashing (1QS). Randomized compilation (RC) was performed at either the physical or logical level. All experiments used a repeat-until-success limit of 3. {}$^*$Timing information obtained during normal operation.}
\bgroup
\setlength{\tabcolsep}{0.35em}
\scalebox{0.8}{
\begin{tabular}{@{}cllc w{c}{1.2em} ccccc w{c}{1.2em} ccccc w{c}{1.2em} ccc@{}}
\toprule\toprule
\multirow{2}{*}{Version}&\multirow{2}{*}{Date (2026)}& \multirow{2}{*}{Runtime}                     & \multirow{2}{*}{\textbf{Pass}} &\qquad & \multicolumn{5}{c}{Circuit options}           &\qquad & \multicolumn{5}{c}{Compiler options}                                      &\qquad & \multicolumn{3}{c}{Results} \\ \cmidrule(r){6-10} \cmidrule(l){12-16} \cmidrule(l){18-20}
                            &&                                              &                                &       & Width & Depth & Size & Circuits & Total shots &       & RC       & RTR           & LR             & DD             & 1QS         && $\hat{\polarization}$ & 95\% CI & Discard      \\ \midrule
v3&August 29                     & {\SI{73}{\minute}$^*$}                            & \cmark                         &       & 4     & 5     & 40   & 500      & 500         &       & Physical & {Full} & \texttt{True}  & \texttt{True}  & \texttt{True}&& 0.730 & $[0.699, 1.0]$ & 0\% \\
v2&June 28                     & \SI{239}{\minute}                            & \cmark                         &       & 4     & 5     & 40   & 500      & 500         &       & Physical & {Full} & \texttt{True}  & \texttt{True}  & \texttt{True}&& 0.682 & $[0.646, 1.0]$ & 0\% \\
v2&June 27                     & \SI{62}{\minute} \SI{38}{\second}$^*$            & \cmark                         &       & 4     & 4     & 32   & 200      & 200         &       & Physical & {Full} & \texttt{True}  & \texttt{True}  & \texttt{True}&& 0.740 & $[0.688, 1.0]$ & 0\% \\
v2&August 12                   & ---                                          & \cmark                         &       & 8     & 1     & 16   & 100      & 100         &       & Physical & {Full} & \texttt{True}  & \texttt{True}  & \texttt{True}&& 0.836 & $[0.770, 1.0]$ & 0\% \\ 
v1&June 11                     & ---  & \xmark                         &       & 4     & 4     & 32   & 10       & 500         &       & Logical  & Partial       & \texttt{True}  & \texttt{False} & \texttt{True}&& 0.565 & $[0.455, 1.0]$ & 0\% \\
v1&June 09                     & --- & \cmark                         &       & 4     & 3     & 24   & 20       & 1000        &       & Logical  & Partial       & \texttt{True}  & \texttt{False} & \texttt{True}&& 0.771 & $[0.745, 1.0]$ & 0\% \\ \bottomrule\bottomrule
\end{tabular}
}
\egroup
\label{tab:[[7,1,3]]-helios}
\end{table*}

\begin{table*}
\caption{
\textbf{Physical experiments run on Quantinuum \helios.}
PS denotes leakage postselection.
A result passes when the lower endpoint of its 95\% confidence
interval exceeds $1/\sqrt{e}$.
Polarization estimates and confidence bounds are rounded to four
decimal places. SPAM experiments were performed for
2000 shots, at the time of the earliest submission for each width.
}
\bgroup
\setlength{\tabcolsep}{0.35em}
\scalebox{0.8}{
\begin{tabular}{@{}llcc w{c}{1.2em} ccccc w{c}{1.2em} ccccc@{}}
\toprule \toprule
\multirow{2}{*}{Date (2026)}
& \multirow{2}{*}{Runtime}
& \multicolumn{2}{c}{\textbf{Pass}}
& \qquad
& \multicolumn{5}{c}{Circuit options}
& \qquad
& \multicolumn{5}{c}{Results}
\\
\cmidrule(r){3-4}
\cmidrule(r){6-10}
\cmidrule(l){12-16}
& & Raw & PS
&
& Width & Size & Circuits & Shots/circuit & Total shots
&
& $\hat{\polarization}$ & 95\% CI & $\hat{\polarization}_{\mathrm{PS}}$
& 95\% CI (PS) & Leakage discard
\\
\midrule
May 29
& \SI{424}{\minute}
& \cmark & \cmark
&
& 8 & 1408 & 20 & 200 & 4000
&
& 0.6677 & $[0.6326,1]$ & 0.7373 & $[0.7013,1]$ & 15.875\%
\\
June 02
& \SI{284}{\minute}
& \cmark & \cmark
&
& 16 & 1504 & 20 & 200 & 4000
&
& 0.6547 & $[0.6214,1]$ & 0.7378 & $[0.7036,1]$ & 20.000\%
\\
June 03
& \SI{397}{\minute}
& \xmark & \cmark
&
& 16 & 1824 & 20 & 200 & 4000
&
& 0.6240 & $[0.5863,1]$ & 0.6791 & $[0.6407,1]$ & 23.850\%
\\
June 05
& ---
& \xmark & \cmark
&
& 32 & 1600 & 20 & 200 & 4000
&
& 0.6412 & $[0.5986,1]$ & 0.7323 & $[0.6890,1]$ & 31.325\%
\\
June 28
& ---
& \cmark & \cmark
&
& 32 & 1408 & 20 & 200 & 4000
&
& 0.6720 & $[0.6332,1]$ & 0.7582 & $[0.7199,1]$ & 26.475\%
\\
May 20
& ---
& \xmark & \cmark
&
& 64 & 1408 & 20 & 200 & 4000
&
& 0.6270 & $[0.5791,1]$ & 0.7569 & $[0.7070,1]$ & 38.950\%
\\
June 28
& ---
& \cmark & \cmark
&
& 64 & 1280 & 20 & 200 & 4000
&
& 0.6749 & $[0.6328,1]$ & 0.7959 & $[0.7533,1]$ & 35.550\%
\\
May 07
& ---
& \xmark & \cmark
&
& 98 & 1372 & 20 & 200 & 4000
&
& 0.5158 & $[0.4662,1]$ & 0.7071 & $[0.6500,1]$ & 51.375\%
\\
July 10
& ---
& \cmark & \cmark
&
& 98 & 881 & 20 & 200 & 4000
&
& 0.6957 & $[0.6508,1]$ & 0.8613 & $[0.8173,1]$ & 40.500\%
\\
\bottomrule\bottomrule
\end{tabular}
}
\egroup
\label{tab:physical-quops-helios}
\end{table*}

\section{Google experiments}
\label{sec:Google}
This section details the QUOPS experiments on Google \willow, a 105-qubit processor \cite{Google-Quantum-AI-and-Collaborators2025-ad}. All experiments and simulations were performed using a physical-qubit architecture. 

\subsection{Circuit construction}
\label{subsec:Google circuit construction}
The QUOPS experiments and simulations were designed to map out the QUOPS capability region of \willow, while keeping overall circuit count low. The circuit sizes were therefore chosen iteratively based on simulations of the process polarization of the next-smallest circuit for a given width, with an initial exponential spacing in circuit depth. When the process polarization of a circuit approached the threshold, we increased the density of sizes to better estimate the maximizing circuit shape near the boundary. Once a circuit size was reached where the measured process polarization fell below the threshold, we returned to an exponential spacing of circuit shapes until the polarization fell close to zero.

As permitted by the QUOPS benchmark, we allow for optimized layouts, routing, and compilation of the QUOPS circuits to match the gate set and connectivity of a given device. This is important as the \willow processor does not have all-to-all connectivity, and thus must efficiently route circuits to maximize QUOPS. All compilation was done with built in \texttt{cirq} \cite{Cirq_Developers_2026} methods. The CNOT gates in the QUOPS circuits were compiled to CZ gates, and all single qubit gates were compiled to PhasedXZ gates.

These compiled QUOPS circuits were then further randomly compiled to create the reference circuits and reversed reference circuits used in MCFE, disallowing all other forms of compilation. From these circuit parts, the ensembles $M_1$, $M_2$, and $M_3$ were constructed and executed in simulation to estimate the process fidelity of the compiled QUOPS circuits using the approximate estimator in Eq.~\eqref{eq:approximate_estimator}.

\subsection{Experiments}
\label{subsec:Google physical experiments}

Two experiments were run on  Google \willow. The first explored the full capability region and found the circuit shape at which the QUOPS score was maximized. The second optimized the QUOPS rate at the maximizing circuit shape. For the full capability region experiment, 100 QUOPS circuits were sampled per unique circuit shape $(w, s)$, corresponding to $\kappa = 200$ mirror circuits for each shape. For each circuit, 1000 shots were collected. We additionally required 200 reference circuits for each width $w$, also executed with 1000 shots.

For the experiment that optimized the QUOPS rate, 25 QUOPS circuits were sampled at the maximizing circuit shape of $(6, 216)$ with 500k shots each. Alongside this, 5 reference circuits were executed with 200 shots for each width. This experiment did pass the success threshold and achieved the same QUOPS as the capability region experiment, but the experiment parameters were optimized to maximize the QUOPS rate at the cost of runtime.

\subsection{Simulations}
\label{subsec:Google physical simulations}
\begin{figure}
    \centering
    \includegraphics[width=\linewidth]{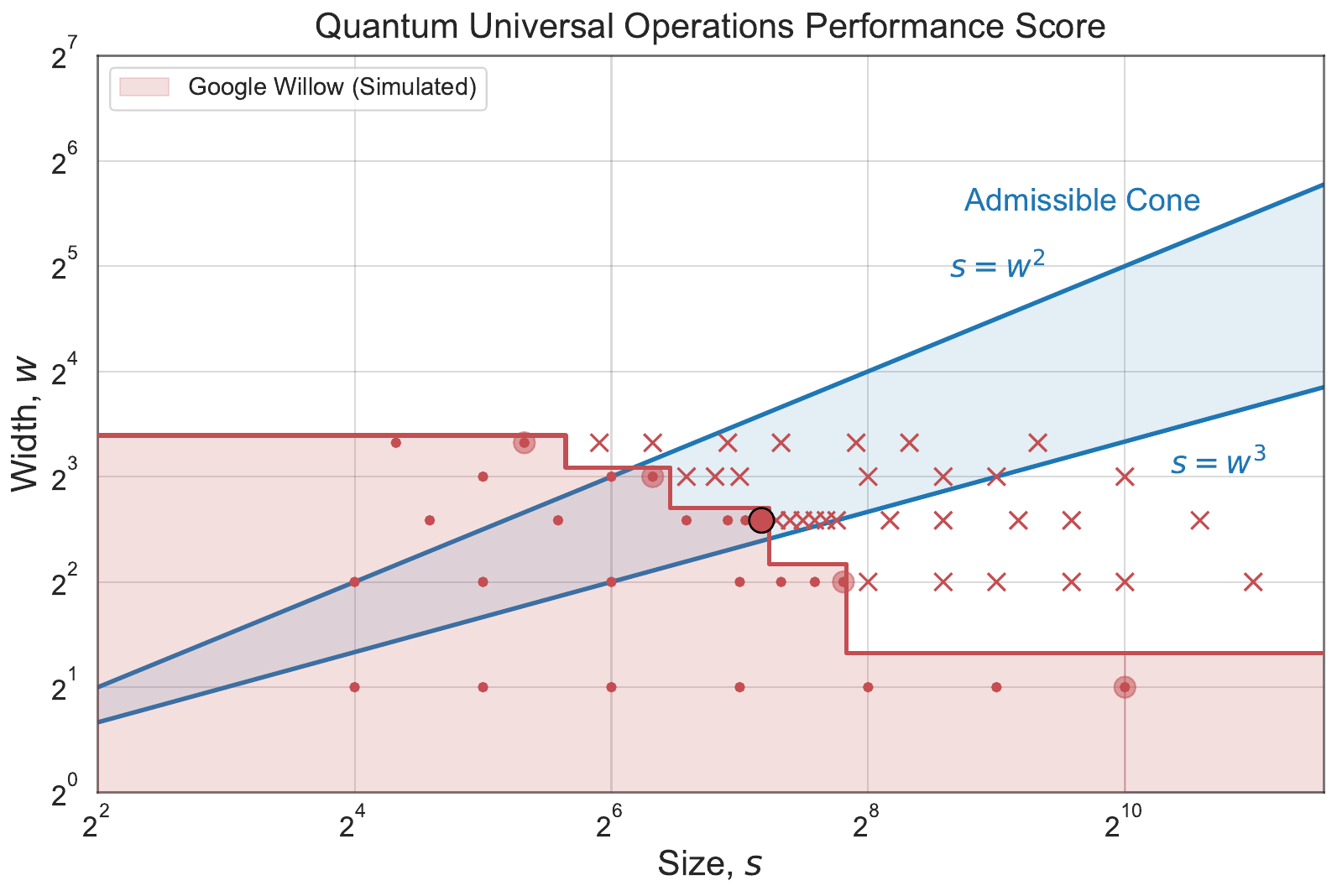}
    \includegraphics[width=\linewidth]{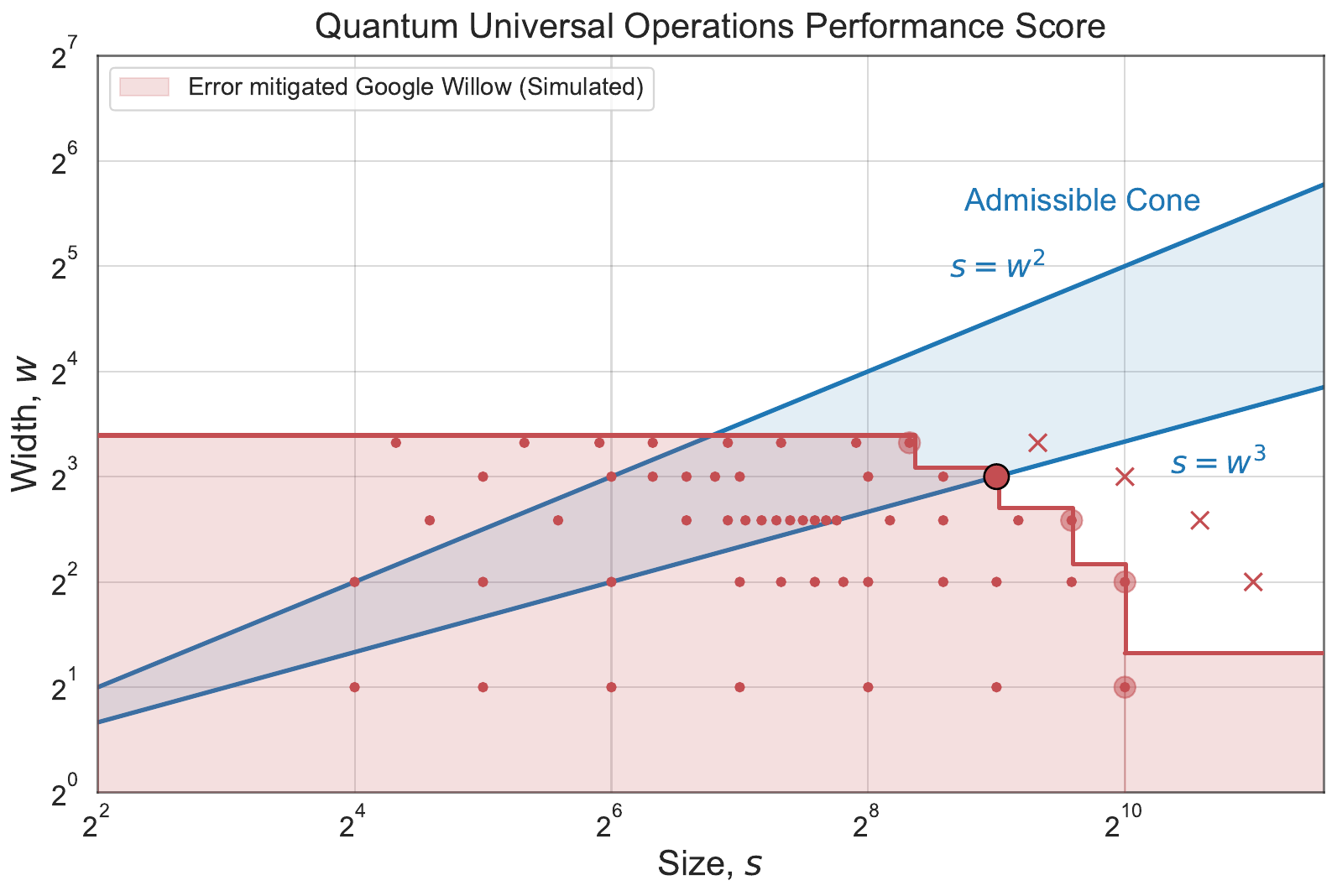}
    \caption{\textbf{Simulated capability regions for \willow.} It achieves a QUOPS of $Q = 144$ at $(6, 144)$ with a QUOPS rate of $\Omega = 7.69 \times 10^{6}$ QUOPS/s and an error mitigated QUOPS of $Q^* = 512$ at $(8, 512)$ with a QUOPS rate of $\Omega^* = 8.42 \times 10^{5}$ QUOPS/s. }
    \label{fig:google_willow_sim}
\end{figure}

Our simulations of the Google \willow processor constructed a noise model using the device noise properties as provided by the \texttt{cirq} package. The simulated capability region of the device is shown in Figure \ref{fig:google_willow_sim}. Simulations result in a maximizing circuit shape of $(w,s) = (6,144)$, for a QUOPS of $Q = 144$ and a QUOPS rate of $\Omega = 7.69 \times 10^{6}$ QUOPS/s. An error mitigated capability region extends to larger sizes as expected, resulting in a QUOPS of $Q^* = 512$ at $(8, 512)$, an overhead of $80$, and a QUOPS rate of $\Omega^* = 8.42 \times 10^{5}$ QUOPS/s. This modest decrease in the QUOPS rate for a circuit shape with a much lower polarization can be explained by the overhead being dominated by circuit loading time, rather than execution time. The capability region in simulation is truncated at $w=10$, due to the computational cost of simulating wider circuits.

\subsection{Estimating the QUOPS rate}
\label{subsec:Google QUOPS/s}
In experiment the QUOPS rate of the Google \willow device is measured using many more shots than was used in the initial experiment to find the QUOPS score maximizing circuit shape. For measuring the rate, $5 \times 10^5$ shots were used, as this amount maximizes throughput for the relevant overheads of Google \willow. In particular, this minimizes the contribution that circuit switching has on the execution time of the circuits, optimizing for the fast repetition rate of the device.

This rate measurement excludes overheads from interleaving jobs from other users, but includes other unavoidable overheads such as classical control latency, waveform compilation time, and the time for measuring and resetting qubits between shots. The wall clock time in experiment also included the time to execute a small number of SPAM reference circuits, which slightly decreases the reported rate.

For the simulated results, the QUOPS rate of Google \willow is reported as an estimate using measurements of the average time between shots  ($t_{shot} = 9.83 \times 10^{-6}$ seconds), the average time between circuits in a batch ($t_{circuit} = 2.87 \times 10^{-2}$ seconds), and the average time between batches of circuits ($t_{batch} = 0.57$ seconds). These measurements were taken from running QUOPS circuits on the device, and correspond to the average time needed to execute circuits of shape $(6, 216)$.

Because these duration estimates are specific to a circuit shape, using them to extrapolate QUOPS rates at other shapes will under- or over-estimating the time contribution from circuit depth. However, this contribution is be small relative to the overheads for switching circuits and batches, which dominates this particular architecture. As such we report estimated QUOPS rates in simulation using the above durations as similar data was not collected for other circuit shapes.

The estimated wall-clock time to execute all the sampled QUOPS circuits of a given shape can be written as:
\begin{equation}
    \tau_{wall}(w,s) = t_{batch} + \kappa  \left( t_{circuit} + K t_{shot} \right)
    \label{eq:google_duration_estimate}
\end{equation}
where $K$ is the number of shots per circuit (1000) and $\kappa$ is the number of mirror circuits per shape (40 for simulation) or twice the number of sampled QUOPS circuits per shape. The QUOPS rate can then be estimated directly from the estimated runtime of the transpiled mirror circuits:
\begin{equation}
    \Omega(w,s) = \frac{2s \hat{\polarization}^2_{w,s}}{\tau_{wall}(w, s)}(K \kappa)
    \label{eq:google_rate_estimate}
\end{equation}

\section{IBM Q experiments}
\label{sec:IBM Q}
This section details the QUOPS experiments on \boston, and also presents simulations of those QUOPS experiments using IBM's noise model for \boston.

\subsection{Circuit construction}
\label{subsec:IBM circuit construction}
The QUOPS circuits for our IBM experiments were chosen in a similar manner as in the Google experiments as described in Section \ref{subsec:Google circuit construction}. The experiments used 100 circuits per circuit shape, and the simulations used 20. Circuit routing and compilation was carried out by the transpiler in \texttt{qiskit}, with the CNOT gates being compiled to native CZ gates and the arbitrary single qubit rotations being compiled to native Z-rotations. The layout and routing both utilize the \textit{sabre} algorithm, with the optimization level set to 3. After transpilation, the circuits are copied back to \texttt{pyGSTi} for mirroring before being recompiled identically back to \texttt{qiskit} once more with barriers between all layers of the circuit to prevent trivial compilation. In experiment, we included an additional layer of dynamical decoupling applied right before execution, as this manner of compilation does not break any of the assumptions of MCFE.

\subsection{Experiments}
\label{subsec:IBM physical experiments}
Our IBM Q experiments were performed on the 156-qubit \boston Heron R3 processor \cite{ibmquantum} using a physical-qubit architecture. These experiments were performed with 100 circuits per circuit shape and 200 reference circuits per width, as in the Google \willow experiments.

\subsection{Simulations}
\label{subsec:IBM physical simulations}
\begin{figure}
    \centering
    \includegraphics[width=\linewidth]{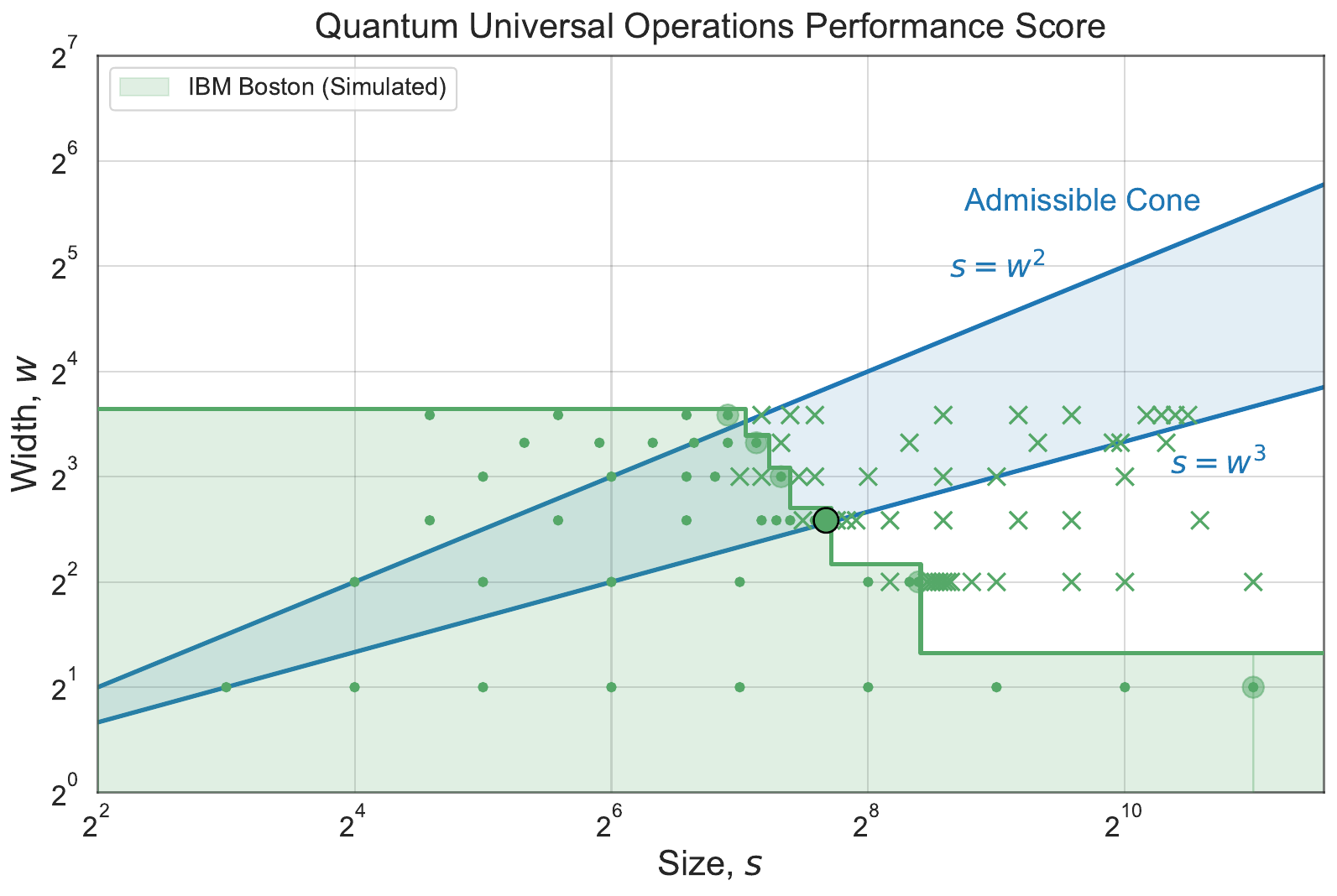}
    \includegraphics[width=\linewidth]{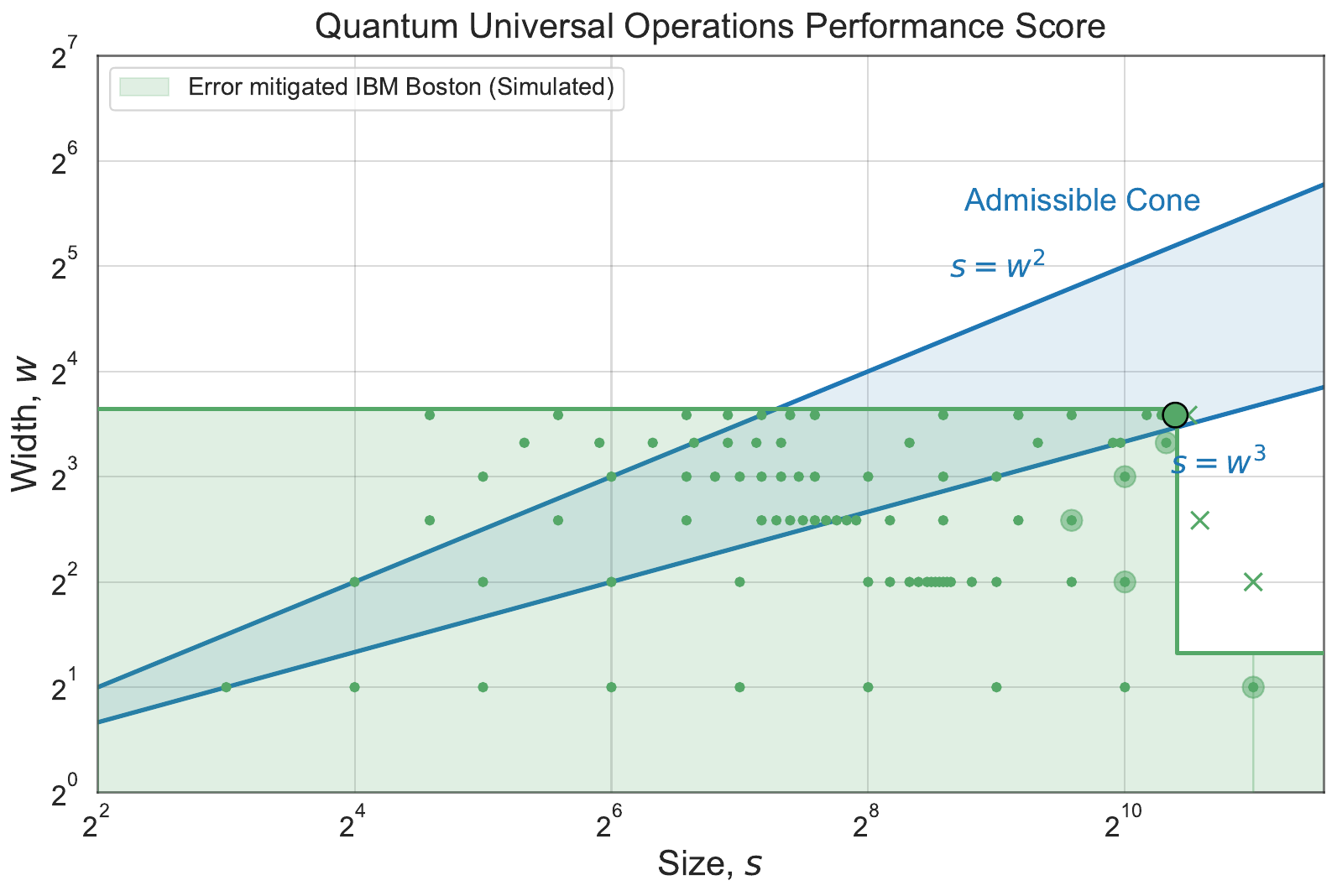}
    \caption{\textbf{Simulated capability regions for \boston.} In simulation, it achieves a QUOPS of $Q = 204$ and a QUOPS rate of $\Omega = 5.72 \times 10^{5}$. It achieves an error mitigated QUOPS of $Q^* = 1344$ at $(12, 1344)$, corresponding to an overhead of $3629$ and an estimated QUOPS rate of $\Omega^* = 2.08 \times 10^{3}$.}
    \label{fig:ibm_boston_sim}
\end{figure}

Our simulations of the \boston device utilize the noise models built into \texttt{qiskit\_ibm\_runtime}. We simulated 20 unique QUOPS circuits per circuit shape $(w, s)$ with 1000 shots per circuit and 20 reference circuits for each width $w$. In simulation, the device achieves a QUOPS of $Q = 204$ at $(6, 204)$ and a corresponding estimated QUOPS rate of $\Omega = 5.72 \times 10^5$, higher than the measured rate in experiment (see Fig.~\ref{fig:ibm_boston_sim}). The error mitigated capability regions extends, resulting in an error mitigated QUOPS of $Q^* = 1344$ at $(12, 1344)$, corresponding to an overhead of $3629$ and a QUOPS rate of $\Omega^* = 2.08 \times 10^3$. The capability region in simulation is truncated at $w=12$, as simulating wider circuits was too costly. 

\subsection{Estimating the QUOPS rate}
\label{subsec:IBM QUOPS/s}
The estimated QUOPS rate of this device uses an approximation in the experiment. In experiment, the manner in which circuits were sent to the device included interleaving of circuits of different circuit shapes, making it infeasible to extract the precise duration of execution for circuits of a particular shape. As a result, we calculate the mean QUOPS rate of a batch of $B$ circuits as twice the sum of the QUOPS for all circuits in the batch divided by the batch wall-clock time:
\begin{equation}
    \Omega_{raw}(B) = \sum_{i=1}^{B} = \frac{2s_{i}}{\tau_{wall}(B)}
    \label{eq:ibm_rate_approx}
\end{equation}
We multiply this raw QUOPS rate by the measured process polarization of a particular certain shape to report the QUOPS rate for that shape.

We note this approximation intermixes the QUOPS rates of different shaped circuits, however we expect the dominant contributions to QUOPS rate for the \boston device to come from classical overheads of loading circuits onto the device and resetting qubits in between shots, not the runtime of the circuits themselves, mitigating the effects of this approximation.

As \texttt{qiskit} only exposes some timing information, there are two different wall-clock times for a batch of circuits. The first is derived from the timestamps of when the batch began running on the device and when the batch finished. The second is to use the QPU usage time reported in the job. The former includes some classical overheads that we wish to exclude from our analysis, such as the addition of dynamical decoupling to the circuits, which could be done before submission as part of more general circuit compilation. As such, we report QUOPS rate metrics using the smaller usage time of the QPU, which results in faster rates.

In simulation, we estimate runtime using available duration estimates for gates and routing to determine how long a given circuit would take to execute on a backend. This method assumes each sampled QUOPS circuit has a variable execution time of $t_{run_{i}}$, a repetition delay $t_{delay}$ of 250 microseconds between executions, a reset duration $t_{reset}$ of 2.2 microseconds, and an overhead of two seconds for loading a batch of $B$ transpiled QUOPS circuits. The estimated duration can be written in terms of the total wall-clock runtime of the sampled QUOPS circuits of a given shape:
\begin{equation}
    \tau_{wall}(w,s) = \frac{2\kappa}{B} + K\sum_{i=1}^{\kappa}\{2t_{run_{i}} + t_{delay} + t_{reset}\}
    \label{eq:ibm_duration_estimate}
\end{equation}
where $K$ is the number of shots per circuit (1000) and $\kappa$ is the number of sampled QUOPS circuits per shape (100 for experiment, 20 for simulation). We assume that a full job of 500 circuits comprises a single batch for loading, setting $B=500$. The factor of two in front of $t_{run}$ accounts for the increased length of the mirror circuit compared to the test circuit.

The QUOPS rate can then be estimated directly from the estimated runtime of the transpiled circuits as in \eqref{eq:omega_operational}:
\begin{equation}
    \Omega(w,s) = \frac{2s \hat{\polarization}^2_{w,s}}{\tau_{wall}(w, s)}(K \kappa)
    \label{eq:ibm_rate_estimate}
\end{equation}

\begin{figure*}
    \centering
    \includegraphics[width=\linewidth]{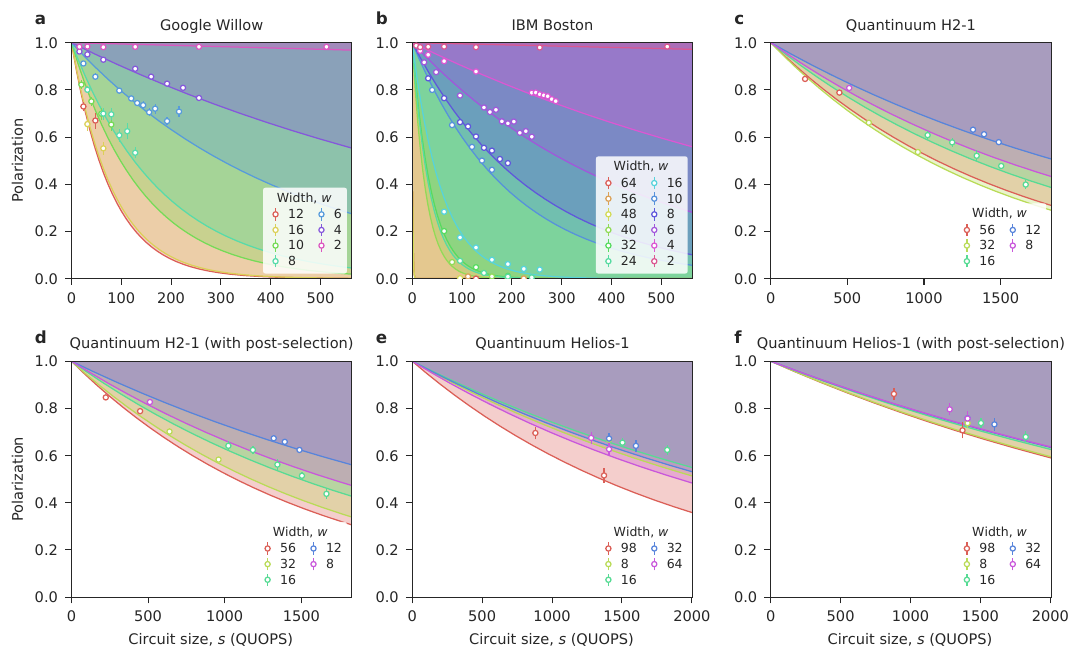}
    \caption{\textbf{Exponential fits for extrapolated regions.} The exponential fits used to obtain the extrapolated error-mitigated capability regions presented in the main text. Here we show the 95\% one-sided confidence intervals on the mean polarization $\bar{\polarization}_{w,s}$, obtained using maximum likelihood estimation and a Gaussian model for estimated polarization uncertainty, for all fits performed to obtain the results in the main text. These 95\% confidence intervals are the regions between $e^{-r_w^{\textrm{upper}} s}$ and 1 as a function of $s$, where $r_w^{\textrm{upper}}$ is the 95\% one-sided confidence interval upper bound on the decay rate of circuit polarization versus $s$. Each panel shows results for one quantum computing system, and within each panel the separate curves correspond to the different circuit widths. Points with 1-$\sigma$ error bars are the estimated polarizations.}
    \label{fig:exponential_fits}
\end{figure*}

\section{Extrapolated capability regions}
This section explains the \emph{extrapolated error-mitigated capability regions} that are presented in the main text. These capability regions use a lower polarization threshold of $\alpha = 0.01$, corresponding to a sample overhead $O$ of up to $O = 1/\alpha^2 = 10^4$. Here we provide the estimation procedure used to produced those extrapolated capability regions. Note that these extrapolated regions extend to larger circuit shapes than are included within the experimental dataset and they might not accurately represent true capabilities if certain assumptions (discussed below) are violated. They should therefore be interpreted with caution and do not carry the same evidential weighted as directly measured regions.

Extrapolated capability regions are based on the following assumption: at a fixed circuit width $w$, the mean process polarization $\bar{\polarization}_{w,s}$ decays exponentially with circuit size $s$. Specifically, 
\begin{equation}
    \bar{\polarization}_{w,s} =  e^{-r_w s}, \label{eq:exp_model}
\end{equation}
where $r_w$ is a width-dependent decay parameter. We estimate $r_{w}$ by fitting the experimentally estimated mean process polarizations $\hat{\bar{\polarization}}_{w,s}$ at a fixed width to Eq.~\eqref{eq:exp_model}. This fit assumes the following model:
\begin{equation}
    \hat{\bar{\polarization}}_{w,s} \sim \mathcal{N}(e^{-r_w s},\sigma_s^2), \label{eq:gauss_model}
\end{equation}
where $\mathcal{N}(\mu,\sigma^2)$ is a normal distribution with mean $\mu$ and standard deviation $\sigma$, and $\sigma_s$ is the standard deviation of $\hat{\bar{\polarization}}_{w,s}$ (estimated via a nonparametric bootstrap). We fit $r_w$ using maximum likelihood estimation, with $r_w$ constrained to $r_w \in [0,1]$, by minimizing the log-likelihood function $\ell(r_w)$ implied by Eqs.~\eqref{eq:exp_model}-\eqref{eq:gauss_model}, producing an estimate $\hat{r}_w$.

To create an extrapolated capability region with 95\% confidence, to match the statistical significance demanded of the directly estimated capability regions, we compute a one-sided 95\% upper confidence interval bound $r_w^{\mathrm{upper}}$ on $r_w$ at each width. $r_w^{\mathrm{upper}}$ is the largest $r_w$ value that is not excluded by a one-sided likelihood-ratio test at 5\% significance level. This is the largest $r_w$ satisfying
\begin{equation}
    2[\ell(\hat{r}_w)-\ell(r_w)] < \chi^2_{1,\,0.95},
\end{equation}
where \(\chi^2_{1,\,0.95}\) is the 95th percentile of the chi-squared distribution with one degree of freedom. 

The extrapolated capability region with polarization threshold $\alpha$ is the set of all circuit shapes such that
\begin{equation}
e^{-r_w^{\textrm{upper}} s} > \alpha.
\end{equation}
All extrapolated capability regions in the main text use $\alpha = 0.01$, and are therefore extrapolated error-mitigated capability regions.

The extrapolated capability regions rely on the assumption that $\bar{\polarization}_{w,s}$ follows an approximately exponentially decay with $s$, for fixed $w$. We conjecture that this will often be a reasonable approximation in practice, but we note that this is heuristic. Because QUOPS circuits can (and typically must) be compiled---and there are few limits on the compilation strategy applied---the theories showing that the polarizations of random circuits decay exponentially in circuit depth (and so size) at fixed width do not apply here.

Figure~\ref{fig:exponential_fits} shows the 95\% one-sided confidence intervals on $\bar{\polarization}_{w,s}$ for all fits performed to obtain the results in the main text. These 95\% confidence intervals are the regions between $e^{-r_w^{\textrm{upper}} s}$ and 1 as a function of $s$, separated by quantum computing system (the panels of Figure~\ref{fig:exponential_fits}) and by width (the separate curves in each panel). 

The extrapolated error-mitigated QUOPS score $Q^*_{\alpha}$ is extracted from the extrapolated error-mitigated capability region in the same way as with a directly measured capability regions: $Q^*_{\alpha}$ is the largest shape in the intersection of the extrapolated region and the admissible cone. Extrapolated error-mitigated QUOPS scores $Q^*_{\alpha}$ are also reported alongside extrapolated QUOPS rates in the main text. These extrapolated rates take the raw QUOPS rate measured at the $Q$-maximizing circuit shape and attenuate them by $e^{-2\hat{r}_w Q_{\alpha}^*}$, which is the fit model's estimate of the square of the circuit polarization at shape $(w,Q_{\alpha}^*)$, where $w$ is the circuit width at which $Q^*_{\alpha}$ is achieved.  Here $Q$ is the directly-measured standard QUOPS score, i.e., defined using the standard threshold of $\alpha = 1/\sqrt{e}$. This process for extrapolating the QOUPS rate assumes that the raw QUOPS rate's dependence on circuit shape is small compared to the contribution of the circuit polarization to the QUOPS rate. This likely results in estimates of the QUOPS rate that have the correct order of magnitude when $\alpha = 0.01$, as the raw rate is attenuated by $10^{-4}$, but it is a heuristic extrapolation.

\section{Estimating QUOPS scores from quantum volume data}
\label{sec:historical_qv}

This section explains how we inferred approximate QUOPS scores for IBM and Quantinuum processors from 2018 to 2025 using historical quantum volume~\cite{Cross2019-ku} data, as shown in Figure~\ref{fig:main:capability:historical}. We used quantum volume (QV) results since there is a long, cross-platform record of QV results and due to similarities between QV and QUOPS. A quantum volume of $2^{n}$ certifies that a processor can execute square circuits of $n$ qubits and $n$ layers, each layer consisting of Haar-random $\mathrm{SU}(4)$ gates on a random pairing of the qubits, with heavy-output probability above $2/3$. In QUOPS terms these are circuits on the square-circuit boundary $s = w^{2}$ of the admissible cone (Section~\ref{ssec:quops_score}), but in a different gate set and with a different success criterion. The conversion to a QUOPS score therefore has two ingredients: a translation between the QV and QUOPS success thresholds, and a translation between the resources consumed by a QV circuit and by a QUOPS circuit on the architecture in question. We use two different procedures for the two platforms, because different data are available for each. For IBM processors only the QV values themselves were available, and so we match compiled gate counts. For Quantinuum processors the heavy-output data are available to us, and we infer an effective two-qubit gate fidelity from them. In both cases the estimates are approximate and are intended to establish trends over years rather than precise scores.

\subsection{Relating success thresholds}
\label{ssec:qv_threshold}

The heavy outputs of a circuit are the bit strings whose ideal probability exceeds the median ideal probability, and $h_{\textrm{ideal}}$ denotes the total ideal probability of the heavy outputs. For Haar-random circuits the ideal output distribution is Porter--Thomas, for which $h_{\textrm{ideal}} \to (1+\ln 2)/2 \approx 0.847$ as $n$ grows~\cite{Cross2019-ku}. We model the effect of noise by a global depolarizing channel with polarization $\polarization$, under which the output distribution is $\polarization\,p_{\textrm{ideal}} + (1-\polarization)/2^{n}$ and the measured heavy-output probability is
\begin{equation}
    h = \tfrac{1}{2} + \polarization\,\big(h_{\textrm{ideal}} - \tfrac{1}{2}\big).
    \label{eq:qv_heavy_output_polarization}
\end{equation}
The QV pass criterion $h > 2/3$ therefore corresponds, for large $n$, to a polarization threshold of \cite{Hines2024-ae}
\begin{equation}
    \polarization_{\textrm{QV}} = \frac{2/3 - 1/2}{(1+\ln 2)/2 - 1/2} = \frac{1}{3\ln 2} \approx 0.48 ,
    \label{eq:qv_polarization_threshold}
\end{equation}
which is lower than the QUOPS threshold of $1/\sqrt{e}\approx 0.61$. If circuit polarization decays exponentially with the number of gates of the dominant error source, then the number of such gates that can be executed at the QUOPS threshold is a fraction 
\begin{equation}
    \eta = \frac{\ln(1/\sqrt{e})}{\ln \polarization_{\textrm{QV}}} = \frac{1/2}{\ln(3\ln 2)} \approx 0.68
    \label{eq:qv_threshold_factor}
\end{equation}
of the number that can be executed at the QV threshold. We ignore the additional statistical requirement of the QV protocol (that $h > 2/3$ with at least two-sigma confidence), and we treat reported QV values as point estimates of capability.

\subsection{Method 1: Matching compiled gate counts}
\label{ssec:qv_ibm}

For IBM processors we convert a quantum volume of $2^{n}$ into a QUOPS score by asking what size of width-$n$ QUOPS circuit consumes, after compilation to a similar architecture (we used a degree-4 lattice in all cases), the same two-qubit-gate budget as a width-$n$ QV circuit, adjusted for the threshold factor $\eta$. The procedure is as follows. We fix a target architecture with a two-dimensional square-grid coupling map and the gate set $\{U_{3}, \textrm{CNOT}\}$, and compile all circuits with Qiskit's transpiler at optimization level 3. For each width $n$ we (i) sample 20 random QV circuits of width and depth $n$, compile them, and record the mean number of CNOT gates $\bar{N}^{\textrm{QV}}_{\textrm{CX}}(n)$ in the compiled circuits; (ii) sample 20 random QUOPS circuits of width $n$ at each candidate size $s$ (all sizes that can be formed from complete benchmark layers plus an optional partial final layer), compile them identically, and record $\bar{N}^{\textrm{QUOPS}}_{\textrm{CX}}(n,s)$; and (iii) take as the converted size the candidate whose compiled CNOT count is closest to the threshold-adjusted QV budget,
\begin{equation}
    s^{\star}(n) = \operatorname*{arg\,min}_{s}\Big|\bar{N}^{\textrm{QUOPS}}_{\textrm{CX}}(n,s) - \eta\,\bar{N}^{\textrm{QV}}_{\textrm{CX}}(n)\Big| .
    \label{eq:qv_ibm_matching}
\end{equation}
This assumes that two-qubit gates dominate the error budget on these processors and that polarization decays exponentially in the number of two-qubit gates in the compiled circuit, with the same per-two-qubit-gate decay for both circuit families. The compiled CNOT count increases monotonically with $s$ (up to sampling fluctuations), so the minimization is performed by a bisection search over the candidate sizes.

Table~\ref{tab:qv_ibm} lists the results for $n = 3,\dots,11$. The converted sizes are well described by $s^{\star}(n)\approx 2.5\,n^{2}$, with deviations of at most $20\%$, and in Fig.~\ref{fig:main:capability:historical} we use this smooth approximation,
\begin{equation}
    Q_{\textrm{est}} = 2.5\,\big[\log_2(\textrm{QV})\big]^{2},
    \label{eq:qv_ibm_conversion}
\end{equation}
rather than the individual values, since the residual variation is smaller than the uncertainty introduced by the choice of coupling map, compiler, and the various other approximates in this methodology. The scaling $s^{\star}\propto n^{2}$ is expected. Before routing, a width-$n$ QV circuit contains $n\lfloor n/2\rfloor$ $\mathrm{SU}(4)$ gates, which can be implemented with three CNOT gates each, and a dense width-$n$ QUOPS circuit of size $s$ contains $s/4$ CNOT gates (Section~\ref{ssec:quops_circuits}), which would give $s^{\star}\approx 6\eta\,n^{2}\approx 4.1\,n^{2}$. The smaller prefactor found in compilation is likely due to the costs of routing, since the three CNOT gates of an $\mathrm{SU}(4)$ gate share a single routing whereas each QUOPS CNOT acts on an independent random pair and is routed separately (although there is a competing effect: a single distant CNOT gate can be implemented with a CNOT ladder rather than a SWAP chain, which uses fewer CNOT gates).

\begin{table}[t]
\centering
\begin{tabular}{llllrr}
\toprule\toprule
Date & System & Source & QV & $s^{\star}$ & $Q_{\textrm{est}}$ \\
\midrule
2018-11 & Tokyo & Ref.~\cite{Cross2019-ku} (v1) & $2^{3}$ & 21 & 22 \\
2019-10 & Tokyo & Ref.~\cite{Cross2019-ku} (v2) & $2^{4}$ & 48 & 40 \\
2020-01 & Raleigh & announcement & $2^{5}$ & 62 & 62 \\
2020-08 & Montreal (Falcon r4) & announcement & $2^{6}$ & 96 & 90 \\
2020-12 & Montreal (Falcon r4) & announcement & $2^{7}$ & 117 & 122 \\
2022-04 & Prague (Falcon r10) & announcement & $2^{8}$ & 161 & 160 \\
2022-05 & Prague (Falcon r10) & announcement & $2^{9}$ & 191 & 202 \\
2025-08 & Pittsburgh (Heron r3) & announcement & $2^{10}$ & 240 & 250 \\
2025-08 & Pittsburgh (Heron r3) & announcement & $2^{11}$ & 277 & 302 \\
\bottomrule\bottomrule
\end{tabular}
\caption{\textbf{QUOPS scores estimated from IBM quantum volume results.} $s^{\star}$ is the compiled-gate-count match of Eq.~\eqref{eq:qv_ibm_matching} for $n = \log_2(\textrm{QV})$ and $Q_{\textrm{est}} = 2.5n^{2}$ is the smoothed value plotted in Fig.~\ref{fig:main:capability:historical}.\label{tab:qv_ibm}}
\end{table}

\subsection{Method 2: Inferring two-qubit gate fidelity}
\label{ssec:qv_quantinuum}

For Quantinuum's \texttt{H0}, \texttt{H1-1}, \texttt{H1-2}, \texttt{H2-1}, and \texttt{H2-2} processors we have the raw QV data: for each experiment, the number of qubits $n$, the measured heavy-output fraction $h$, and the mean ideal heavy-output probability $h_{\textrm{ideal}}$ of the sampled circuits~\cite{BaldwinQuantum2022}. These processors have all-to-all connectivity and compile $\mathrm{SU}(4)$ gates into arbitrary-angle $R_{zz}(\theta)$ gates rather than CNOT gates (Section~\ref{ssec:helios_system}), so gate-count matching against a CNOT-based compilation is not appropriate. Instead we infer an effective two-qubit gate fidelity from each QV experiment and convert it to a QUOPS score. The conversion proceeds in four steps.

\emph{(i) Circuit polarization and fidelity.} Inverting Eq.~\eqref{eq:qv_heavy_output_polarization} gives the polarization of the QV circuits, $\polarization = (h - 1/2)/(h_{\textrm{ideal}} - 1/2)$, and hence their process fidelity $F = \polarization + (1-\polarization)/2^{n}$ (Section~\ref{ssec:fidelity_definitions}). Here we use the exact $h_{\textrm{ideal}}$ of each experiment rather than the asymptotic value, which matters for the smallest $n$.

\emph{(ii) SPAM removal.} We attribute a fidelity of $0.995$ per qubit to state preparation and measurement~\cite{Pino2020, Moses_2023} and remove it, $F_{\textrm{gates}} = F/0.995^{\,n}$.

\emph{(iii) Fidelity per arbitrary-angle two-qubit gate.} We attribute the remaining infidelity to the two-qubit gates. A width-$n$ QV circuit compiled to Quantinuum's native gate set contains an effective number
\begin{equation}
    N_{\textrm{arb}}(n) = a\,\lfloor n/2 \rfloor\,(n-1), \qquad a = 2.8,
    \label{eq:qv_quantinuum_gate_count}
\end{equation}
of arbitrary-angle $R_{zz}(\theta)$ gates, and the fidelity per such gate is $f_{\textrm{arb}} = F_{\textrm{gates}}^{\,1/N_{\textrm{arb}}}$. 

\emph{(iv) Fidelity per maximally entangling gate.} A CNOT gate is implemented on these processors by a maximally entangling $R_{zz}(\pi/2)$ gate, whose error is larger than that of a typical arbitrary-angle gate because the error grows with the interaction angle. We take the infidelity of an average arbitrary-angle gate to be a fraction $b$ of the infidelity of the maximally entangling gate~\cite{BaldwinQuantum2022},
\begin{equation}
    1 - f_{2\textrm{q}} = \frac{1 - f_{\textrm{arb}}}{b}, \qquad b = \frac{0.65}{3\times 0.5} \approx 0.43 .
    \label{eq:qv_quantinuum_angle_scaling}
\end{equation}

Finally, a dense QUOPS circuit of size $s$ contains $s/4$ CNOT gates, each implemented directly (no routing is needed with all-to-all connectivity), and with single-qubit gate errors neglected its polarization is $\bar{\polarization}_{w,s} \approx f_{2\textrm{q}}^{\,s/4}$. Setting this equal to $1/\sqrt{e}$ gives the estimated QUOPS score
\begin{equation}
    Q_{\textrm{est}} = \frac{4\ln(1/\sqrt{e})}{\ln f_{2\textrm{q}}} = \frac{2}{\ln(1/f_{2\textrm{q}})} \approx \frac{2}{1 - f_{2\textrm{q}}} .
    \label{eq:qv_quantinuum_conversion}
\end{equation}
Table~\ref{tab:qv_quantinuum} lists the 23 QV experiments and the resulting estimates. Two consistency checks support the procedure. The inferred two-qubit infidelities ($1.4\times 10^{-3}$ for H2-2 in 2025, $2\times 10^{-3}$ for \htwoone in 2024) are close to the independently benchmarked two-qubit gate errors of these processors (Section~\ref{ssec:h2_system}). And the estimate for \htwoone from its April 2025 QV data ($1.1\times 10^{3}$) is close to the QUOPS score of $1320$ measured directly on \htwoone in 2026 (Table~\ref{tab:measured_extrapolated_quops}).

\begin{table}[t]
\centering
\begin{tabular}{llrrrrrr}
\toprule\toprule
System & Date & $n$ & $h$ & $h_{\textrm{ideal}}$ & $\polarization$ & $1-f_{2\textrm{q}}$ & $Q_{\textrm{est}}$ \\
\midrule
H0 & 2020-01-15 & 2 & 0.776 & 0.797 & 0.93 & $3.6\times 10^{-2}$ & 55 \\
H0 & 2020-01-15 & 3 & 0.833 & 0.863 & 0.92 & $2.5\times 10^{-2}$ & 78 \\
H0 & 2020-01-15 & 4 & 0.768 & 0.837 & 0.79 & $2.6\times 10^{-2}$ & 75 \\
H0 & 2020-06-20 & 6 & 0.730 & 0.852 & 0.65 & $2.1\times 10^{-2}$ & 93 \\
H1-1 & 2020-09-29 & 7 & 0.718 & 0.857 & 0.61 & $2.1\times 10^{-2}$ & 96 \\
H1-1 & 2021-03-02 & 9 & 0.733 & 0.852 & 0.66 & $9.4\times 10^{-3}$ & 212 \\
H1-1 & 2021-07-12 & 10 & 0.704 & 0.849 & 0.58 & $8.9\times 10^{-3}$ & 223 \\
H1-2 & 2021-12-17 & 11 & 0.698 & 0.849 & 0.57 & $8.4\times 10^{-3}$ & 236 \\
H1-2 & 2022-04-11 & 12 & 0.690 & 0.848 & 0.55 & $6.7\times 10^{-3}$ & 295 \\
H1-1 & 2022-09-20 & 13 & 0.693 & 0.847 & 0.56 & $6.0\times 10^{-3}$ & 335 \\
H1-1 & 2023-01-11 & 14 & 0.699 & 0.847 & 0.57 & $4.4\times 10^{-3}$ & 453 \\
H1-1 & 2023-01-18 & 15 & 0.691 & 0.847 & 0.55 & $4.4\times 10^{-3}$ & 455 \\
H1-1 & 2023-03-17 & 16 & 0.680 & 0.847 & 0.52 & $4.0\times 10^{-3}$ & 504 \\
H2-1 & 2023-04-18 & 16 & 0.682 & 0.847 & 0.53 & $3.9\times 10^{-3}$ & 518 \\
H1-1 & 2023-05-10 & 17 & 0.685 & 0.847 & 0.53 & $3.5\times 10^{-3}$ & 571 \\
H1-1 & 2023-05-24 & 18 & 0.686 & 0.847 & 0.54 & $2.9\times 10^{-3}$ & 698 \\
H1-1 & 2023-06-08 & 19 & 0.679 & 0.847 & 0.52 & $2.9\times 10^{-3}$ & 692 \\
H1-1 & 2024-04-03 & 20 & 0.692 & 0.847 & 0.55 & $2.1\times 10^{-3}$ & 940 \\
H2-1 & 2024-07-29 & 21 & 0.686 & 0.847 & 0.54 & $2.1\times 10^{-3}$ & 940 \\
H2-1 & 2025-04-03 & 22 & 0.685 & 0.847 & 0.53 & $1.8\times 10^{-3}$ & 1087 \\
H2-1 & 2025-04-22 & 23 & 0.681 & 0.847 & 0.52 & $1.8\times 10^{-3}$ & 1101 \\
H2-2 & 2025-09-02 & 24 & 0.694 & 0.847 & 0.56 & $1.4\times 10^{-3}$ & 1456 \\
H2-2 & 2025-09-04 & 25 & 0.688 & 0.847 & 0.54 & $1.4\times 10^{-3}$ & 1443 \\
\bottomrule\bottomrule
\end{tabular}
\caption{\textbf{QUOPS scores estimated from Quantinuum quantum volume data.} For each QV experiment (quantum volume $2^{n}$), $h$ is the measured heavy-output fraction, $h_{\textrm{ideal}}$ the mean ideal heavy-output probability, $\polarization$ the inferred circuit polarization, $1-f_{2\textrm{q}}$ the inferred infidelity of a maximally entangling two-qubit gate [Eq.~\eqref{eq:qv_quantinuum_angle_scaling}], and $Q_{\textrm{est}}$ the estimated QUOPS score [Eq.~\eqref{eq:qv_quantinuum_conversion}].}
\label{tab:qv_quantinuum}
\end{table}

\subsection{Trend lines}
\label{ssec:qv_trends}
The dashed lines in Fig.~\ref{fig:main:capability:historical} are least-squares fits of $\ln Q$ to a linear function of time.  The Quantinuum fit uses the 23 QV-derived estimates of Table~\ref{tab:qv_quantinuum} together with the direct 2026 measurements on \htwoone ($Q = 1320$) and \helios ($Q = 1504$). The fitted doubling times are $2.1$ years for IBM and $1.4$ years for Quantinuum.

\subsection{Limitations}
\label{ssec:qv_limitations}

Both conversions rest on simplifying assumptions: a global depolarizing noise model, two-qubit gates as the sole error source (with SPAM removed for Quantinuum), exponential decay of polarization with gate count, and fixed conversion constants (the compiled-gate-count ratio for IBM, and $a$, $b$ and the SPAM fidelity for Quantinuum). The IBM analysis assumes a square-grid architecture and a specific compiler, whereas the processors have heavy-hexagonal connectivity and were benchmarked with IBM's own compilation. A reported quantum volume is also a lower bound on the width at which the QV test would pass, and it is a binary pass/fail outcome rather than a point on a capability boundary, so QV-derived estimates are coarser than the Quantinuum heavy-output data, which resolve the margin by which each test was passed. We therefore expect individual estimates to be accurate to within a factor of about two, which is sufficient to identify multi-year trends but not to compare processors at a single point in time. Direct QUOPS measurements, which do not rely on any of these assumptions, are shown with filled markers in Fig.~\ref{fig:main:capability:historical}.

\end{document}